\documentclass[iop,apj,useAMS,usenatbib]{emulateapj-rtx4}

\usepackage{graphicx}
\usepackage{lscape}
\usepackage{amsmath}
\usepackage{longtable}
\usepackage[usenames,dvipsnames]{color}

\newcommand{\herschel}{{\it Herschel}}
\newcommand{\hermes}{HerMES}
\newcommand{\atlas}{H-ATLAS}
\newcommand{\spitzer}{{\it Spitzer}}

\newcommand{\hst}{{\it HST}}
\newcommand{\myr}{${\rm M_{\sun}yr^{-1}}$}
\newcommand{\lsun}{${\rm L_{\sun}}$}

\newcommand{\mumax}{$\mu_{\rm max}$}
\newcommand{\bootes}{Bo\"{o}tes}
\newcommand{\eyelash}{SMMJ2135$-$0102}

\newcommand{\cii}{[\ion{C}{2}]158\micron}
\newcommand{\oiii}{[\ion{O}{3}]52\micron}
\newcommand{\oi}{[\ion{O}{1}]63\micron}
\newcommand{\aco}{CO($J$=1$\to$0)}
\newcommand{\bco}{CO($J$=2$\to$1)}
\newcommand{\cco}{CO($J$=3$\to$2)}
\newcommand{\dco}{CO($J$=4$\to$3)}
\newcommand{\eco}{CO($J$=5$\to$4)}

\def\mnras{MNRAS}
\def\aap{A\&A}
\def\aaps{A\&AS}
\def\aj{AJ}
\def\araa{ARA\&A}
\def\apjl{ApJ}
\def\apj{ApJ}
\def\apjs{ApJS}
\def\pasp{PASP}
\def\nat{Nature}
\def\pasj{PASJ}

\begin{document}

\shorttitle{Candidate lensed submillimeter galaxies in HerMES}
\shortauthors{J.\,L.\ Wardlow et al.}

\title{HerMES: Candidate Gravitationally Lensed Galaxies and Lensing Statistics at Submillimeter Wavelengths}
\author{Julie L. Wardlow\altaffilmark{1$\dag$},
Asantha Cooray\altaffilmark{1,2},
Francesco De Bernardis\altaffilmark{1},
A.~Amblard\altaffilmark{3},
V.~Arumugam\altaffilmark{4},
H.~Aussel\altaffilmark{5},
A.J.~Baker\altaffilmark{6},
M.~B{\'e}thermin\altaffilmark{5,7},
R.~Blundell\altaffilmark{8},
J.~Bock\altaffilmark{2,9},
A.~Boselli\altaffilmark{10},
C.~Bridge\altaffilmark{2},
V.~Buat\altaffilmark{10},
D.~Burgarella\altaffilmark{10},
R.~S.~Bussmann\altaffilmark{8},
A.~Cabrera-Lavers\altaffilmark{11,12,13},
J.~Calanog\altaffilmark{1},
J.M.~Carpenter\altaffilmark{2},
C.M.~Casey\altaffilmark{14},
N.~Castro-Rodr{\'\i}guez\altaffilmark{11,12},
A.~Cava\altaffilmark{15},
P.~Chanial\altaffilmark{5},
E.~Chapin\altaffilmark{16},
S.C.~Chapman\altaffilmark{17},
D.L.~Clements\altaffilmark{18},
A.~Conley\altaffilmark{19},
P.~Cox\altaffilmark{20},
C.D.~Dowell\altaffilmark{2,9},
S.~Dye\altaffilmark{21},
S.~Eales\altaffilmark{22},
D.~Farrah\altaffilmark{23,24},
P.~Ferrero\altaffilmark{11,12},
A.~Franceschini\altaffilmark{25},
D.T.~Frayer\altaffilmark{26},
C.~Frazer\altaffilmark{1},
Hai~Fu\altaffilmark{1},
R.~Gavazzi\altaffilmark{27},
J.~Glenn\altaffilmark{28,19},
E.A.~Gonz\'alez~Solares\altaffilmark{17},
M.~Griffin\altaffilmark{22},
M.A.~Gurwell\altaffilmark{8},
A.I.~Harris\altaffilmark{29},
E.~Hatziminaoglou\altaffilmark{30},
R.~Hopwood\altaffilmark{18},
A.~Hyde\altaffilmark{18},
E.~Ibar\altaffilmark{31},
R.J.~Ivison\altaffilmark{31,4},
S.~Kim\altaffilmark{1},
G.~Lagache\altaffilmark{7},
L.~Levenson\altaffilmark{2,9},
L.~Marchetti\altaffilmark{25},
G.~Marsden\altaffilmark{32},
P.~Martinez-Navajas\altaffilmark{11,12},
M.~Negrello\altaffilmark{25},
R.~Neri\altaffilmark{20},
H.T.~Nguyen\altaffilmark{9,2},
B.~O'Halloran\altaffilmark{18},
S.J.~Oliver\altaffilmark{23},
A.~Omont\altaffilmark{27},
M.J.~Page\altaffilmark{33},
P.~Panuzzo\altaffilmark{5},
A.~Papageorgiou\altaffilmark{22},
C.P.~Pearson\altaffilmark{34,35},
I.~P{\'e}rez-Fournon\altaffilmark{11,12},
M.~Pohlen\altaffilmark{22},
D.~Riechers\altaffilmark{2},
D.~Rigopoulou\altaffilmark{34,36},
I.G.~Roseboom\altaffilmark{23,4},
M.~Rowan-Robinson\altaffilmark{18},
B.~Schulz\altaffilmark{2,37},
D.~Scott\altaffilmark{32},
N.~Scoville\altaffilmark{2},
N.~Seymour\altaffilmark{38,33},
D.L.~Shupe\altaffilmark{2,37},
A.J.~Smith\altaffilmark{23},
A.~Streblyanska\altaffilmark{11,12},
A.~Strom\altaffilmark{17},
M.~Symeonidis\altaffilmark{33},
M.~Trichas\altaffilmark{8},
M.~Vaccari\altaffilmark{25,39},
J.D.~Vieira\altaffilmark{2},
M.~Viero\altaffilmark{2},
L.~Wang\altaffilmark{23},
C.K.~Xu\altaffilmark{2,37},
L.~Yan\altaffilmark{2},
M.~Zemcov\altaffilmark{2,9}}
\altaffiltext{1}{Dept. of Physics \& Astronomy, University of California, Irvine, CA 92697}
\altaffiltext{2}{California Institute of Technology, 1200 E. California Blvd., Pasadena, CA 91125}
\altaffiltext{3}{NASA, Ames Research Center, Moffett Field, CA 94035}
\altaffiltext{4}{Institute for Astronomy, University of Edinburgh, Royal Observatory, Blackford Hill, Edinburgh EH9 3HJ, UK}
\altaffiltext{5}{Laboratoire AIM-Paris-Saclay, CEA/DSM/Irfu - CNRS - Universit\'e Paris Diderot, CE-Saclay, pt courrier 131, F-91191 Gif-sur-Yvette, France}
\altaffiltext{6}{Department of Physics and Astronomy, Rutgers, The State University of New Jersey, 136 Frelinghuysen Rd, Piscataway, NJ 08854}
\altaffiltext{7}{Institut d'Astrophysique Spatiale (IAS), b\^atiment 121, Universit\'e Paris-Sud 11 and CNRS (UMR 8617), 91405 Orsay, France}
\altaffiltext{8}{Harvard-Smithsonian Center for Astrophysics, 60 Garden Street, Cambridge, MA 02138}
\altaffiltext{9}{Jet Propulsion Laboratory, 4800 Oak Grove Drive, Pasadena, CA 91109}
\altaffiltext{10}{Laboratoire d'Astrophysique de Marseille - LAM, Universit\'e Aix-Marseille \& CNRS, UMR7326, 38 rue F. Joliot-Curie, 13388 Marseille Cedex 13, France}
\altaffiltext{11}{Instituto de Astrof{\'\i}sica de Canarias (IAC), E-38200 La Laguna, Tenerife, Spain}
\altaffiltext{12}{Departamento de Astrof{\'\i}sica, Universidad de La Laguna (ULL), E-38205 La Laguna, Tenerife, Spain}
\altaffiltext{13}{GTC Project, E-38205 La Laguna, Tenerife, Spain}
\altaffiltext{14}{Institute for Astronomy, University of Hawaii, 2680 Woodlawn Drive, Honolulu, HI 96822}
\altaffiltext{15}{Departamento de Astrof\'isica, Facultad de CC. F\'isicas, Universidad Complutense de Madrid, E-28040 Madrid, Spain}
\altaffiltext{16}{Herschel Science Centre, European Space Astronomy Centre, Villanueva de la Ca\~nada, 28691 Madrid, Spain}
\altaffiltext{17}{Institute of Astronomy, University of Cambridge, Madingley Road, Cambridge CB3 0HA, UK}
\altaffiltext{18}{Astrophysics Group, Imperial College London, Blackett Laboratory, Prince Consort Road, London SW7 2AZ, UK}
\altaffiltext{19}{Center for Astrophysics and Space Astronomy 389-UCB, University of Colorado, Boulder, CO 80309}
\altaffiltext{20}{Institut de RadioAstronomie Millim\'etrique, 300 Rue de la Piscine, Domaine Universitaire, 38406 Saint Martin d'H\`eres, France}
\altaffiltext{21}{School of Physics and Astronomy, University of Nottingham, NG7 2RD, UK}
\altaffiltext{22}{School of Physics and Astronomy, Cardiff University, Queens Buildings, The Parade, Cardiff CF24 3AA, UK}
\altaffiltext{23}{Astronomy Centre, Dept. of Physics \& Astronomy, University of Sussex, Brighton BN1 9QH, UK}
\altaffiltext{24}{Department of Physics, Virginia Tech, Blacksburg, VA 24061}
\altaffiltext{25}{Dipartimento di Astronomia, Universit\`{a} di Padova, vicolo Osservatorio, 3, 35122 Padova, Italy}
\altaffiltext{26}{NRAO, PO Box 2, Green Bank, WV 24944}
\altaffiltext{27}{Institut d'Astrophysique de Paris, UMR 7095, CNRS, UPMC Univ. Paris 06, 98bis boulevard Arago, F-75014 Paris, France}
\altaffiltext{28}{Dept. of Astrophysical and Planetary Sciences, CASA 389-UCB, University of Colorado, Boulder, CO 80309}
\altaffiltext{29}{Department of Astronomy, University of Maryland, College Park, MD 20742-2421}
\altaffiltext{30}{ESO, Karl-Schwarzschild-Str. 2, 85748 Garching bei M\"unchen, Germany}
\altaffiltext{31}{UK Astronomy Technology Centre, Royal Observatory, Blackford Hill, Edinburgh EH9 3HJ, UK}
\altaffiltext{32}{Department of Physics \& Astronomy, University of British Columbia, 6224 Agricultural Road, Vancouver, BC V6T~1Z1, Canada}
\altaffiltext{33}{Mullard Space Science Laboratory, University College London, Holmbury St. Mary, Dorking, Surrey RH5 6NT, UK}
\altaffiltext{34}{RAL Space, Rutherford Appleton Laboratory, Chilton, Didcot, Oxfordshire OX11 0QX, UK}
\altaffiltext{35}{Department of Physical Sciences, The Open University, Milton Keynes MK7 6AA,UK}
\altaffiltext{36}{Department of Astrophysics, Denys Wilkinson Building, University of Oxford, Keble Road, Oxford OX1 3RH, UK}
\altaffiltext{37}{Infrared Processing and Analysis Center, MS 100-22, California Institute of Technology, JPL, Pasadena, CA 91125}
\altaffiltext{38}{CSIRO Astronomy \& Space Science, PO Box 76, Epping, NSW 1710, Australia}
\altaffiltext{39}{Astrophysics Group, Physics Department, University of the Western Cape, Private Bag X17, 7535, Bellville, Cape Town, South Africa}

\label{firstpage}

\begin{abstract}
  We present a list of 13 candidate gravitationally lensed
  submillimeter galaxies (SMGs) from 95\,deg$^2$ of the {\it Herschel}
  Multi-tiered Extragalactic Survey, a surface density of $0.14 \pm
  0.04\,\rm{deg}^{-2}$. The selected sources have 500\,\micron\ flux
  densities ($S_{500}$) greater than 100\,mJy.  Gravitational lensing
  is confirmed by follow-up observations in 9 of the 13 systems
  ($70\%$), and the lensing status of the four remaining sources is
  undetermined.
  We also present a supplementary sample of 29
  ($0.31\pm0.06\,\rm{deg}^{-2}$) gravitationally lensed SMG
  candidates with $S_{500}=80$--100\,mJy, which are expected to
  contain a higher fraction of interlopers than the primary
  candidates.
  The number counts of the candidate lensed galaxies are consistent
  with a simple statistical model of the lensing rate, which uses a
  foreground matter distribution, the intrinsic SMG number counts, and
  an assumed SMG redshift distribution.  The model predicts that
  32--74\% of our $S_{500}\ge100$\,mJy candidates are strongly
  gravitationally lensed ($\mu\ge2$), with the
  brightest sources being the most robust; this is consistent with the
  observational data.
  Our statistical model also predicts that, on average, lensed
  galaxies with $S_{500}=100$\,mJy are magnified by factors of
  $\sim9$, with apparently brighter galaxies having
  progressively higher average magnification, due to the shape of
    the intrinsic number counts. 65\% of the sources are
  expected to have intrinsic 500\,\micron\ flux densities less than
  $30$\,mJy.
  Thus, samples of strongly gravitationally lensed SMGs, such as those
  presented here, probe below the nominal \herschel\ detection limit at
  500\,\micron. They are good targets for the detailed study
  of the physical conditions in distant dusty, star-forming galaxies,
 due to the lensing magnification, which can lead to spatial
    resolutions of $\sim0.01\arcsec$ in the source plane.
\end{abstract}


\section{Introduction}

Gravitational lensing increases the angular size and integrated flux
of affected sources.  It is exploited to investigate the mass
distribution of the foreground lensing structures as well as the
properties of the background lensed galaxies \citep[see reviews
by][]{Bartelmann10, Treu10R}.

The magnification provided by gravitational lensing makes it an
effective tool for identifying and studying intrinsically faint and
typically distant galaxies \citep[e.g.][]{Stark07, Richard08,
  Richard11}.  The flux boost from lensing yields an improved
detection and the associated spatial enhancement increases the ability
to investigate the internal structure of distant galaxies to levels
otherwise unattainable with the current generation of instrumentation
\citep[e.g.][]{Riechers08, Swinbank10, Swinbank11, Gladders12}.
Furthermore, gravitational lensing probes the total mass of the
foreground deflectors, including the relative content of dark and
luminous mass. In combination with dynamical studies, lensing mass
reconstruction allows one to obtain the density profile of the dark
matter in individual lensing galaxies down to $\sim10$\,kpc scales
\citep[e.g.][]{MiraldaEscude95, Dalal02, Metcalf02, Rusin05, Treu04}.

The statistics of galaxy-galaxy lensing events, particularly the
abundance of strongly lensed sources from a sample of galaxies with a
known redshift distribution, can be used to constrain the global
cosmological parameters, such as the cosmological constant. For
example, the 28 lensed quasars in the Sloan Digital Sky Survey (SDSS)
Quasar Lens Search (SQLS) led to an estimate of $\Omega_\Lambda=0.74
\pm 0.17$, assuming a spatially flat Universe \citep[][]{Oguri12}.  A
systematic search for large samples of lensed galaxies could provide
constraints on cosmological parameters that are competitive with other
cosmological probes \citep[e.g.][]{Marshall09}.

The past decade has seen dedicated optical and radio imaging and
spectroscopic surveys for background sources strongly lensed by single
foreground galaxies (so-called strong galaxy-galaxy lensing).  These
include the Sloan Lens ACS Survey (SLACS; \citealt{Bolton06}), SQLS
\citep{Oguri06}, Strong Lens Legacy Survey (SL2S;
\citealt{More12,Ruff11}), and the BOSS Emission Line Lens Survey
(BELLS, \citealt{Brownstein12}) in the optical and the Cosmic Lens All
Sky Survey (CLASS; \citealt{Browne03}).  Unfortunately, in order to
convert the lensing rate into a test of cosmological models,
cosmological studies require that the selection function be simple and
easy to describe.  Furthermore, the initial selection of candidates is
often inefficient or the resulting targets are biased to low
redshifts. For example, at radio wavelengths $\sim0.2\%$ of the
initial targets of CLASS are lensed \citep{Browne03}, although this
can be improved to $\sim2\%$ by including SDSS information in the
initial target selection \citep{Jackson07}.  SDSS-based lensing
searches are more efficient, but are limited to lensed galaxies with
$z\la0.7$ due to the survey depth \citep{Treu04, Koopmans06,
  Bolton08a, Bolton08b, Auger10b, Brownstein12}.

It has long been proposed that large samples of gravitationally
lensed, high-redshift galaxies can be efficiently selected by
searching for bright sources in wide area blank-field submillimeter
surveys \citep[e.g.][]{Blain96, Perrotta02, Negrello07, Paciga09}.
The unique advantage of selecting bright submillimeter sources as
lensed galaxy candidates lies in the efficiency of this technique and
the low contamination of samples.  The number counts of distant
submillimeter galaxies (SMGs) have intrinsically steep slopes at
bright flux densities \citep[e.g.][]{Barger99, Blain99, Coppin06,
  Scott06, Patanchon09, Weiss09, Glenn10, Oliver10, Clements10}.
Thus, a population of apparently bright 500\,\micron\ sources is
expected to be dominated by gravitationally lensed sources, local
late-type galaxies and flat spectrum radio quasars \citep[or
blazars;][]{Negrello07}.  The two contaminants can be easily removed
by cross-identifying the bright submillimeter sources with shallow
all-sky optical and radio surveys \citep[e.g.][]{Negrello10}.

With the launch of the \herschel\ Space
Observatory\footnote{\herschel\ is an ESA space observatory with
  science instruments provided by European-led Principal Investigator
  consortia and with important participation from NASA.}
\citep{Pilbratt10} blank-field submillimeter surveys of
hundreds of square degrees are being undertaken for the
first time \citep[e.g.][]{Oliver12, Eales10}, making systematic
searches for strong galaxy-galaxy lensing practical.  The first
systematic \herschel\ survey for lensed SMGs was undertaken with
\herschel-Astrophysical Terahertz Large Area Survey (H-ATLAS;
\cite{Eales10}) Science Demonstration Phase (SDP) data. In that study
\citet{Negrello10} identified a total of 12 sources with 500\,\micron\
flux density, $S_{500}$,$>100$\,mJy in 14.4\,deg$^2$. Seven of the
sources are associated with $z<0.1$ late-type galaxies or radio-loud
blazars and the five remaining sources were confirmed as systems
undergoing strong galaxy-galaxy lensing. Thus, \citet{Negrello10}
showed that one can reach $\sim100\%$ efficiency in the identification
of strongly lensed galaxies, simply based on an observed submillimeter
flux cut and existing all-sky survey data.  The five \atlas\ lensed
systems have since been subject to extensive analysis and follow-up
effort \citep{Frayer10, Hopwood11, Lupu10, Omont11, Valtchanov11}.

Additional gravitationally lensed SMGs have since been identified in
other extragalactic surveys with the Spectral and Photometric
Imaging Receiver (SPIRE; \citealt{Griffin10, Swinyard10}) on \herschel.
These include the $z=2.957$ \hermes\ source, HLock01
\citep[Section~\ref{sec:l1}; ][]{Conley11, Gavazzi11, Riechers11,
  Scott11}, and the \atlas\ galaxies ID141 ($z=4.24$; \citealt{Cox11}),
and HATLAS12--00 ($z=3.259$; \citealt{Fu12}).  In addition,
\citet{Harris12} used \aco\ linewidths and integrated luminosities to
show that 11 lensed galaxies from \atlas\ are, on average, magnified by
factors of 10, with a range of $\sim 3$--20 for individual sources.
\citet{GonzalezNuevo12} recently showed that \herschel-SPIRE data can
be used to identify fainter lensed galaxies, although the selection
process is necessarily more involved, and will not be discussed further
here.  Outside of \herschel, a recent South Pole Telescope (SPT;
\citealt{Carlstrom11}) survey of the cosmological millimeter background
identified 13 discrete sources in
  87~deg$^2$ that are consistent with gravitationally lensed galaxies
at high redshift. These sources are 
  detected at $>4.5\sigma$ ($\sim 15$~mJy) at 1.4~mm, have 1.4 to
  2.0~mm spectral indices consistent with thermal dust emission, and
  were not detected by IRAS \citep{Vieira10}.

In order to build-up large samples of strongly lensed SMGs it is
clearly necessary to test the supposition that high efficiency can be
reached with only a flux cut and existing shallow optical and radio
data.  In this paper, we present a systematic survey for a sample of
strongly lensed SMGs in $\sim95\,{\rm deg}^2$ of the \herschel\
Multi-tiered Extragalactic
Survey\footnote{http://hermes.sussex.ac.uk/} \citep[\hermes;][]{
  Oliver12} data.  In the $\sim95\,\rm{deg}^2$ of \hermes\ blank-field
data 13 principal and 29 supplementary candidate lensed SMGs are
identified.  We describe a simple statistical lensing model,
consisting of a foreground matter distribution and background SMGs,
and show that the observed lensed number counts are consistent with
the model prediction.

We have begun a follow-up multi-wavelength campaign to further
understand the nature of the candidate lensed SMGs. Detailed
observations of nine of the sources, are presented, and these
establish that they are all gravitationally lensed.  Follow-up data is
available for four of the 29 secondary candidates, of which only one
is confirmed to be lensed; one is an intrinsically luminous galaxy;
one is a blend of multiple sources in the \herschel\ beam; and the
nature of the other is unclear.

In this paper we focus on the ensemble properties of the primary
lensed candidate list and consider statistics such as the lensing
rate, number counts, and submillimeter color and redshift
distributions of the lensed SMGs.  In order to facilitate community
participation in the follow-up observations we also present the
catalogs of our primary and secondary candidate lensed sources.
Future publications will present detailed analysis of individual
systems, including lensing mass models and properties of both the
foreground and background galaxies, similar to the detailed
presentation in \citet{Fu12}.  The paper is organized as follows: the
selection of the candidate gravitationally lensed galaxies is
described in Section~\ref{sec:sample}, and their basic properties are
discussed in Section~\ref{sec:prop}.  In Section~\ref{sec:model} we
present a simple statistical model of the lensing rate and discuss the
model predictions for the population of strongly lensed 500-\micron\
selected sources. Follow-up data are described in
Section~\ref{sec:followup} and lensed candidates are discussed on a
source-by-source basis in Section~\ref{sec:indi}.  Individual
supplementary candidate gravitationally lensed galaxies are discussed
in Appendix~\ref{app:sources}.  Throughout this paper we use J2000
coordinates and $\Lambda$CDM cosmology with $\Omega_{\rm M}=0.27$,
$\Omega_{\Lambda}=0.73$ and $H_{0}=71\,{\rm km\,s^{-1}\,Mpc^
  {-1}}$. All photometry is on the AB magnitude system where $23.9\
{\rm m_ {AB}}=1\,\mu{\rm Jy}$.  For the purposes of our analysis we
consider ``strong'' lensing as lensing in which the magnification
factor, $\mu$, is $\ge2$.

\section{Identification of Candidate Lensed Galaxies}
\label{sec:sample}

\begin{deluxetable*}{lccc|cc|cc|cc|cc}
\tablewidth{0pt}
\tablecaption{Sources with $S_{500}\ge80$mJy in \hermes\ blank-fields.
 \label{tab:fields}}
\startdata
\hline\hline
\multicolumn{4}{c}{} & \multicolumn{2}{|c}{Lens candidates} & \multicolumn{2}{|c}{Blazars} & \multicolumn{2}{|c}{Local spirals} & \multicolumn{2}{|c}{Faint candidates}\\
\multicolumn{4}{c}{} & \multicolumn{2}{|c}{{\tiny ($S_{500}\ge100$\,mJy)}} & \multicolumn{2}{|c}{{\tiny ($S_{500}\ge80$\,mJy)}} & \multicolumn{2}{|c}{{\tiny ($S_{500}\ge80$\,mJy)}} & \multicolumn{2}{|c}{{\tiny ($S_{500}=80$--100\,mJy)}}\\
Field$^a$ & RA$^a$ & Dec$^b$ &Area$^c$ & ${\rm N}$ & Density & ${\rm N}$ & Density & ${\rm N}$ & Density & ${\rm N}$ & Density\\
 & & & (${\rm deg}^{2}$) &  & (${\rm deg}^{-2}$) &   & (${\rm deg}^{-2}$)&   & (${\rm deg}^{-2}$)& & (${\rm deg}^{-2}$)\\
\hline
ELAIS-S1 SWIRE  & $00^{\rm h}35^{\rm m}03^{\rm s}$ & $-43\degr34\arcmin42\arcsec$ & 8.6  & 0 & $<0.1 $      & 1 & $0.1\pm0.1 $ & 4  & $0.5\pm0.2 $ &  1 & $0.1\pm0.1 $    \\
XMM-LSS SWIRE   & $02^{\rm h}20^{\rm m}36^{\rm s}$ & $-04\degr31\arcmin27\arcsec$ & 21.6 & 3 & $0.1\pm0.1 $ & 0 & $<0.05$      & 16 & $0.7\pm0.2 $ & 10 & $0.5\pm0.2 $    \\
CDFS SWIRE      & $03^{\rm h}32^{\rm m}05^{\rm s}$ & $-28\degr16\arcmin35\arcsec$ & 12.9 & 3 & $0.2\pm0.1 $ & 1 & $0.1\pm0.1 $ & 6  & $0.5\pm0.2 $ &  3 & $0.2\pm0.1 $    \\
ADFS            & $04^{\rm h}43^{\rm m}29^{\rm s}$ & $-53\degr51\arcmin09\arcsec$ & 8.6  & 0 & $<0.1 $      & 0 & $<0.1 $      & 7  & $0.8\pm0.3 $ &  2 & $0.2\pm0.2 $    \\
COSMOS HerMES   & $10^{\rm h}00^{\rm m}28^{\rm s}$ & $+02\degr12\arcmin55\arcsec$ & 3.3  & 0 & $<0.3 $      & 0 & $<0.3 $      & 2  & $0.6\pm0.4 $ &  0 & $<0.3 $         \\
Lockman SWIRE   & $10^{\rm h}48^{\rm m}00^{\rm s}$ & $+58\degr09\arcmin02\arcsec$ & 18.2 & 4 & $0.2\pm0.1 $ & 0 & $<0.1$       & 16 & $0.9\pm0.2 $ &  7 & $0.4\pm0.1 $    \\
EGS HerMES      & $14^{\rm h}20^{\rm m}19^{\rm s}$ & $+52\degr48\arcmin56\arcsec$ & 3.1  & 0 & $<0.3 $      & 0 & $<0.3 $      & 0  & $<0.3 $      &  1 & $0.3\pm0.3 $    \\
\bootes\ NDWFS  & $14^{\rm h}32^{\rm m}45^{\rm s}$ & $+34\degr10\arcmin10\arcsec$ & 11.3 & 3 & $0.3\pm0.2 $ & 0 & $<0.1 $      & 12 & $1.1\pm0.3 $ &  2 & $0.2\pm0.2 $    \\
FLS             & $17^{\rm h}15^{\rm m}52^{\rm s}$ & $+59\degr23\arcmin15\arcsec$ & 7.3  & 0 & $<0.1 $      & 0 & $<0.1 $      & 12 & $1.6\pm0.5 $ &  3 & $0.4\pm0.2 $    \\
\hline
Total           &        \nodata                &          \nodata             & 94.8 &13 & $0.14\pm0.04$& 2 & $0.02\pm0.01$& 75 & $0.79\pm0.09$& 29 & $0.31\pm0.06$
\enddata
\tablecomments{
$^a$~Field names correspond to those in \citet{Oliver12}.
$^b$~Coordinates are for the center of the \hermes\ field of view.
$^c$~The total area of pixels with any 500\,$\mu$m coverage; same as $\Omega_{\rm max}$ in \citet{Oliver12}. This area is larger than the nominal HerMES coverage due to turn-arounds and exact scan designs.
}
\end{deluxetable*}

Candidate strongly lensed galaxies are selected from \hermes\
blank-field data\footnote{Publicly available \hermes\ data can be
  retrieved from http://hedam.oamp.fr/HerMES/} (see \citealt{Oliver12}
for a full description of the \hermes\ survey). The total area
considered for the candidate selection is $94.8\,\rm{deg^{2}}$ in nine
independent fields (see Table\,\ref{tab:fields}).  We employ the
\hermes\ SPIRE imaging and photometry at 250, 350, and 500\,\micron\
in these fields.  Sources are selected from \hermes\ catalogs
(\citealt{Smith12}; Wang et al.\ in preparation), which are extracted
from the HerMES maps (\citealt{Levenson10}).

We now briefly summarize details of the source detection and
extraction procedure.  Source detection is performed by {\sc
  StarFinder} \citep{Diolaiti00} at 250\,\micron. {\sc StarFinder}
models the data as the summation of beam-smoothed point sources and
iteratively detects, fits, and removes sources with decreasing
brightnesses.  {\sc StarFinder} was designed to detect point sources
in crowded fields, which results in a program that is good at
deblending sources that are close together.  This is important for our
purposes because the blending of multiple sources in the SPIRE beam
has the potential to mimic a single bright source, although we show in
Section~\ref{sec:blend} that blending is not a concern for our
sample. The opposite effect -- whereby lensed galaxies with large
separations between components are mistaken for blended sources -- is
not a significant issue for this work. This is because we only
consider blank-field data, and not the regions around massive galaxy
clusters. As such, most of the candidates are expected to be
galaxy-galaxy lenses (cf.\ HLock01), for which separations
$\ga5\arcsec$ (significantly smaller than the SPIRE beam: 18, 25 and
36\arcsec\ full-width at half maximum (FWHM) at 250, 350 and 500\,\micron, respectively) are rare
\citep{Treu10}.

The 250, 350 and 500\,\micron\ flux densities of {\sc StarFinder}
250-\micron\ selected sources are extracted using the \hermes\ {\sc
  XID} pipeline \citep{Roseboom10,Roseboom12}.  {\sc XID} allocates
flux to sources on the basis of positional priors, which in this case
are provided from {\sc StarFinder} at 250\,\micron.  Therefore, the
source extraction algorithm uses the positional information at
250\,\micron, where the point spread function (PSF) is 18\arcsec\
FWHM, to deblend sources at 350 and 500\,\micron.  The use of the
250\,\micron\ positions as priors is not expected to bias our results
against red, or high-redshift, galaxies because we are interested in
the apparently brightest sources. For example, an Arp~220-like galaxy
with $S_{500}=80$\,mJy at $z=6$ will have $S_{250}\sim30$\,mJy and be
detected in the 250\,\micron\ catalogs. Furthermore, the final step of
the {\sc XID} algorithm is to extract sources from the residual maps,
so there is no a priori requirement for a 250\,\micron\ detection for
inclusion in the catalog.  Extended sources, such as local late-type
galaxies, are fragmented into multiple components by the {\sc
  StarFinder+XID} process outlined above. Therefore, for sources that
are extended in the SPIRE beam, we make use of \hermes\ {\sc
  SUSSEXtractor} \citep{Smith12} flux densities, which are measured in
large apertures.

The selection of candidate gravitationally lensed galaxies is
performed at 500\,\micron, which is the SPIRE wavelength with the
fewest expected contaminants \citep[e.g.][]{Negrello07, Negrello10}.
Sources that are bright at 500\,\micron\ are typically either local
($z\la0.1$) late-type galaxies \citep[e.g.][]{Dunne00a, Serjeant05},
blazars \citep[e.g.][]{DeZotti05, GonzalezNuevo10}, or gravitationally
lensed SMGs \citep[see also][]{Negrello07, Negrello10}. Blazars are
considered contaminants because the submillimeter emission is
dominated by synchrotron radiation primarily from the radio jets.
There may also be a contribution from intrinsically luminous SMGs;
this contribution is strongly dependent on the 500\,\micron\ selection
limit (Section~\ref{sec:modpredict})

We begin by identifying all sources with $S_{500}\ge80$\,mJy in the
{\sc StarFinder+XID} catalogs.  At $z\ga1$ this flux density cut is
equivalent to selecting only the most apparently far-infrared luminous
sources (with far-infrared luminosity, $L_{\rm IR}$,
$\ga6\times10^{12}{\rm L_{\sun}}$ for an Arp220 spectral energy
distribution (SED), or $L_{\rm IR}\ga10^{13}{\rm L_{\sun}}$ for an M82
SED or optically-thin modified blackbody with $T_{\rm D}=35$\,K and
$\beta=1.5$).  Thus, if they are not amplified by gravitational
lensing, these sources are undergoing some of the most extreme growth
in the Universe (with star-formation rate, SFR $\ga1000\,{\rm
  M_{\sun}yr^{-1}}$; \citealt{Kennicutt98}).

\begin{figure}
   \centering
\includegraphics[width=8.5cm]{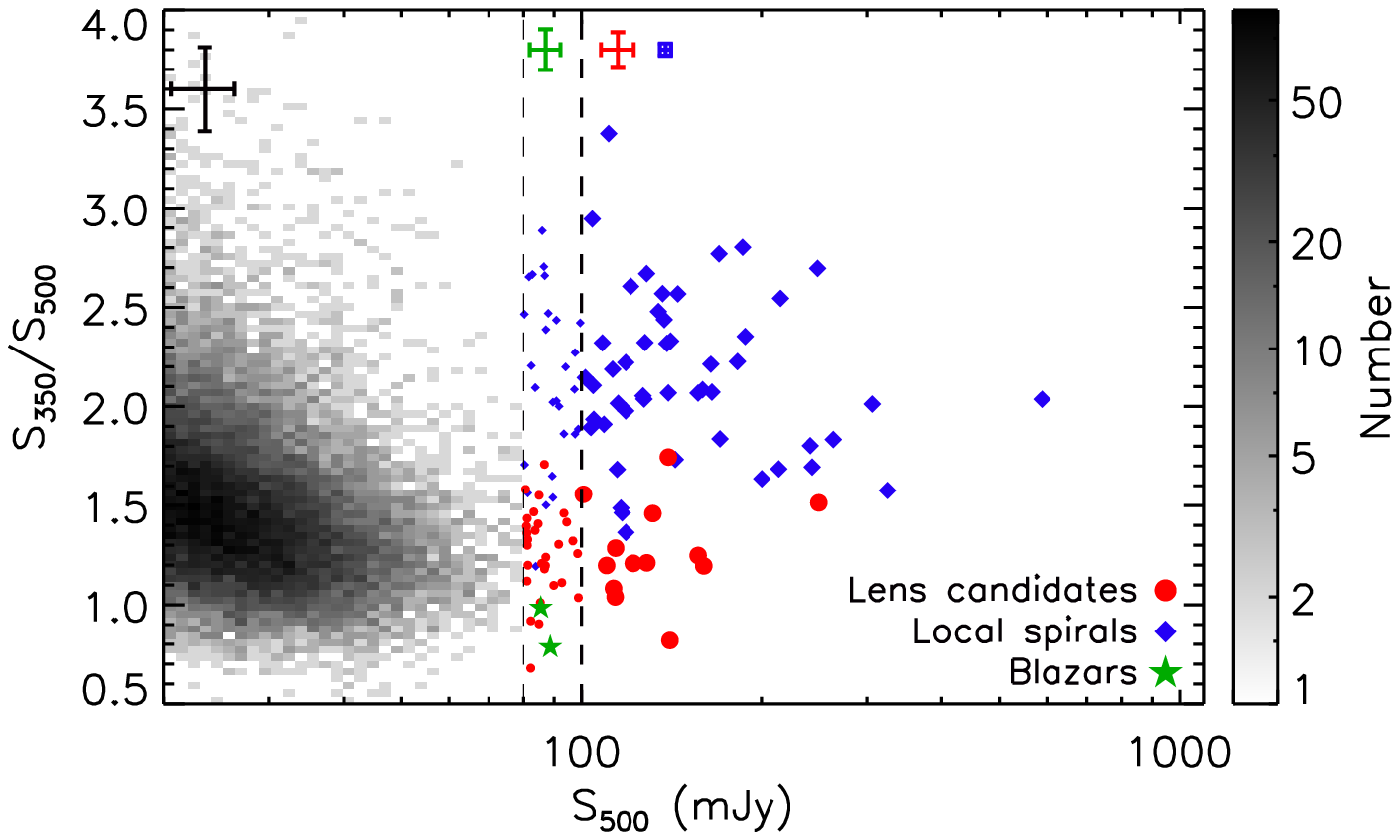}
\includegraphics[width=8.5cm]{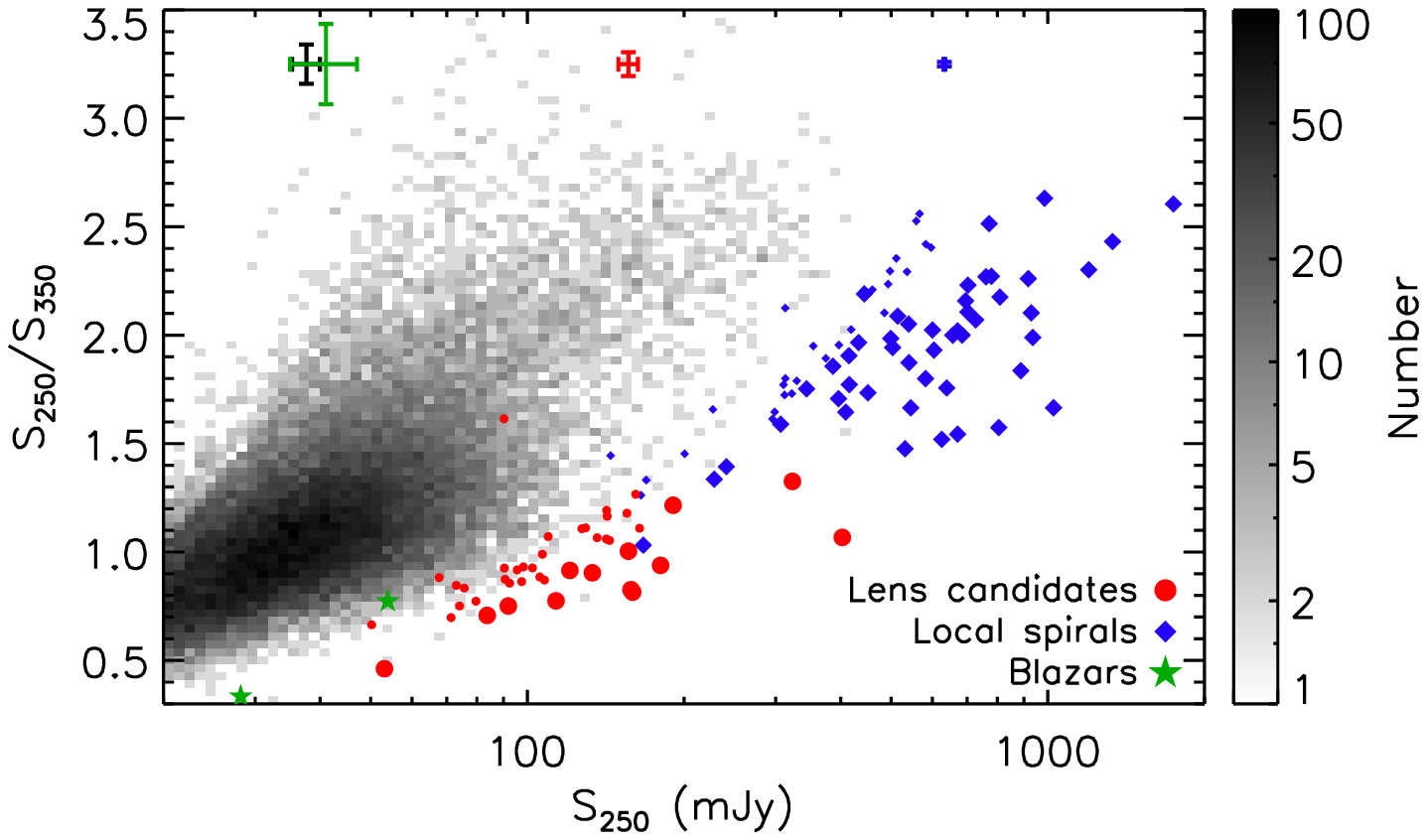}
\includegraphics[width=8.5cm]{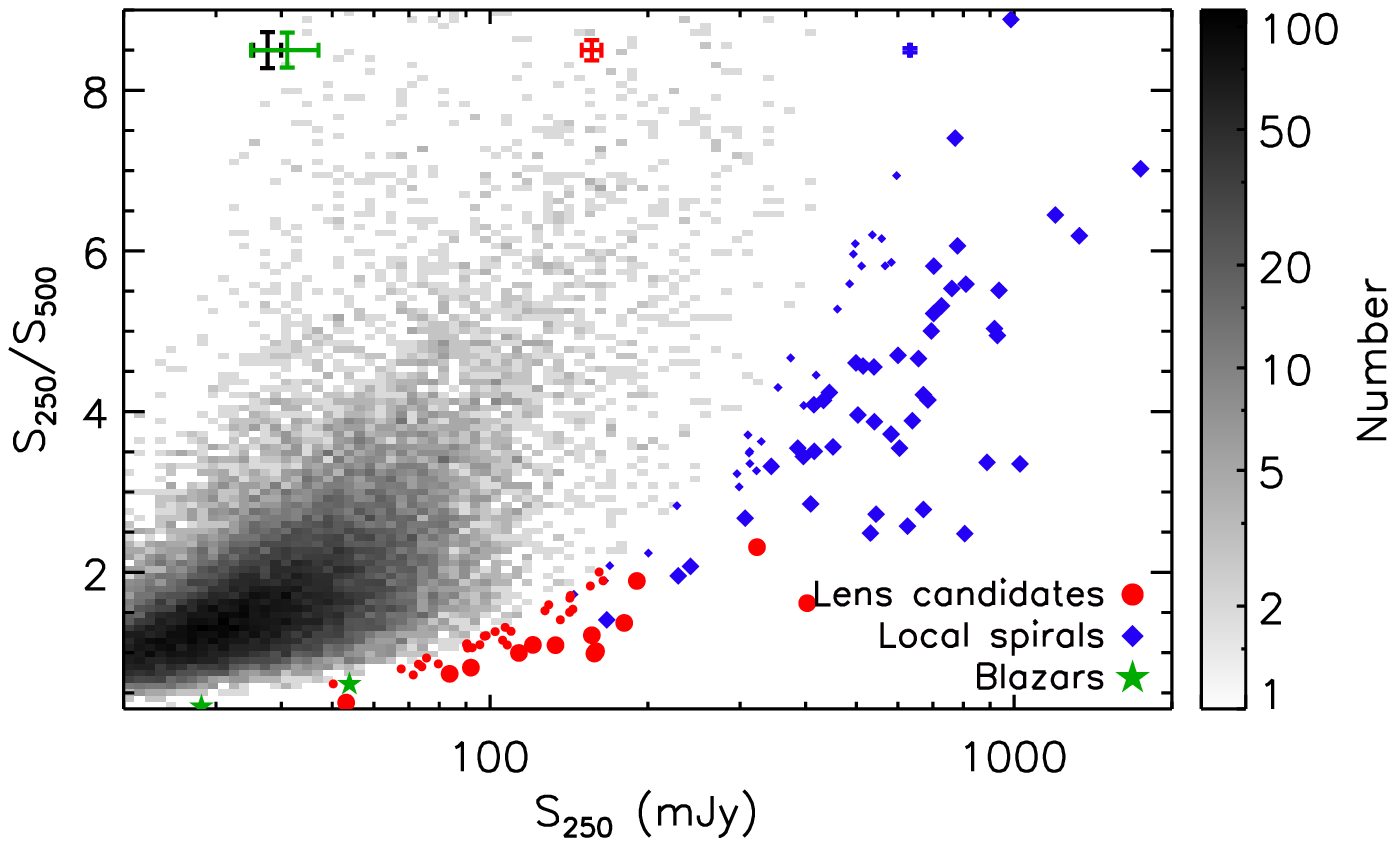}
\caption{
  SPIRE color-flux density plots for sources detected in \hermes\
  blank fields.  Sources brighter than 80\,mJy at 500\,\micron\ are
  highlighted and classified as blazars, local late-type
  galaxies or candidate gravitationally lensed SMGs (see
  Section~\ref{sec:sample} for details). Large and small symbols
  correspond to sources with $S_{500}>100$\,mJy (robust lensed candidates) and 
  $S_{500}=80$--100\,mJy (supplementary lensed candidates), respectively. Grayscale data
  represent the density of all \hermes\ sources in these fields.
  Candidate gravitationally lensed SMGs have redder SPIRE colors than
  local late-type galaxies, indicating that they are typically higher
  redshift sources. Median error bars for the individual
  populations are shown at the top of each figure, at the median flux
  density of each population.  We note that the apparent offset in SPIRE
  color-flux density space of the highlighted sources compared to the bulk of
  the \hermes\ population is due to our selection of the brightest
  sources. Indeed, in SPIRE color-color space no such offset is
  apparent (Fig.~\ref{fig:colcol}).}
   \label{fig:colflux}
\end{figure}

\begin{figure}
   \centering
\includegraphics[width=8.5cm]{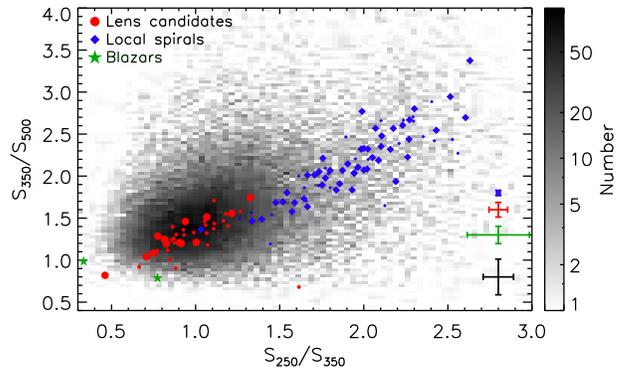}
\caption{SPIRE color-color diagram for sources detected in \hermes\
  blank fields. Candidate lensed galaxies, blazars and bright
  local spirals are highlighted; large and small symbols signify
  sources with $S_{500}>100$\,mJy and $S_{500}=80$--100\,mJy,
  corresponding to the principal and supplementary samples,
  respectively. The grayscale data show the density of all \hermes\
  sources in these fields. The lensing
  candidates have submillimeter colors that are consistent with rest of
  the SPIRE population. Median error bars for each population are
  shown at the bottom-right.}
   \label{fig:colcol}
\end{figure}

Local late-type galaxies are identified by searching the NASA/IPAC
Extragalactic Database\footnote{http://www1.ned.ipac.caltech.edu/}
(NED) at the 250\,\micron\ positions of the $S_{500}\ge80$\,mJy
sources.  There are 75 bright local late-type galaxies, with $z<0.1$,
in the 94.8 deg$^2$ used for this study (Table~\ref{tab:fields}).
These sources form a separate population from traditional SMGs, which
are dusty, star-bursting spheroidals, primarily at $z>0.5$
\citep{Lagache03,Negrello07}.  There is the chance of alignment
between a background SMG and a local late-type galaxy. However, these
local spirals are rare ($0.8\,{\rm deg}^{-2}$; table~\ref{tab:fields})
and the distance ratio of the background and foreground populations is
such that they are not expected to act as strong gravitational
lenses. Therefore, by removing the local late-type galaxies from the
sample we are not inadvertently removing a significant number of
distant gravitationally lensed sources. Indeed, the local spirals have
different distributions of submillimeter colors to the lensed
candidates (Figs~\ref{fig:colflux} and \ref{fig:colcol}), which is
indicative of two distinct populations at different redshifts.

\begin{deluxetable*}{llcccccc}
\tablewidth{0pt}
\tablecaption{Blazars with $S_{500}\ge80$ mJy in the HerMES fields
 \label{tab:agn}}
\tablehead{
\colhead{} & \colhead{}  & \colhead{} & \colhead{} &\colhead{} & \colhead{Observed} & \colhead{Predicted} & \colhead{}\\
\colhead{HerMES source} & \colhead{Name$^a$} & \colhead{$S_{250}$} & \colhead{$S_{350}$} &\colhead{$S_{500}$} & \colhead{radio flux} & \colhead{radio flux$^b$} & \colhead{$z$} \\
\colhead{} & \colhead{} & \colhead{(mJy)} & \colhead{(mJy)} & \colhead{(mJy)}&
\colhead{(mJy)} & \colhead{(mJy)} & 
}
\startdata
1HERMES S250 J003017.4$-$422443 & [HB89]0027$-$426$^c$         & $54\pm6$ & $70\pm5$ & $89\pm5$ & $419\pm23$$^d$ & 0.5 & 0.495$^c$ \\
1HERMES S250 J032752.0$-$290908 & NVSSJ032752$-$290912$^e$     & $28\pm6$ & $84\pm4$ & $86\pm5$ & $23.1\pm1.1$$^f$ & 0.1  & \nodata
\enddata
\tablecomments{
$^a$ Name is the name of the associated blazar from NED. 
$^b$ The predicted radio flux is calculated from the far-infrared radio
correlation, assuming that the radio emission is from star-formation
(see text for details). 
$^c$ \citet{Hewitt89,Wright94,Massaro09,Jackson02}. 
$^d$ The radio flux measurement for [HB89]0027$-$426 is at 4.85\,GHz.
$^e$ \citet{Condon98}. 
$^f$ The radio flux measurement for NVSSJ032752$-$290912 is at 1.4\,GHz.
}
\end{deluxetable*}

Blazars are identified by searching for \hermes\ sources with
associated radio emission in shallow, wide-area surveys (e.g.\ the
NRAO VLA Sky Survey (NVSS), \citealt{Condon98}, VLA Faint Images of
the Radio Sky at Twenty-Centimeters (FIRST), \citealt{Becker95,
  White97}, Parkes Radio Catalog, \citealt{Wright94}), which we access
using NED.  There are two bright \hermes\ sources with
$S_{500}>80$\,mJy, that are associated with bright radio emission;
these are 1HERMES~S250~J003017.4$-$422443 and
1HERMES~S250~J032752.0$-$290908 (Table~\ref{tab:agn}).

We determine whether the radio emission could be the result of
star-formation, rather than Active Galactic Nuclei (AGN) activity, by
comparing the observed radio flux with that predicted from the
far-infrared/radio correlation.  The calculation is performed for
$q_{\rm IR}=2.40\pm0.24$ \citep{Ivison10a}, where
\begin{equation}\label{eqn:qir}
q_{\rm IR}= \log_{10}{(L_{\rm IR}/3.75\times10^{12}{\rm W})} - \log_{10}{(L_{1.4}/{\rm WHz^{-2}})} \,.
\end{equation}
$L_{\rm IR}$ is the rest-frame 8--1000\,\micron\ luminosity,
determined from the \hermes\ photometry, and $L_{1.4}$ is rest-frame
1.4\,GHz radio luminosity density. Radio luminosities are
$K$-corrected assuming $S_{\nu}\propto\nu^{\alpha}$, with
$\alpha=-0.8$.

A 4.85\,GHz flux density of 0.5\,mJy is predicted for
1HERMES~S250~J003017.4$-$422443 at $z=0.495$, which is $\sim900$ times
lower than the observed radio flux density of 420\,mJy.  Thus, we
conclude that 1HERMES~S250~J003017.4$-$422443 is a blazar and it is
excluded from our sample. This classification is confirmed by the
X-ray ({\it ROSAT\/}; \citealt{Voges99}) and gamma ray ({\it Fermi};
\citealt{Abdo09}) detections of 1HERMES~S250~J003017.4$-$422443, and
its identification as a quasar by \citet{Hewitt89}.

The other radio-detected source, 1HERMES~S250~J032752.0$-$290908, does
not have an available archival redshift. Instead, we fit cold dust
SEDs, of the same form as Arp\,220, M82 and HR10 \citep{Silva98}, to
the SPIRE photometry, to estimate $z\sim4.5$, if the far-infrared
emission is star-formation dominated.  This assumption is also
required for the far-infrared/radio correlation and therefore, if the
far-infrared/radio correlation holds then the photometric redshift
should also be reasonable. For $z=4.5$, the predicted radio flux
density for 1HERMES~S250~J032752.0$-$290908 is 0.13\,mJy at 1.4\,GHz,
which is $\sim170$ times lower than observed. Therefore, we conclude
that 1HERMES~S250~J032752.0$-$290908 is also a blazar and the
photometric redshift estimate from the far-IR/sub-mm SED is incorrect.
It is also removed from the sample of candidate gravitationally lensed
SMGs.

Finally, the star Mira (o Ceti), is in the XMM-LSS SWIRE field and has
$S_{500}>80$\,mJy and it too is removed from our analysis \citep[see
also][]{Mayer11}. There are no other submillimeter-luminous stars in
the data. The associations between the HerMES sources with
  $S_{500}\ge80$~mJy and local late-type galaxies, AGN, and Mira are
  confirmed by eye, and a thorough inspection of the data affirms that
  there is no ambiguity as to the validity of any of these
  associations.

Having removed the 78 contaminants described above from the initial
list, 42 sources remain. There are 13 with $S_{500}>100$\,mJy, which
comprise our primary sample of robust candidate gravitationally lensed
\hermes\ galaxies (discussed individually in
Section~\ref{sec:indi}). The remaining 29 sources have
$S_{500}=80$--100\,mJy and make up the supplementary sample.  The
  division at 100~mJy between the robust and supplementary sample is
  made on the basis of the 500\micron\ number counts
  (Fig.~\ref{fig:numcount}) and is supported by the fraction of lensed
  sources that is predicted by modelling
  (Section~\ref{sec:moddetail}). The distribution of the candidates
between the nine survey fields is shown in Table~\ref{tab:fields}, and
a list of all the candidates and their 250, 350 and 500\,\micron\ flux
densities are presented in Table~\ref{tab:cand}. Additional archival
and follow-up infrared and radio photometry is listed in
Table~\ref{tab:allwaves}. 

The surface density of the main ($S_{500}>100$\,mJy) and supplementary
samples ($S_{500}=80$--100\,mJy) of lensing candidates is
$0.14\pm0.04\,\rm{deg}^{-2}$ and $0.31\pm0.06\,\rm{deg}^{-2}$,
respectively.  For comparison, \citet{Negrello10} identified five
candidate lensed galaxies with $S_{500}>100$\,mJy in $14.4\,{\rm
  deg}^{2}$ ($0.35\pm0.16\,{\rm deg}^{-2}$) of the \atlas\ SDP area
(all five are confirmed to be lensed), and 13 candidate lensed
galaxies were identified at $>4.5\sigma$ ($\sim15$\,mJy) at 1.4\,mm in
87deg$^2$ of SPT data (0.15\,deg$^{-2}$; \citealt{Vieira10}). The
number density of \hermes\ lens candidates is lower than the \atlas\
SDP area from \citep{Negrello10}, but the difference is not
statistically significant.


\section{Properties of candidate strongly lensed SMGs}
\label{sec:prop}

Having identified 13 robust and 29 supplementary candidate
gravitationally lensed SMGs, we next consider the basic properties of
these sources, including their submillimeter colors
(Section~\ref{sec:colors}), redshifts (Section~\ref{sec:nz}) and
apparent luminosities (Section~\ref{sec:lirz}).

\subsection{SPIRE colors}
\label{sec:colors}

Fig.~\ref{fig:colflux} shows \herschel\ SPIRE 250, 350 and
500\,\micron\ color-flux density plots for sources in \hermes\
blank-fields. Candidate lensed SMGs presented in this paper are
highlighted, and local late-type galaxies and blazars that are
brighter than 80\,mJy at 500\micron\ are also identified.  As
discussed in Section~\ref{sec:sample}, the {\sc StarFinder+XID}
catalogs are used for blazars and candidate lensed galaxy flux densities,
whereas the flux densities of local late-type galaxies are from the
{\sc SUSSEXtractor} catalog, which is more reliable for extended
sources.

The principal sample of lensed SMG candidates has median
$S_{250}/S_{350}=0.90$ ($\sigma=0.22$), $S_{350}/S_{500}=1.21$
($\sigma=0.24$) and $S_{250}/S_{500}=1.09$ ($\sigma=0.51$). These
colors are comparable to the background SMG population
($S_{250}/S_{350}=1.07$, $S_{350}/S_{500}=1.51$ and
$S_{250}/S_{500}=1.60$, with $\sigma=0.37$, $1.02$ and $1.67$,
respectively). There is a hint that the candidate lensed sources may
be slightly redder than the background. As expected, the local spiral
galaxies have significantly bluer colors than both the candidate lensed
galaxies and the background SPIRE population, with median
$S_{250}/S_{350}=1.97$ ($\sigma=0.33$), $S_{350}/S_{500}=2.08$
($\sigma=0.42$) and $S_{250}/S_{500}=4.14$ ($\sigma=1.48$) for the
$S_{500}\ge100$\,mJy subset.  These colors also indicate that the
candidate lensed galaxies lie at higher redshifts than local spiral
galaxies with similar $500\,\micron$ flux densities.

The candidate lensed sources, local spirals, and blazars appear
offset in $S_{250}$--$S_{250}/S_{350}$ and $S_{250}$--$S_{250}/S_{500}$
color-flux density spaces compared to the background of \hermes\ sources
(Fig.~\ref{fig:colflux}). This is because the highlighted sources are
bright at 250\,\micron, which is a direct result of the flux selection
at 500\,\micron. Indeed, the color-color diagram
(Fig.~\ref{fig:colcol}) shows that the lensing candidates have colors
consistent with the \hermes\ background population, while local spiral
galaxies have colors that are bluer than the majority of sources.

There are two lensed galaxy candidates that have bluer submillimeter
colors than the majority.  Both of these sources are known to be
strongly gravitationally lensed (H\bootes03, \citealt{Borys06},
Section~\ref{sec:b10}; and HLock01, \citealt{Conley11},
Section~\ref{sec:l1}). It is possible that differential
  magnification could affect the submillimeter colors of these galaxies
  \citep[e.g.][]{Hezaveh12, Serjeant12}, although cold dust dominates
  the emission at 250--500\micron\ so the effect is likely to be minor.
  We conclude that although the candidate gravitationally lensed
galaxies are typically redder at submillimeter wavelengths than local
late-type galaxies, there are some exceptions, and a color selection is
not sufficient to identify local interlopers. Instead, the removal of
interlopers requires the additional information that is provided by
optical and radio data, which can be provided by existing shallow
surveys (see Section~\ref{sec:sample}).

\subsection{Redshift distribution}
\label{sec:nz}

\begin{figure}
\includegraphics[width=8.5cm,clip]{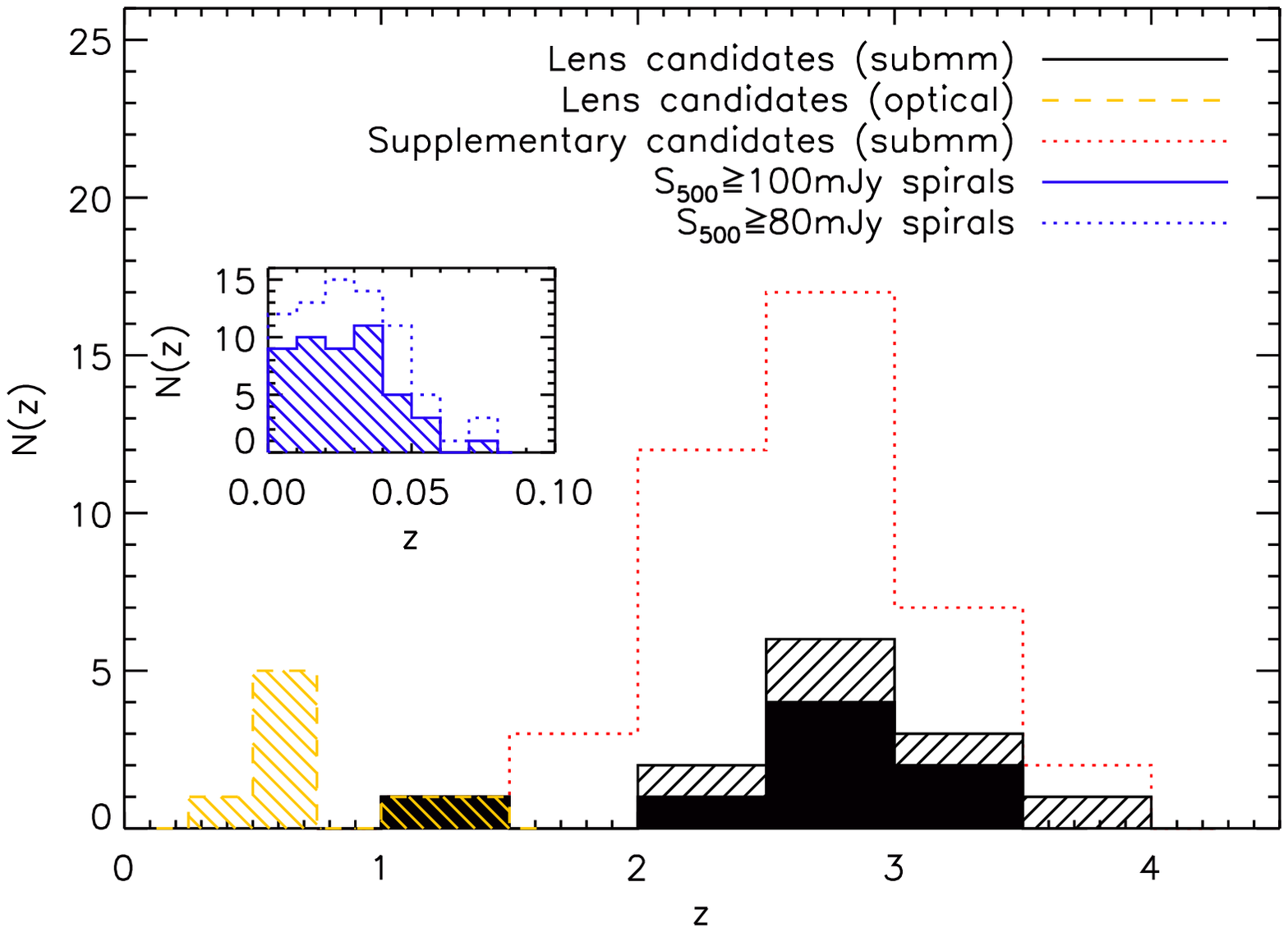}
\includegraphics[width=8.5cm,clip]{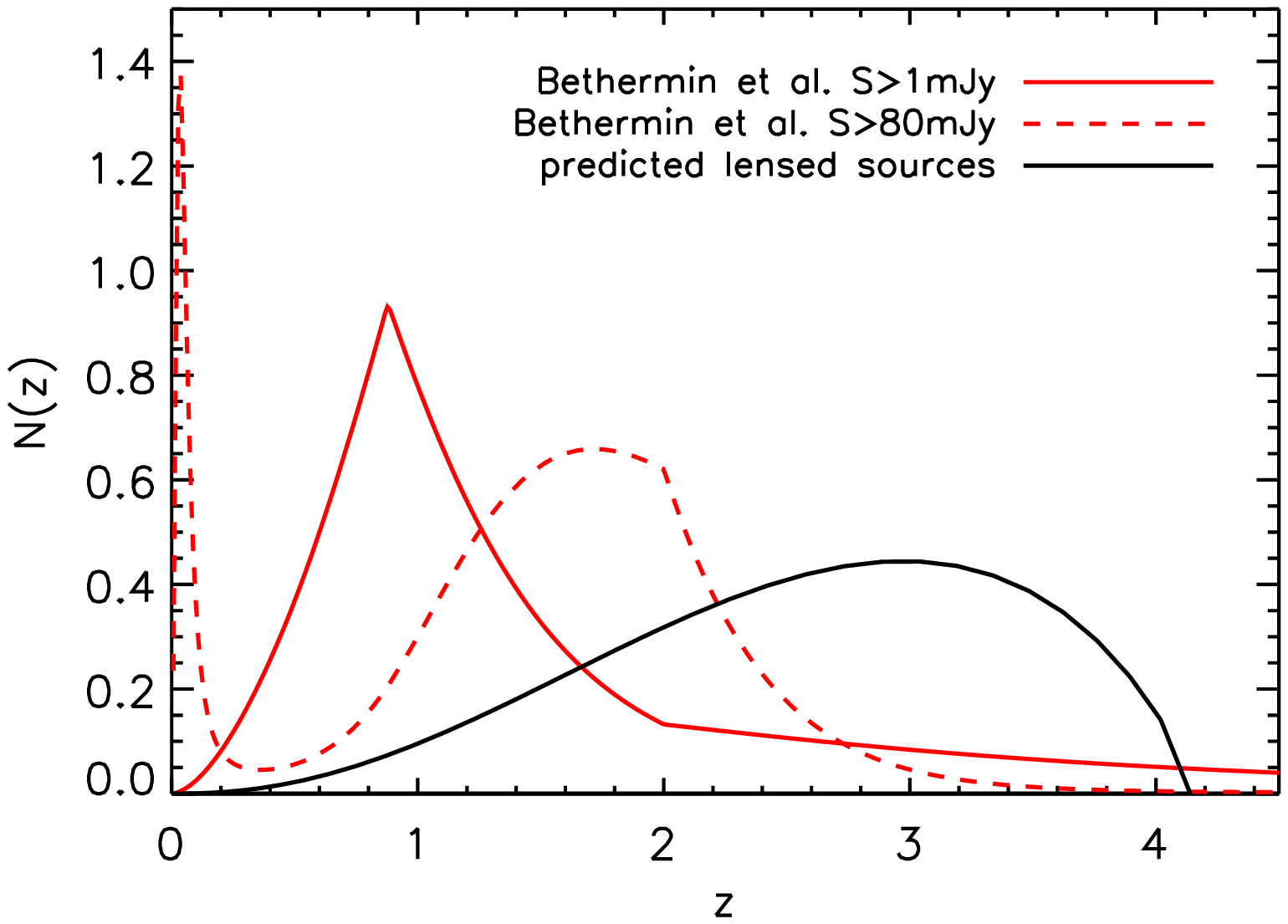}
\caption{ {\it Top:} Redshift distribution of \hermes\ lens candidates
  and (inset) bright local late-type galaxies.  Where possible,
  submillimeter redshifts are derived from the detection of one or
  more CO emission lines (solid histogram for the robust sample), and
  otherwise submillimeter photometric redshifts are used. Optical
  redshifts trace the confirmed foreground lenses. The background SMGs
  are typically at $z_{\rm submm}=2.8$ (mean and median) and are
  lensed by sources at $z_{\rm opt}\sim0.7$ (mean $z_{\rm opt}=0.72$;
  median $z_{\rm opt}=0.60$).  We note that the redshift distribution
  of the supplementary sample of lens candidates is consistent with
  the principal sample.
The local late-type galaxies are all at $z<0.1$, and peak at
$z=0.028$, confirming that they are easily identified in shallow
all-sky surveys (Section~\ref{sec:sample}).  {\it Bottom:} Normalized
redshift distribution of theoretical SMG populations, and the
prediction from our model for the distribution of gravitationally
lensed galaxies (Section~\ref{sec:model}). The predicted redshift
distribution of strongly lensed SMGs is not a
random sampling of the input populations due to the optical depth to
lensing. The model predicts that the redshift distribution of lensed
SMGs peaks at $z\sim3$, which is in broad agreement with available
data, although confirmation requires a larger sample.  }
  \label{fig:nz}
\end{figure}

We have shown that the candidate lensed SMGs have redder submillimeter
colors than local spiral galaxies, which is indicative of a higher
redshift population. We next consider the full redshift distributions
of these sources.

There are four main ways to calculate the redshifts of SMGs. The most
reliable is through the detection of submillimeter emission lines, the
brightest of which are CO transitions. The second method is to
calculate submillimeter photometric redshifts from the 250, 350 and
500\,\micron\ photometry, and any available longer wavelength data
(Table~\ref{tab:allwaves}).  Finally, optical or near-infrared
photometry or spectroscopy of the counterparts can be
utilized. However, if an SMG is gravitationally lensed then the
foreground deflector will usually dominate the short-wavelength
flux. In this case, if the foreground lens is misidentified as the SMG
then the optical redshift will be lower than the submillimeter
redshift.

We are currently undertaking an extensive radio and millimeter
spectroscopic follow-up campaign, targeting CO emission lines in the
candidate gravitationally lensed galaxies (see Section~\ref{sec:zdata}
for details).  Confirmed (multiple-line) redshifts have been obtained
for five of the candidate lensed SMGs, and single-line redshifts for a
further four (Table~\ref{tab:cand}; Riechers et al., in prep.). The
single-line CO redshifts are guided by the photometric redshifts in
determining the most likely identification of the line emission.

Submillimeter photometric redshifts are calculated from $\chi^2$
template fitting to the available submillimeter and millimeter
data. The SED of \eyelash\ (the Cosmic Eyelash; \citealt{Ivison10b,
  Swinbank10}) is used as the reference template, because
\citet{Harris12} showed that it is representative of \herschel\ lensed
galaxies with Robert C.\ Byrd Green Bank Telescope (GBT) detections of
\aco. The analysis is restricted to a single template because
\citet{Harris12} also showed that $\chi^2$ fitting to the SPIRE data
alone is unable to effectively select between multiple SEDs.  We
caution that the template choice results in a potential bias in the
submillimeter photometric redshifts, due to the assumption that each
source has an intrinsic SED (and dust temperature, $T_{\rm D}$) that
is similar to \eyelash. We assign each photometric redshift an error
of $\pm0.25$, which was shown empirically by \citet{Harris12} to
account for statistical errors, the uncertainty in the choice
of SED template, and the implicit assumption
that differential magnification between these wavelengths is
unimportant. The redshift error is not required
for our analysis,  and is only used for display purposes on Fig.~\ref{fig:lirz}.

In Fig.~\ref{fig:nz} (upper panel) we show the redshift distributions
of the candidate lensed galaxies, as derived from submillimeter and
optical data; CO redshifts are preferred where available. The optical
redshift distribution peaks at $z_{\rm opt}=0.60$ (median; mean
$z_{\rm opt}=0.72$), and as expected, the submillimeter-derived
redshifts peak at a higher value -- $z_{\rm submm}=2.8$ (median and
mean). Thus, on average the \hermes\ lensed SMGs lie at $z\sim2.8$ and
are magnified by sources at $z\sim0.7$. We note that submillimeter
redshifts of the supplementary sample are similar to those of the
primary lensed candidates.  Inset in Fig.~\ref{fig:nz} (upper panel),
we also show the distribution of local late-type galaxies, identified
in Section~\ref{sec:sample}. All of the submillimeter-luminous local
galaxies are at $z<0.1$, with a median of $z=0.028$, confirming that
they are easily distinguished from the SMGs.

For comparison, the lower panel of Fig.~\ref{fig:nz} shows the
redshift distribution of unlensed submillimeter sources in the
\citet{Bethermin11} model. Both models are
depend on the flux limits assumed, so we show sources with
$S_{500}\ge80$\,mJy and $S_{500}\ge1$\,mJy separately.  We note that,
as observed in the \hermes\ data, at the bright flux limit the models
contain a population of $z\ll1$ spiral galaxies in addition to the
higher redshift SMGs. Fig.~\ref{fig:nz} (lower panel) also contains
the redshift distribution 
of strongly gravitationally lensed SMGs as
predicted by our model (Section~\ref{sec:modpredict}). The predicted
redshift distribution peaks at $z\sim3$, in broad
agreement with the \hermes\ candidates. The redshift distribution of
lensed sources is not a random sampling of the parent SMG population
because the optical depth to lensing is also important.

\subsection{Apparent luminosity distribution}
\label{sec:lirz}

\begin{figure}
\includegraphics[width=8.5cm,clip]{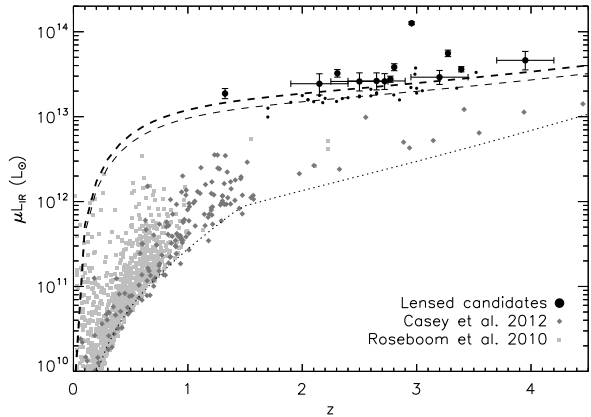}
\caption{Apparent rest-frame $8$--$1000$\,\micron\ infrared luminosity
  as a function of redshift for HerMES sources, {\it assuming no
    lensing magnification}. Large and small symbols represent our
  principal and supplementary candidates, respectively.  Due to the
  selection of the brightest 500-\micron\ sources, the candidate
  gravitationally lensed galaxies appear significantly more luminous
  than a comparison sample of ``normal'' (assumed not to be lensed)
  \hermes\ sources (\citealt{Roseboom10}, Casey et al.\ in prep.).
  The curved lines represent the {\it approximate} selection limits for
  the samples: the two upper lines {\it (dashed)} represent the
  candidate lensed SMG selection and are the luminosities of an
  optically-thin, $T_{\rm D}=35$\,K
  modified blackbody with observed 500\,\micron\  flux densities of 100 and 80\,mJy
  (thick and thin, respectively, corresponding to the principal and
  supplementary candidates, respectively).
  The lower line {\it (dotted)} represents the selection of sources in
  the two comparison samples, as described in the text.
  Error bars are omitted from the supplementary candidates and the
  comparison samples for clarity.  
}
\label{fig:lirz}
\end{figure}

We next consider the apparent far-infrared luminosities of the
\hermes\ candidate gravitationally lensed SMGs.  In
Fig.~\ref{fig:lirz} we show the distribution of redshift against
apparent infrared (rest-frame $8$--$1000$\,\micron) luminosity (i.e.\
assuming no lensing magnification) for the candidate gravitationally
lensed sources.  The infrared luminosities are calculated by fitting
an optically-thin modified blackbody, with fixed dust emissivity,
$\beta=1.5$, to the available submillimeter and millimeter data
(Table~\ref{tab:allwaves}).

For comparison, we also show $\sim1300$ \hermes\ sources with optical
and near-infrared redshifts from a dedicated optical spectroscopic
follow-up program (Casey et al.\ in prep.) and those in the \bootes\
field with archival optical spectroscopic and photometric
redshifts. The optical counterparts for both comparison samples are
identified according to the method described in \citet{Roseboom10} and
we require that the sources are detected at $>3\sigma$ in all three
SPIRE bands (250, 350 and 500\,\micron), so that the infrared
luminosities can be reliably determined.

We also show the approximate selection limits on Fig.~\ref{fig:lirz},
which for the lensed candidate samples are calculated assuming $T_{\rm
  D}=35$\,K and $S_{500}=100$\,mJy ($S_{500}=80$\,mJy for the
supplementary sample). The selection of the comparison sources is
complicated because it is driven by different wavelengths at different
redshifts. In this case we follow Casey et al.\ (in prep.) and use the
minimum of three optically-thin modified blackbodies -- with $T_{\rm
  D}=20$\,K and $S_{250}=15$\,mJy, $T_{\rm D}=30$\,K and
$S_{350}=15$\,mJy, and $T_{\rm D}=50$\,K with $S_{500}=15$\,mJy.

The candidate gravitationally lensed sources are at $z\sim1$--4 and
have apparent infrared luminosities between $1.9\times10^{13}$ and
$1.3\times10^{14}\,\mu^{-1}$\lsun, with a median of
$2.9\times10^{13}\,\mu^{-1}$\lsun, where $\mu$ is the magnification
factor from gravitational lensing. This median infrared luminosity
corresponds to a star-formation rate of $\sim5500\,\mu^{-1}$\myr\
\citep{Kennicutt98}, which represents some of the most extreme
star-formation episodes in the Universe, unless the amplification from
gravitational lensing is large.  We note that, despite the flux
selection, some of the supplementary lensed candidates are more
infrared-luminous than some members of the principal sample because of
the variation in $T_{\rm D}$ between the galaxies.

The comparison sample of \hermes\ sources with known redshifts has a
median infrared luminosity of $6.9\times10^{10}$\,\lsun. However, these
values are not directly comparable with the candidate lensed SMGs
because $95\%$ of the comparison sample are at $z<1$. The trend to
low redshift is not indicative of the redshift distribution of
SPIRE-selected galaxy populations, but is due to the
inherent biases in archival surveys (see Casey et al.\ in prep.\ for a
detailed discussion).
If we only consider the 73 sources in the comparison sample with
$z>1$, the median infrared luminosity is $1.5\times10^{12}$\,\lsun\
(${\rm SFR}\sim260$\,\myr), which is still significantly lower than the
candidate gravitationally lensed sources. As illustrated in
Fig.~\ref{fig:lirz}, this difference is a direct result of a
combination of our source selection technique, which identifies
candidates on the basis of bright 500\,\micron\ flux densities
(Section~\ref{sec:sample}), and the relatively flat $K$-correction for
$z\ga0.5$ at 500\,\micron.

\section{Lensing Statistics at submillimeter Wavelengths}
\label{sec:model}

Submillimeter-selected populations have steep number counts at the
  bright end, and therefore the gravitational lensing of SMGs leads to
  a change in the observed counts
  \citep[e.g.][]{Blain96, Negrello07, Jain11}. The intrinsic slope of
  the luminosity function of the population
  targeted determines whether lensing will affect the observed counts
  and luminosity function. For example, gravitational lensing does not
  significantly affect the observed luminosity functions of
  radio-selected sources \citep{Peacock82}.

 In this section we
use the \hermes\ observed 500\,\micron\ number counts, to 
  constrain a 
statistical model of the effect of flux boosting. The model is described
in Section~\ref{sec:moddetail} and in Section~\ref{sec:modpredict} we
use it to predict additional properties of the lensed galaxies,
including the expected mean magnification and fraction of candidates
that are strongly lensed.  In Section~\ref{sec:blend} we show that the
effect of source blending -- whereby multiple SMGs contribute to the
flux in a single SPIRE beam -- is negligible.

Similar analyses have recently been undertaken for sources
  detected in the SPT and BLAST surveys, using both analytic models
  \citep{Lima10a, Lima10b} and ray-tracing simulations
  \citep{Hezaveh11}. Both ray-tracing \citep{Lapi12} and analytical
  modeling \citep{Short12} of  lensed \atlas\ sources
  have also been performed. Likewise, \citet{Paciga09} used constraints
  from existing SCUBA data to predict the number of lensed sources
  likely to be detected in upcoming 850\,\micron\ SCUBA-2 surveys. We
  reiterate that a substantial fraction of the brightest 500\,\micron\
  sources are local late-type galaxies and thus a lensing model need
  not account for all of the bright sources \citep[cf.][]{Lima10b}.

\subsection{Modeling the lensed SMG population}
\label{sec:moddetail}
 
Existing literature discusses the details of the lensing calculations
that are undertaken here \citep[e.g.][]{Perrotta02, Negrello07}.
Therefore, in this section we provide a summary of the calculations
performed and refer to the appropriate papers for the details. 

Briefly, the calculation consists of the following processes and
assumptions:
\begin{itemize}\itemsep-1pt
\item {\bf Foreground mass profile:} Consider foreground masses with
  Navarro, Frenk \& White (NFW; \citeyear{Navarro97}) or Single
  Isothermal Sphere (SIS) profiles. The effect of the choice of mass
  profile is discussed in Section~\ref{sec:modpredict}.
\item {\bf Spatial distribution of foreground lenses:} Determine the
  comoving number density of foreground lenses as a function of mass
  and redshift, using the \citet{Sheth99} relation.
\item {\bf Redshift distribution of SMGs:} Use the model of
  \citet{Bethermin11}, with $S_{500}>1$\,mJy, to
  trace the redshift distribution of the intrinsic (unlensed)
  population of SMGs (Fig.~\ref{fig:nz}; lower panel). 
\item {\bf Strongly lensed area:} Calculate the fraction of the sky
  ($f_{\mu}$) that is strongly lensed ($\mu>\mu_{\rm min}$, for
  $\mu_{\rm min}=2$), using the profile and distribution of foreground
  masses and the redshift distribution of SMGs.
\item {\bf Intrinsic population:} Assume that the shape of the
  intrinsic (unlensed) number counts has the form of a
  \citet{Schechter76} function. Although other choices may be
    suitable, this function is adequately describes the observed data
    with only three free parameters.
\item {\bf Perform the magnification:} Integrate and apply the lensing
  probability to the intrinsic flux distribution to determine the net
  effect of lensing. The limits of the integration are set to
  $\mu_{\rm min}=2$ (for strong lensing) and $\mu_{\rm max}=50$ (for
  feasible sizes of the submillimeter emission regions).
\item {\bf Fitting:} Use a Monte Carlo Markov Chain (MCMC)
  minimization technique to fit the total \hermes\ counts with the
  four model components: unlensed SMGs, lensed SMGs, blazars and local
  late-type galaxies.
\item {\bf Predictions:} Use the fitted model to predict properties of
  strongly lensed SMGs, including their number counts and mean
  magnification (see Section~\ref{sec:modpredict})
\end{itemize}

We begin by considering the properties of the foreground (lensing)
structures. The effect of gravitational lensing on the submillimeter
number counts can be quantified when specific assumptions about the
foreground masses are made. Virialized dark matter halos have a NFW
universal density ($\rho$) profile, which is a function of radius,
$r$:
\begin{eqnarray}
\rho(r)=\frac{\rho_{\rm s}}{(cr/r_{\rm vir})(1+cr/r_{\rm vir})^2}\,,
\end{eqnarray}
where $r_{\rm vir}$ is the virial radius and $\rho_{\rm s}$ is the
characteristic density \citep{Navarro97} . The halo concentration parameter, $c$,
describes how centrally concentrated the mass is, and 
can be obtained from a
fit to simulations \citep[e.g.][]{Bullock01}:
\begin{eqnarray}
c(M_{\rm vir},z)=\frac{9}{1+z}\left(\frac{M_{\rm vir}}{M_*}\right)^{-0.13}\,,
\end{eqnarray}
where $M_*$ is the mass value such that $\sigma(M_*)=\delta_{\rm c}$ and
$\sigma^2(M)$ is the variance of the linear density field.  $M_{\rm
  vir}$ is the virial mass of a dark matter halo and $\delta_c$ is the
critical density contrast for the spherical collapse model
\citep{Gunn72, Lacey93}.  Since
$\delta_{\rm c}$ has only a very mild redshift and cosmology dependence, here we
take it to be fixed at the value for a matter dominated Universe,
i.e.\ $\delta_{\rm c}=1.686$.

While the NFW profile is the observed density profile of dark matter
halos, gravitational lensing captures the total mass.  Existing
lensing analysis of individual galaxy- and group-scale lenses show
that the total density profile is more consistent with an SIS
parameterization \citep[e.g.][]{Kochanek95, Koopmans06,
  Gavazzi07, Koopmans09, Barnabe10, Ruff11, Bolton12}.  Thus, as an
alternative choice, we also consider the SIS density profile:
\begin{eqnarray}
\rho(r)=\frac{\sigma^2_v}{2\pi Gr^2}\,,
\end{eqnarray}
where $\sigma_v^2$ is the line of sight velocity dispersion, calculated
 following \citet{Perrotta02}. The SIS
profile offers many advantages due to the simplicity of its form, and
\citet{Perrotta02} showed that the SIS and NFW profiles provide
similar results for statistical magnifications.  Indeed, as discussed
in Section~\ref{sec:modpredict}, we find that the uncertainties in the
overall modeling are such that we cannot reliably distinguish
the lensing by NFW halos and SIS spheres using the HerMES data
presented here.  However,  detailed analysis of individual systems may be able
to distinguish one profile over another on a case-by-case basis.
Given the limited statistics we do not account for sub-halo
masses \citep[e.g.][]{Oguri06}, the effect edge on spirals as
  lenses \citep[e.g.][]{Blain99b},  or more complicated issues such as
galaxy/dark matter halo ellipticity and external shear.
Our predictions are in agreement with the current data, but in wider area
surveys with more statistics it may become feasible to constrain these
additional effects. We refer the reader to \citet{Oguri06} for more
detail of these effects in calculations of gravitational lensing rates.

The projected density field, $\Sigma$, around each mass profile, is
obtained by integrating over the parallel component, $r_{\|}=\chi$, of
the position vector $\bar{x}=(r_{\|},r_{\bot})$, where the
perpendicular coordinate is $r_{\bot}=D_{\rm A}(\chi)\theta$, with
$D_{\rm A}$ being the angular diameter distance, $\theta$ the angular
coordinate in the lens plane, and $\chi$ the comoving radial
distance. The convergence term, $\kappa$, associated with the isotropic
light distortion from gravitational lensing, can then be defined in
terms of the critical density, $\Sigma_{\rm crit}$ \citep{Lima10b}:
\begin{eqnarray}
\kappa(\theta)&=&\frac{\Sigma(\theta)}{\Sigma_{\rm crit}}\,, \\
{\Sigma_{\rm crit}}&=&\frac{a}{4\pi G}
                       \frac{D_{\rm A}(\chi_{\rm s})}{D_{\rm A}(\chi) D_{\rm A}(\chi_{\rm s} - \chi)}\,,
\end{eqnarray}
where $a=(1+z)^{-1}$ is the cosmological scale factor and $\chi_{\rm s}$ is
the comoving radial distance to the source.  The anisotropic distortion
from lensing is measured by the shear,
$\vec{\gamma}=\gamma_1+i\gamma_2$, where $\gamma_i$ are the two components of the shear matrix.
The magnitude of shear depends on the shape of the density profile of the
foreground mass. For the NFW profile it is \citep{Takada03}
\begin{eqnarray}
\gamma_{\rm NFW}(\theta)= \frac{M_{\rm vir} f(c) c^2}{2\pi r_{\rm vir}^2}\frac{g(c\theta/\theta_{\rm vir})}{\Sigma_{\rm crit}} \,,
\end{eqnarray}
where $g(x)$ is a function that can be calculated analytically (see
\citealt{Lima10b} for details)  and $f(c)=(\ln(1+c)-c/(1+c))^{-1}$.

For the SIS profile all the quantities can be derived analytically and
the resulting shear is:
\begin{eqnarray}
\gamma_{\rm SIS}(\theta)=4\pi\sigma_{v}^2\frac{D_{\rm A}(\chi_{\rm s} - \chi)}{D_{\rm A}(\chi_{\rm s})}\frac{1}{2\theta}.
\end{eqnarray}

Gravitational lensing causes a flux amplification due to an increase in
the angular size of source galaxies. This effect is measured by the
magnification factor, $\mu$, which is given by the inverse determinant of the
Jacobian matrix describing the transformation between source and image
angular coordinates. In terms of the convergence and shear we have:
\begin{eqnarray}\label{eq:mag}
\mu(\theta)=\frac{1}{\left[1-\kappa(\theta)\right]^2-|\gamma(\theta)|^2}\,.
\end{eqnarray}
A halo with mass, $M$, and at redshift, $z_{\rm l}$, magnifies a
source at redshift, $z_{\rm s}$, in an elliptical region of the image
plane with area, $\Delta\Omega_\mu(z_{\rm l},z_{\rm s},M,\mu_{\rm
  min})$, within which the magnification is larger than $\mu_{\rm
  min}$:
\begin{eqnarray}
\Delta \Omega_{\mu}(\mu_{\rm min}) = \int_{\mu(\theta)>\mu_{\rm min}} \frac{d\theta^2}{\mu(\theta)}\,,
\end{eqnarray}
where we set $\mu_{\rm min}=2$ for strong gravitational lensing.

For a known distribution of foreground masses it is possible to define
a fraction, $f_{\mu}$, of the sky where $\mu>\mu_{\rm min}$ due to all
the halos above a certain mass and in a redshift range. $f_{\mu}$ will
therefore depend on the comoving number density of halos, defined as:
\begin{equation}\label{comov_den}
    \frac{dn}{d\ln M_{\rm vir}}=\frac{\rho_m}{M_{\rm vir}}f(\nu)\frac{d\nu}{d\ln M_{\rm vir}}\,,
\end{equation} 
where $\nu = \delta_{\rm c}/\sigma(M_{\rm vir})$ and $\sigma^2(M)$ is the
variance in a top hat of radius, $r$. Here we use the \citet{Sheth99}
relation:
\begin{eqnarray}
\nu f(\nu) = A\sqrt{{2 \over \pi} a\nu^2 } [1+(a\nu^2)^{-p}] \exp[-a\nu^2/2]\,.
\end{eqnarray} 
We take the parameter values $p=0.3$, $a=0.75$ and $A\simeq0.3222$ as
normalization constant \citep{Cooray02}. We have then:
\begin{equation}
\begin{split}
f_{\mu}=\int_{0}^{z_{\rm s}} dz_l \frac{D_{\rm A}^2(z_{\rm l})}{H(z_{\rm l})}
         \int_{M_{\rm min}}^{M_{\rm max}} d\ln M_{\rm vir}
          \int_{0}^{\infty} dz_{\rm s}P(z_{\rm s})\times\\
          \times\Delta \Omega_{\mu}(z_{\rm l},z_{\rm s},M_{\rm vir},\mu_{\rm min})
          \frac{dn(z_{\rm l},M_{\rm vir})}{d\ln M_{\rm vir}}\end{split},
\end{equation}
where $P(z_{\rm s})$ is the redshift distribution of the background
SMGs, and we set $M_{\rm min}=10^{12}h^{-1}{\rm M_{\sun}}$ as the lower
limit of the mass integral. The upper limit of the mass integral,
  $M_{\rm max}$, is set to $10^{15}h^{-1}{\rm M_{\sun}}$ and therefore
  the effect of groups and clusters, including galaxies embedded in groups and clusters, is implicitly included in the
  calculations.

We use $P(z_{\rm s})$ from the model of \citet{Bethermin11} for
$S_{500}>1$\,mJy (Fig.~\ref{fig:nz}; lower panel). This model, of
  the intrinsic \hermes\ SMG redshift distribution (Fig.~\ref{fig:nz}; lower panel), contains faint galaxies that are
  mostly concentrated at $z \sim 1$, with a decreasing tail
  thereafter. However, we note that the details of the choice of the
redshift distribution of the background sources does not significantly
affect our conclusions (see also Section~\ref{sec:modpredict}).  While
the SMG population peaks at $z\sim1$--3, we note that the redshift
distribution of lensed galaxies does not directly trace the background
population since it is weighted by the optical depth to lensing, which
is a function of redshift and increases in a non-linear manner, so
most of the SMGs that are lensed are expected to be at $z\sim3$, as is
observed (Fig.~\ref{fig:nz}).

\begin{figure}
\includegraphics[width=8.5cm,clip]{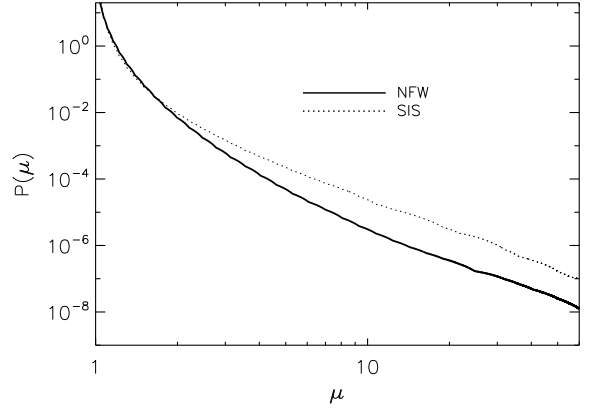}
\caption{ Magnification probability density, $P(\mu)$, for NFW and SIS
  mass profiles for a source at $z=2$.  At low magnifications the
  magnification probability of two profiles are comparable, but they
  becomes increasingly divergent at $\mu\ga3$, where higher
  magnifications are expected for the SIS model. This means that if
  all other parameters are fixed, the NFW model predicts fewer of the
  most apparently luminous sources (see also Fig.~\ref{fig:nfwsis}).
  The curves are normalised such that the enclosed area
    integrates to one, although the demagnification from weak lensing
    at $\mu\sim1$ is not considered.} 
  \label{fig:pmu}
\end{figure}

The lensing probability (the probability of having a magnification
larger than $\mu$) is given by \citep{Lima10b}:
\begin{eqnarray}
P(>\mu)=1-e^{-f_{\mu}}\,,
\end{eqnarray}
from which we can calculate the probability distribution of magnification,
$P(\mu)=-dP(>\mu)/d\mu$, for \hermes\ sources.

Fig.~\ref{fig:pmu} presents the magnification probability distribution
for foreground NFW and SIS profiles lensing a source at $z=2$. This
confirms that few SMGs are strongly gravitationally lensed, and those
that are magnified by factors $\ga10$ are rare.  As expected (see also
discussion in \citealt{Perrotta02}) higher magnifications are more
frequent for the SIS compared to the NFW profiles. However, at lower
magnifications, which are generally more likely, the SIS and NFW
profiles have similar $P(\mu)$.  We note that the NFW and SIS
  profiles do not encompass all possible galaxy mass profiles. For
  example, edge-on disks can significantly affect the magnification
  attained \citep{Blain99b}. However, the current data is insufficient
  to statistically distinguish between the even the simple NFW and SIS profiles
  (Section~\ref{sec:modpredict}) and therefore, further data are
  required to determine the distribution of mass profiles affecting
  strongly lensed SMGs.

\begin{figure*}[ht]
   \centering
\includegraphics[width=15cm,clip]{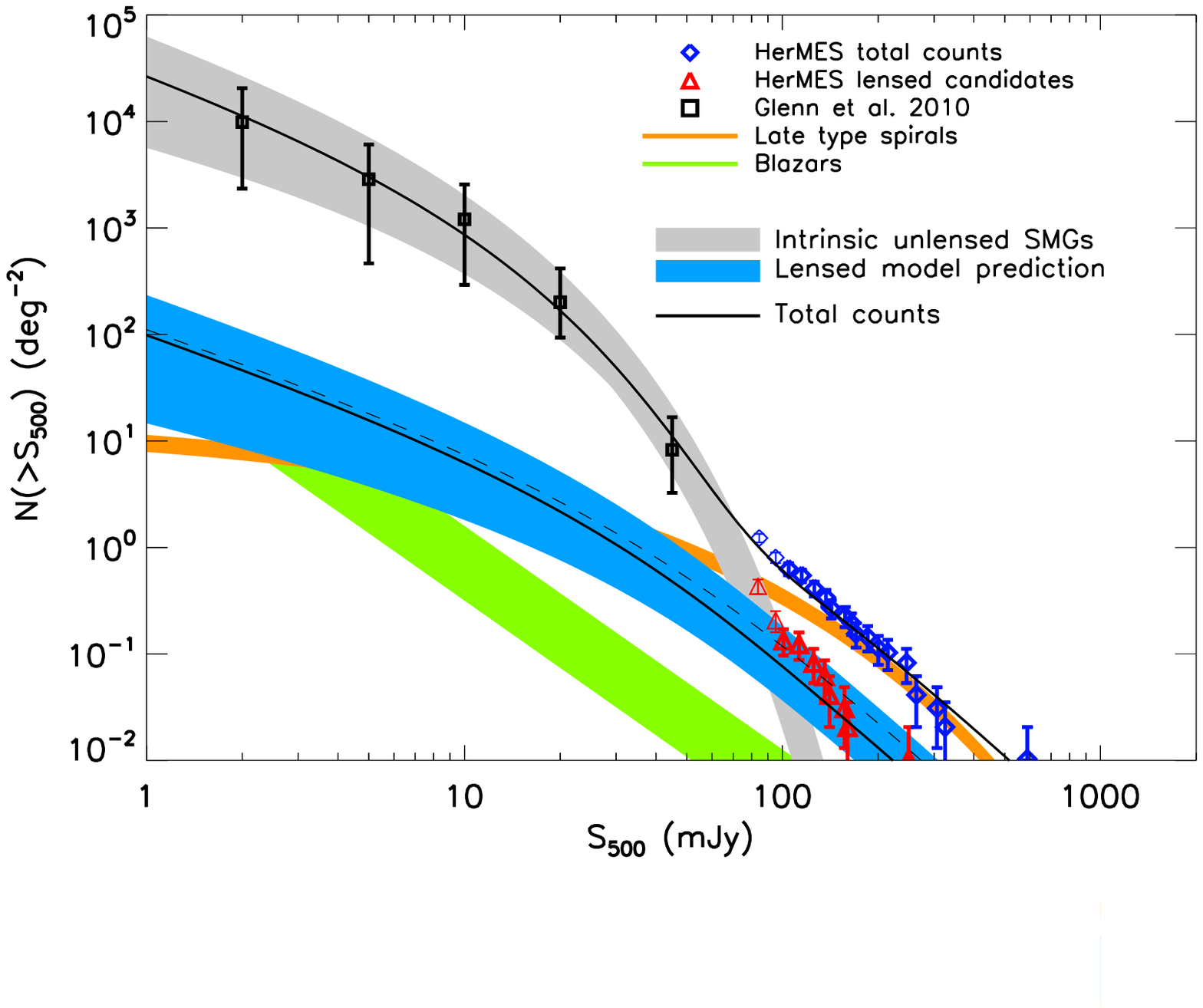}
\caption{Cumulative 500\,\micron\ number counts for HerMES blank-field
  catalogs \citep{Oliver10} and $P(D)$ analysis \citep{Glenn10}; the
  number counts of the candidate strongly gravitationally lensed
  sources in HerMES (Table~\ref{tab:cand}) are also shown.  Shaded
  regions represent the 68\% uncertainty in the model components and
  prediction for the NFW lensing model. The intrinsic number counts,
  and the contributions from blazars and local spiral galaxies
  are constrained by the available data. However, the number counts of
  the lensed sources is a prediction; in this case the shaded region
  captures the uncertainty in the modeled statistical lensing rate,
  due to a combination of uncertainties in intrinsic counts and the
  redshift distribution.  The solid line inside
  the prediction is the best-fit solution for the model based on the
  NFW profile. The dashed line is the prediction if we instead use SIS
  profile for the deflectors {\it and hold all other parameters
    fixed}. Note that when using the SIS profile and minimizing over
  all the available parameters the result is consistent with that for
  the model that utilizes the NFW profile (see Fig.~\ref{fig:nfwsis}).
  The model prediction agrees with the data for the number of
  candidate lensed SMGs with $S_{500}\ga100$\,mJy, but at
  $S_{500}\sim80$\,mJy we observe marginally more supplementary lens
  candidates than predicted by the model. This is consistent with the
  supposition that the supplementary lens candidates have a lower
  fidelity than the principal sample and a higher fraction that are
  unlensed, intrinsically luminous galaxies.  }
   \label{fig:numcount}
\end{figure*}

Lensing magnification increases the solid angle, $\Omega$, and the
integrated flux, $S$, of the lensed sources by a factor of $\mu$. The
number counts of SMGs have a steep slope at bright fluxes, so lensing
magnification results in an increase in the number density of sources
above a fixed flux limit.  By definition, the observed and intrinsic
(unlensed) quantities are related by:
\begin{eqnarray}
S_{\rm obs}=\mu S\,, \\
d\Omega_{obs}=\mu d\Omega \,.
\end{eqnarray}
For a given magnification the intrinsic differential number
density, $dn/dS$, is modified as \citep{Refregier97}:
\begin{eqnarray}
\frac{dn}{ds}\rightarrow\left.
\frac{1}{\mu^2}\frac{dn}{dS}\right|_{S=S_{\rm obs}/\mu}
\end{eqnarray}
Following the formalism and the discussion of \citet{Lima10b} the
observed differential number counts for a population is then:
\begin{eqnarray}\label{eq:dnobsdSobs}
\frac{d n_{\rm obs}(S_{\rm obs})}{d S_{\rm obs}} &=& \frac{1}{\langle \mu \rangle}\int d\mu \frac{P(\mu)}{\mu} \frac{dn}{dS}\left(\frac{S_{\rm obs}}{\mu}\right) \,, 
\end{eqnarray}
where
\begin{eqnarray}
\langle \mu\rangle = \int d\mu \ \mu P(\mu)\,.
\end{eqnarray}
and the cumulative number counts are:
\begin{eqnarray}
n_{\rm obs}(>S_{\rm obs}) &=& \frac{1}{\langle \mu \rangle}\int d\mu P(\mu) n\left(>\frac{S_{\rm obs}}{\mu}\right) \,.
\end{eqnarray}

The limits of these integrals are set to $\mu_{\rm min}=2$ and
$\mu_{\rm max}=50$, as previously discussed.  The maximum
magnification, $\mu_{\rm max}$, is determined by the size of the
background source \citep{Peacock82}.  \citet{Perrotta02} calculated
that \mumax$=10$--30 for 1--10\,kpc sources at $z=1$--4, and SMGs in
lensing simulations by \citet{Serjeant12} are observed with
$\mu\le33$, although most have $\mu<10$.  The star-formation regions
in SMGs are thought to be a few to 10\,kpc in extent
\citep[e.g.][]{Tacconi06, Younger08, HaileyDunsheath10, Ferkinhoff11},
although some may be as small as $\sim100$\,pc
\citep{Swinbank10}. Furthermore, the gas reservoirs in SMGs are often
more extended than the star-formation \citep{Ivison11, Riechers11b}.
Therefore, the choice of appropriate maximum magnification factor is
not straightforward, and $\mu_{\rm max}=50$ is chosen as a
  conservative limit. We have have also considered $\mu_{\rm max}=20$
  and $\mu_{\rm max}=30$ and find that the choice of higher $\mu_{\rm
    max}$ mainly affects the bright-end of the predictions from the
  model, since high values of $\mu$ are required to lens SMGs to the
  brightest apparent fluxes. Thus, a lower value of $\mu_{\rm max}$
  would reduce the bright-end of the mean magnification in
  Fig.~\ref{fig:mean} (top panel) and reduce the numbers of extremly
  bright lensed sources ($S_{500}\ga300$~mJy) such that observed
  sources with those flux denisities are typically local late-type
  galaxies (top panel of Fig.~\ref{fig:fracmu}). As additional data
  are obtained, it is likely that $\mu_{\rm max}$ will be constrained
  by observational results.

The intrinsic number counts of SMGs are most commonly parameterized as
Schechter \citeyearpar{Schechter76} functions or broken power-laws
\citep[e.g.][]{Coppin06, Weiss09, Lindner11}.  These two forms diverge
only at the faint and bright ends \citep[see e.g.][]{Paciga09}, and
cannot be distinguished with existing data. At the bright end the
broken power-law has a flatter slope than the Schechter function and
allows for a higher number of intrinsically luminous
sources. Consequently, models with the intrinsic number counts
parameterized by a broken-power law predict fewer strongly lensed sources than
models that make use of the Schechter function parameterization.  As
current data do not allow it we do not attempt to discriminate between
these models, although we note that additional constraints may be
possible as studies of the brightest submillimeter sources progress
(e.g. Section~\ref{sec:indi}).  For simplicity, we make use of a three
parameter Schechter function, characterized by a flux density,
$S^{\prime}$, beyond which the distribution is steeper and
gravitational lensing is more effective at boosting the observed
number of sources \citep{Schechter76}:
\begin{eqnarray}\label{eq:schechter}
\frac{d n(S)}{d S}=\left(\frac{N}{S^{\prime}}\right)\left(\frac{S}{S^{\prime}}\right)^\alpha e^{-S/S^{\prime}} \, .
\end{eqnarray}
Here, $\alpha$ is the slope of the counts below the characteristic flux
density, $S^{\prime}$, and $N/S^{\prime}$ is the normalization.

The model described above is fit to the \hermes\ 500\,\micron\ number
counts (Fig.~\ref{fig:numcount}).  We run a MCMC code and calculate
the $\chi^2$ for different combinations of the Schechter distribution
parameters, requiring the total 500\,\micron\ counts to fit the low
flux data points (up to $45$\,mJy) from the $P(D)$ analysis
\citep{Glenn10}.  For bright sources, the number counts presented here
are consistent with the analyses of \hermes\ SDP data by
\citet{Oliver10} and \citet{Glenn10}, which are also shown in
Fig.~\ref{fig:numcount}.  

To run the MCMC analysis we used a modified version of the {\sc
  cosmoMC} \citep{Lewis02} package, fixing the input cosmology to the
WMAP-7 best fit from \citet{Komatsu10}.  The convergence diagnostic is
based on the Gelman and Rubin R statistic \citep{Gelman92}.  As
described above, the minimum and maximum halo mass are fixed to
$10^{12}$ and $10^{15}\,h^{-1}M_{\sun}$, respectively, and the minimum
and maximum magnification are set to $\mu=2$ and 50,
respectively. Thus, $N$, $S^{\prime}$ and $\alpha$, from the Schechter
function description of the intrinsic 500\,\micron\ number counts, are
the only free parameters in the model.

The model is constrained by the total number counts, and the
numbers of blazars and local spiral galaxies.  We do not attempt
to constrain the intrinsic flux densities or redshift distribution of \hermes\
galaxies with the lensing counts. We simply show that, down to the
present accuracy, and with existing models, our predicted lensing
counts are consistent with the sample of \hermes\ candidate lensed SMGs
(Section~\ref{sec:modpredict} and Fig.~\ref{fig:numcount}). In the
future, when additional data are available and the nature of a higher
fraction of candidates is confirmed, it may be possible to constrain
the model further.

Unless explicitly stated, all the results presented in this paper are
for NFW profiles, although in Fig.~\ref{fig:nfwsis} we also consider the
effect of instead using the SIS profile.

\subsection{Model predictions}
\label{sec:modpredict}

\begin{figure}
\includegraphics[width=8.5cm,clip]{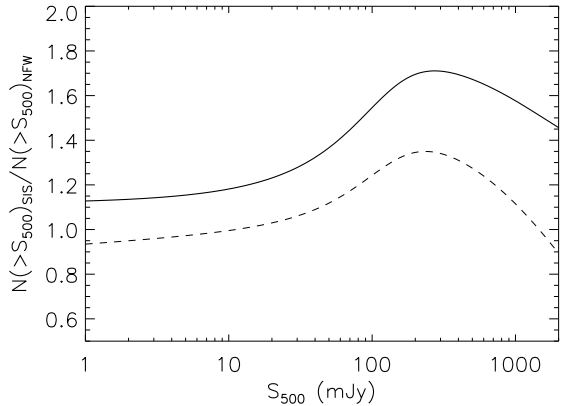}
\caption{The ratio of the number of strongly gravitationally lensed
  sources predicted by the NFW and the SIS models, as a function of
  apparent 500\,\micron\ flux density.  The solid line shows the case
  in which both models use the best-fit parameters from the NFW
  minimization and only the shape of the deflector profiles is
  changed. In this case the number of lensed sources predicted by the
  SIS profile always exceeds that from the NFW profile, although at
  500\,\micron\ flux densities below $\sim30$\,mJy the differences are
  minimal. The difference rapidly increases for sources with
  $S_{500}\sim30$--$200$\,mJy, and above this limit peaks at
  $N(>S_{500})_{\rm SIS}/N(>S_{500})_{\rm NFW}\sim1.7$ for
  $S_{500}\sim200$~mJy. Note that if,
  instead of fixing the parameters to the values from the NFW
  minimization, the parameters of the SIS model are derived from
  fitting that model, the prediction of the number of lenses sources is
  comparable to that from the NFW model (dashed line).  }
\label{fig:nfwsis}
\end{figure}

We next use the statistical model of galaxy-galaxy lensing, described
in Section~\ref{sec:moddetail} and constrained by the observed \hermes\
500-\micron\ number counts, to make predictions about the prevalence
and population of lensed \hermes\ SMGs.

In Fig.~\ref{fig:numcount} we show the model, the total observed
number counts in \hermes\ blank-fields, and the candidate strongly
gravitationally lensed sources from Table~\ref{tab:fields}. The
contribution to the model from unlensed (intrinsic) SMGs, local
late-type galaxies and blazars are also shown in
Fig.~\ref{fig:numcount}.  The model number counts for these three
components are constrained by observational data, and the shaded
regions represent the range of parameters that correspond to the 68\%
confidence limits. The blazars are modeled as a power law, and the
local late-type galaxies as a power law with an exponential cutoff at
80\,mJy, i.e.\ $dN/dS\propto S^{\alpha}{\rm exp}(-S_0/S)$, with
$S_0=80$\,mJy.  The predicted distribution of strongly lensed SMGs is
derived from an intrinsic Schechter function, using the model
described in Section~\ref{sec:moddetail}.

The model shown in Fig.~\ref{fig:numcount} corresponds to NFW mass
profiles for the foreground lenses. If SIS profiles are used instead,
and the model parameters recalculated, we find that prediction for the
number counts of strongly lensed sources is comparable to the NFW
result (Fig.~\ref{fig:nfwsis}, dashed line). Hence, it appears that
the choice between the two profiles is not important in fitting the
current data. We investigate this conclusion further by using the SIS
profile with parameters derived from the best-fit NFW model, and show
the resulting prediction for the counts of lensed sources in
Fig.~\ref{fig:numcount}.  The ratio of the predictions from the NFW
model, and the SIS model with NFW parameters is shown in
Fig.~\ref{fig:nfwsis}.  

\begin{figure}
  \includegraphics[width=8.5cm,clip]{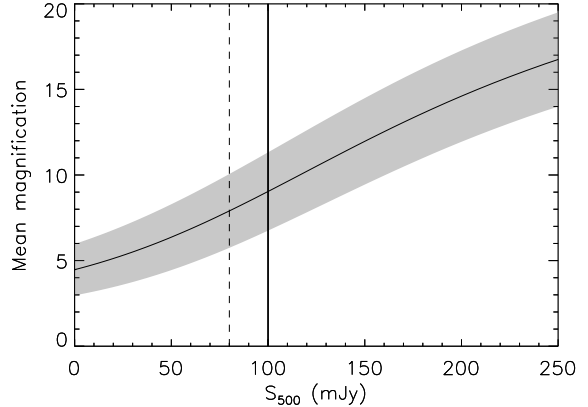}
  \includegraphics[width=8.5cm,clip]{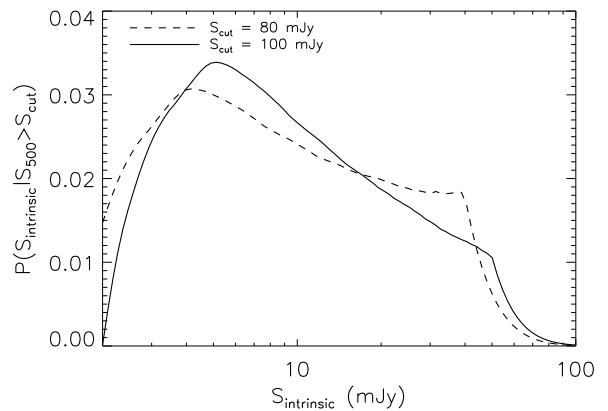}
  \caption{
{\it Top:} Predicted mean magnification of strongly lensed sources as a
    function of observed 500\,\micron\ flux density. The shaded region represents the
    $68\%$ confidence limits of the model and the solid line is the best-fit result.
    For sources with $S_{500}=100$\,mJy, the typical magnification is 
    predicted to be a factor of 6--11, while for the
    supplementary sample, the typical magnification is only a factor of
    5--9.  
    {\it Bottom:} $P(S_{\rm intrinsic}|S_{500}>S_{\rm cut})$, the
    predicted distribution of the intrinsic (i.e.\ before lensing)
    500~\micron\ flux density of strongly gravitationally lensed
SMGs selected with observed $S_{500}>80$ or
    100\,mJy.  
    The predicted intrinsic flux density distribution of
    the lens candidates peaks at $S_{\rm intrinsic}=5$\,mJy, and 65\%
    of sources have $S_{\rm intrinsic}<30$\,mJy which is the \hermes\ nominal
    detection limit at 500\,\micron\ (90\% completeness;
    Wang et al., in prep.), and thus 65\% of the gravitationally
    lensed SMGs constitute a population that would otherwise be
    undetectable.
    The strongly lensed sources in our supplementary
    sample ($S_{500}\ge80$\,mJy) are predicted to have a similar
    distribution of intrinsic flux densities.  
}
  \label{fig:mean}
\end{figure}

There is minimal difference at the faintest flux densities: when the
profiles are fit separately we find that for
$S_{500}\la30$\,mJy the ratio $N(>S_{500})_{\rm SIS}/N(>S_{500})_{\rm
  NFW}$) is $<1.2$, and an increase in the difference between the
models between $S_{500}\sim30$ and 200\,mJy to a maximum of
$N(>S_{500})_{\rm SIS}/N(>S_{500})_{\rm NFW}\sim1.7$ at
  $S_{500}\sim200$~mJy. The increase in the ratio at bright fluxes is
expected, because the difference between the two mass profiles is most
pronounced at the high-magnification end (see Fig.~\ref{fig:pmu}).  The
sharp cutoff in the number of sources that are intrinsically bright
means that the sources with high apparent flux densities are
typically subject to larger magnification factors (see
Fig.~\ref{fig:mean}; upper panel), which is where the difference in the
probability of magnification between the NFW and SIS profiles is most
pronounced.

There are two underlying reasons for this behavior between NFW and SIS
models.  NFW models under predict the lensing cross-section at the low
mass end where baryons dramatically steepen the total mass
profile. SIS models over predict the lensing cross-section at the high
mass end as the velocity dispersion is based on the virial mass and
not the sub-halo mass.  As shown in Fig.~\ref{fig:numcount} current
data are unable to distinguish between the two profiles. Therefore we
refer to the NFW profile for the remainder of this paper. With
additional data from follow-up programs we may eventually be able to
distinguish between different foreground mass profiles, including
  additional profiles that are not considered here, and provide
further constraints on the models. However, as discussed by
\citet{Perrotta02}, the magnification distribution is dominated by
foreground deflectors with a small range of masses, and therefore,
effects such are expected to be negligible.

In Fig.~\ref{fig:mean} (upper panel) we show the prediction of the
mean magnification of strongly lensed sources, which increases as a
function of apparent 500\,\micron\ flux density.  For sources with
$S_{500}= 100$\,mJy the model predicts that the mean magnification of
the lensed sources is $\mu=9.1\pm2.5$, where the range indicates the
uncertainty in the model. For brighter sources with
$S_{500}=250$\,mJy, the calculated mean magnification is
$16.9\pm2.7$.  Fainter lensed SMGs, such as those in our
supplementary sample ($S_{500}=80$--100\,mJy) have lower
magnifications, with $\mu\sim7$ on average, although there may be
individual examples with much higher amplifications.  We note that the
prediction of relatively low ($\mu\lesssim10$) flux amplifications for
the majority of sources is qualitatively consistent with results from
the simulated lensing of individual SMGs \citep{Serjeant12}.

\begin{figure}
\includegraphics[width=8.5cm,clip]{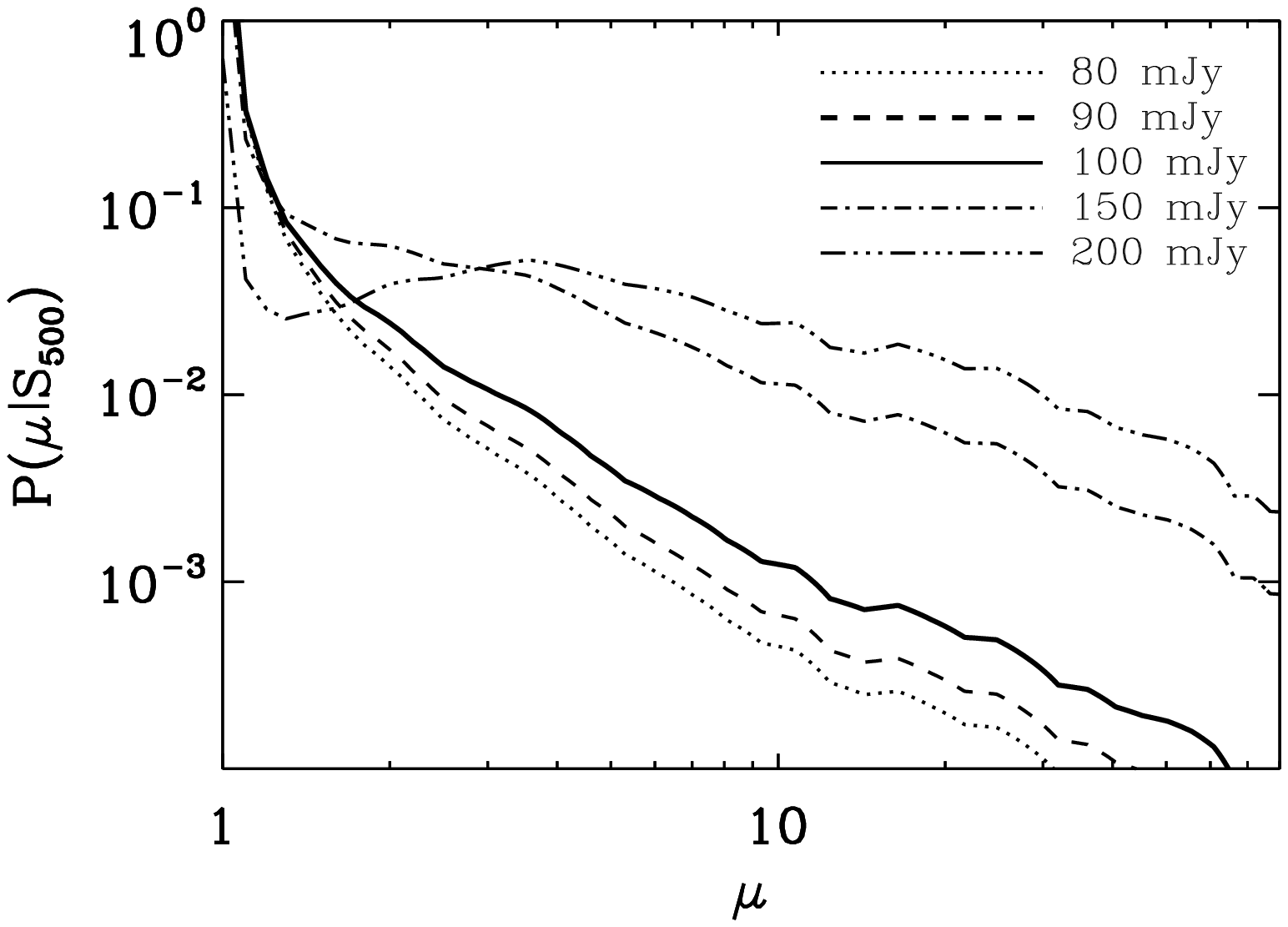}
\includegraphics[width=8.5cm,clip]{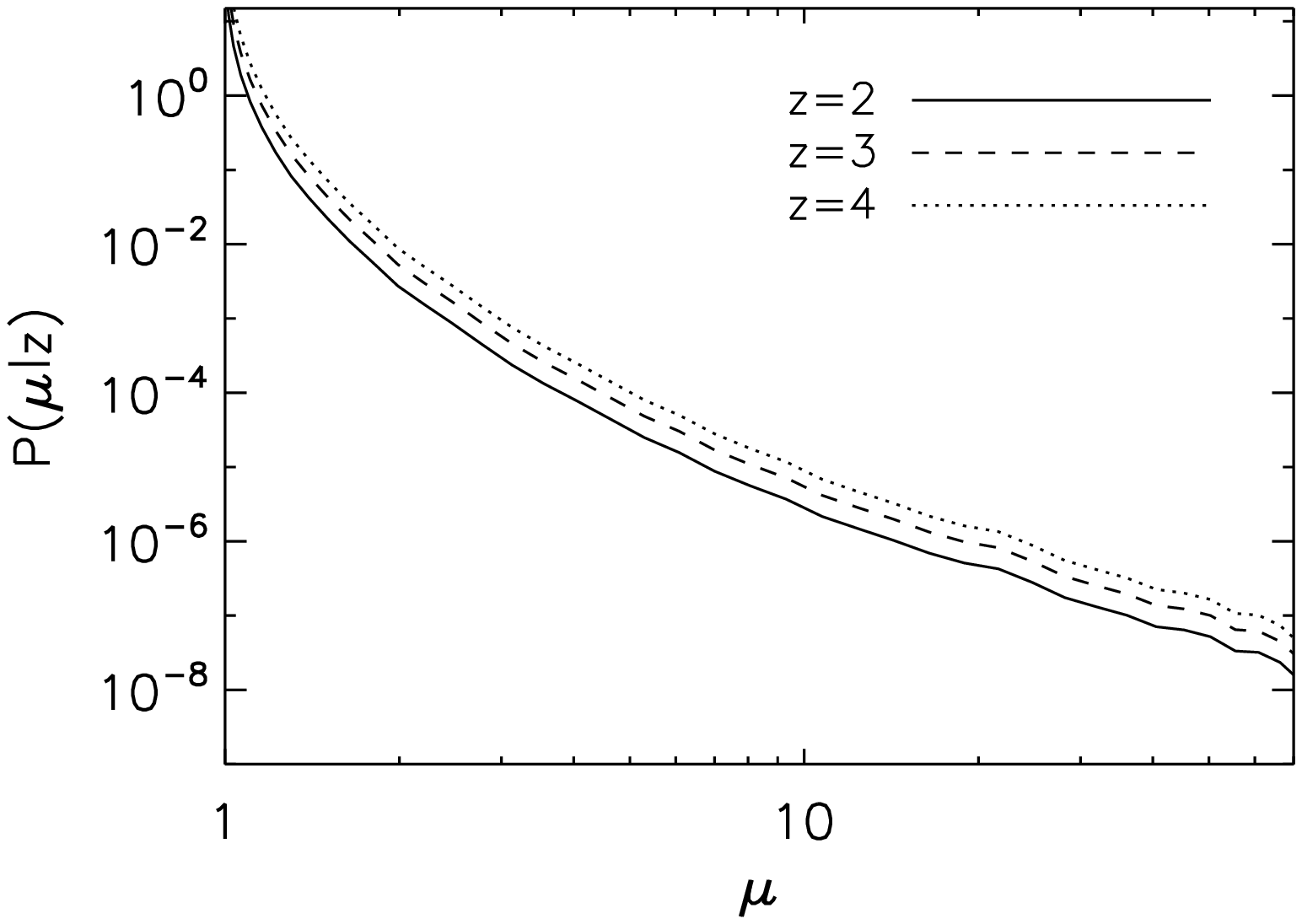}
\caption{ {\it Top:} The conditional magnification probability
  $P(\mu|S_{500})$, i.e.\ the probability of achieving a
  magnification, $\mu$, given an observed flux density, $S_{500}$, for
  SMGs. The thick
  line shows the magnification at $S_{500}= 100$\,mJy, which is the
  selection limit for our lensed candidates.  For $S_{500}\le100$\,mJy
  the magnification probability 
  decreases steeply as $\mu$ increases, but, for
  $S_{500}\ge150$\,mJy, the probability distribution is flatter due to the rarity of sources with intrinsic 500\,\micron\
flux densities $\ga70$\,mJy. 
{\it Bottom:} The conditional magnification probability $P(\mu|z)$,
i.e.\ the probability of achieving a magnification, $\mu$, given a SMG
redshift, $z$. For $z=2$--$4$ there is little dependence on redshift,
demonstrating that the exact redshift distribution of the background
(lensed) sources in the model has a minimal effect on the results.
In both panels the demagnification from weak lensing at
  $\mu\sim1$ is not considered and the curves are normalised to integrate
  to one.}  
  \label{fig:probs}
\end{figure}

Qualitatively, the trend of apparently brighter sources being more
highly magnified is easy to understand, because intrinsically faint
sources are more numerous than bright ones. Therefore, strongly lensed
sources with observed flux densities $S_{500}\la100$\,mJy may be the result of
magnifying one of the numerous sources with intrinsic
$S_{500}\sim30$--50\,mJy by a factor of a few. The dearth of
unlensed sources with $S_{500}\sim100$\,mJy, makes it less likely that
the most apparently luminous sources will have low
magnifications (see also Fig.~\ref{fig:probs}; upper panel).

We also consider the predicted intrinsic (i.e.\ before lensing
amplification) 500\,\micron\ flux density distribution of strongly
lensed sources that are selected with apparent flux densities above a
fixed limit ($S_{\rm cut}$). Fig.~\ref{fig:mean} (lower panel) shows
that strongly lensed sources, with observed $S_{500}>100$\,mJy (such as
the candidates presented here), have a broad distribution of intrinsic
500\,\micron\ flux densities, peaking at $S_{\rm intrinsic}=5$\,mJy,
and 65\% of sources have $S_{\rm intrinsic}<30$\,mJy. The
results are similar for an observed flux density cut of 80\,mJy (such
as our supplementary sample). The sharp turnover in the curves
  that occurs at $S_{\rm intrinsic}\simeq S_{\rm cut}/2$ is caused by
  the combination of our definition of strong lensing ($\mu>2$) and the
  steep decrease in the number counts at the bright-end.
The nominal
\hermes\ detection limit is 30\,mJy at 500\,\micron\ (90\%
completeness; Wang et al., in prep.), and thus $\sim65\%$ of
the strongly lensed SMGs are sources that would otherwise not be
detected. The amplification from gravitational lensing makes these
sources ideal targets for the detailed study of high-redshift,
star-forming galaxies, which would otherwise be prohibitively faint.

Within the framework of the model we can also calculate
$P(\mu|S_{500})$, the probability of a source with a given observed
flux density, being magnified by a factor, $\mu$, as a function of
$S_{500}$ (Fig.~\ref{fig:probs}). Indeed, $P(\mu|S_{500})$,
  provides a measure of the predicted magnification distribution for
  sources with a given apparent 500~\micron\ flux.
Fig.~\ref{fig:probs} (upper panel) shows that for
$S_{500}\le100$\,mJy, the probability of lensing decreases with
increasing magnification and $\mu\gtrsim10$ is rare, having a
probability of $\sim10^{-3}$, among detected galaxies.  However, at
the brightest flux densities, $S_{500}\ge150$, the probability
distribution is flatter, due to the rarity of intrinsically bright
sources.  Note that $\mu_{\rm max}=50$
  (Section~\ref{sec:moddetail}), so for the purposes of calculating
  the predicted number counts we only consider the range $\mu=2$--50.

We also investigate the effect of the background (lensed) source
redshift distribution on the probability of the source being magnified by
$\mu$. Fig.~\ref{fig:probs} (lower panel) shows $P(\mu|z)$ and
demonstrates that for $z=2$--4, the assumed redshift distribution has
only a small effect on our analysis.  Indeed, the prediction of the
model and the results presented in this paper are insensitive to the
exact form of the assumed redshift distribution of SPIRE sources. 
We note that if we use the redshift distribution from the \citet{Valiante09} model instead for the \citet{Bethermin11} model, the difference in the predicted number counts, mean magnification and lensed fractions is within the 68\%
 confidence limits of our results.

\begin{figure}
\includegraphics[width=8.5cm,clip]{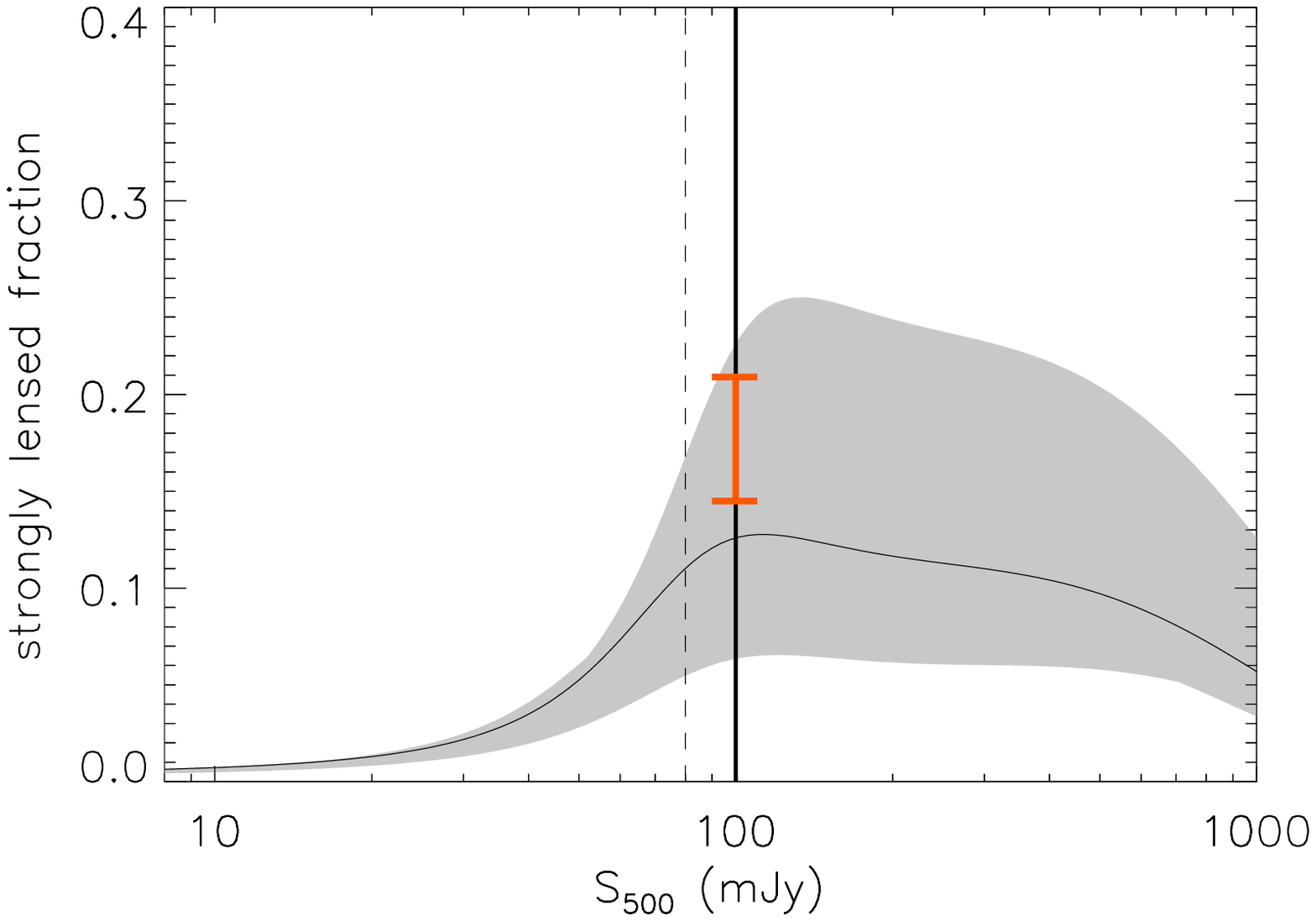}
\includegraphics[width=8.5cm,clip]{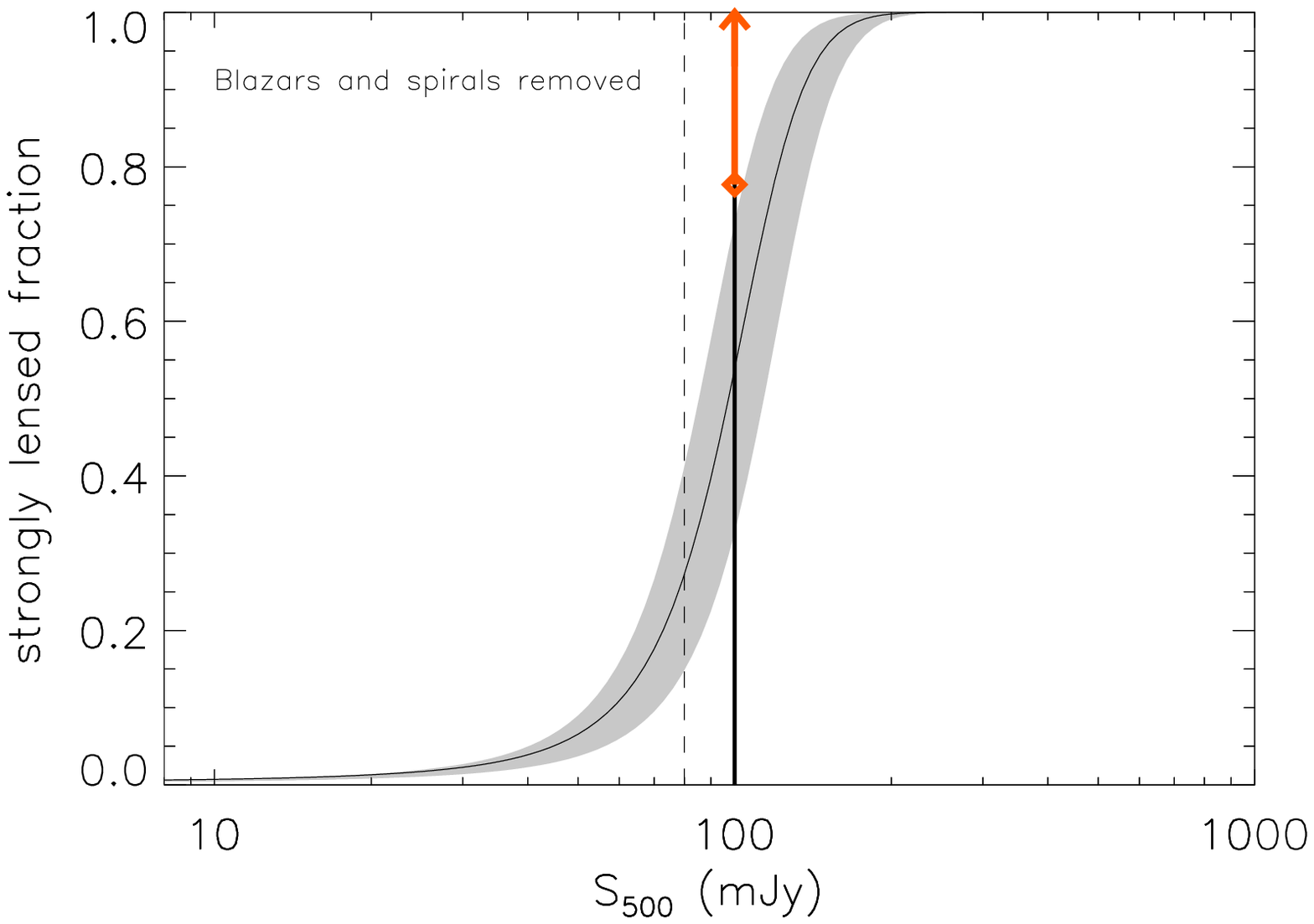}
\caption{ Predicted fraction of strongly lensed
  SMGs, for sources with 500\,\micron\ flux density brighter than
  $S_{500}$. Shaded regions represent the $68\%$ confidence limits and solid curves
  are the best-fit models.  Solid and dashed vertical lines
  demarcate the selection limits of our primary and supplementary lens
  candidates, respectively. 
  {\it Top}: fraction of all 500\,\micron\ sources that are predicted
  to be strongly lensed.  The decrease at $S_{500}\ga100$\,mJy is due to
  the increasing contribution from local late-type galaxies, which dominate the
  number counts at the brightest fluxes.  The error bar is the 
  observed fraction of \hermes\ sources with $S_{500}>100$\,mJy that are
  gravitationally lensed: the range is due to the four lens candidates
  with insufficient data to determine their nature. 
  {\it Bottom}: Fraction of 500\,\micron\ sources that are predicted
  to be strongly lensed, excluding local late-type galaxies or
  blazars. The arrow shows the fraction of the 18 \herschel\
  candidates (13 in \hermes; this paper, five in \atlas;
  \citealt{Negrello10}) that are lensed.  The model predicts that
  32--74\% of the candidates are lensed, which is consistent with the
  available data. For the supplementary candidates, with
  $S_{500}=80$--100\,mJy the predicted lensed fraction drops to
  15--40\%.  }
   \label{fig:fracmu}
\end{figure}

Fig.~\ref{fig:fracmu} shows the predicted fraction of strongly lensed
sources as a function of observed 500\,\micron\ flux density. The
analysis is based on our statistical model only; in
Section~\ref{sec:blend} we show that blending is only a minor effect.

We first consider all the 500-\micron\ sources, and find that the
strongly lensed fraction varies from $<1\%$ for $S_{500}<14$\,mJy to a
peak of 13\% at $S_{500}=105$\,mJy and then declines for the brightest
fluxes (Fig.~\ref{fig:fracmu}; upper panel). The decline for
$S_{500}\ga100$\,mJy is the result of the contribution of local
late-type galaxies -- at these flux densities the 500-\micron\
population is increasingly dominated by local spirals
(Fig.~\ref{fig:numcount}). Observationally, our data contain 49 local
late-type galaxies, 13 lens candidates (nine confirmed;
Section~\ref{sec:indi}) and no blazars with $S_{500}>100$\,mJy. Thus,
8--23\% of \hermes\ sources with $S_{500}>100$\,mJy are
gravitationally lensed, which is in agreement with the model.

If blazars and local spirals are excluded, then the fraction of
strongly lensed sources increases from $<1\%$ for $S_{500}<15$\,mJy to
$100\%$ for $S_{500}\ga200$\,mJy (Fig.~\ref{fig:fracmu}; lower
panel). A correction of $<1\%$ is required for blending
(Section~\ref{sec:blend}), so these values are applicable to \hermes\
lensed candidates.  Smaller catalog flux limits have greater
contamination from unlensed SMGs. For $S_{500}\ge100$\,mJy 32--74\% of
the sources are strongly lensed; for $S_{500}\ge80$\,mJy
(corresponding to our supplementary sources) the strongly lensed
fraction is only 14--40\%.  In Section~\ref{sec:indi} we show that
nine of the 13 \hermes\ candidates are lenses, and the nature of the
remaining four is unknown (Section~\ref{sec:indi}). Similarly,
\citet{Negrello10} confirmed that all five \atlas\
$S_{500}\ge100$\,mJy candidates in their SDP data are gravitationally
lensed. Thus, observationally, $>78\%$ of \herschel\ lensed galaxy
candidates with $S_{500}\ge100$\,mJy are bona fide lenses. These
values are consistent with the predictions from our model, although
the small number of sources dominates observational uncertainties.

\subsection{Could blended sources contaminate lensed SMG selection?}
\label{sec:blend}

\begin{figure}
  \includegraphics[width=8.5cm,clip]{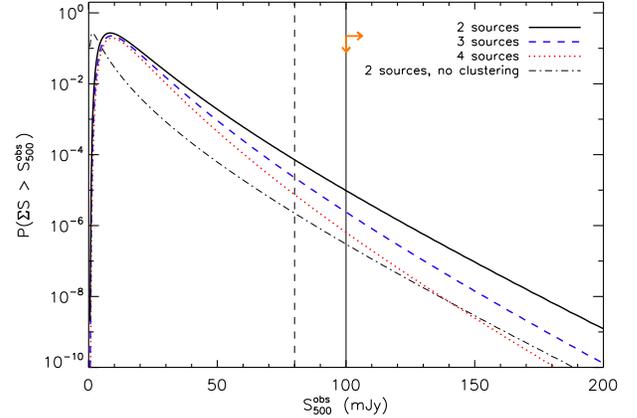}
  \caption{ Probability of multiple sources blending in the \herschel\
    beam and being detected as a single source with a total flux
    density $>S_{500}^{\rm obs}$.  The calculation uses the cumulative
    number counts and assumes that the minimum flux density of the
    components is equal.  We show the results for two, three and four
    blended sources, with clustering from \citet{Cooray10}, and for
    two sources without clustering. Vertical solid and dashed lines
    represent the selection limits of our primary and supplementary
    lensed candidate catalogs, respectively. Arrows denote the limits
    obtained from follow-up observations of lens candidates, which is
    consistent with the model.  Blending is rare in bright cataloged
    sources: lens candidates, with $S_{500}^{\rm obs}\ge100$\,mJy have
    a probability of $\sim10^{-5}$ of being blends of multiple
    equi-flux sources. The number of sources in the blend only has a
    minor effect on the likelihood of occurrence.}
  \label{fig:blend_obs}
\end{figure}

\begin{figure}
  \includegraphics[width=8.5cm,clip]{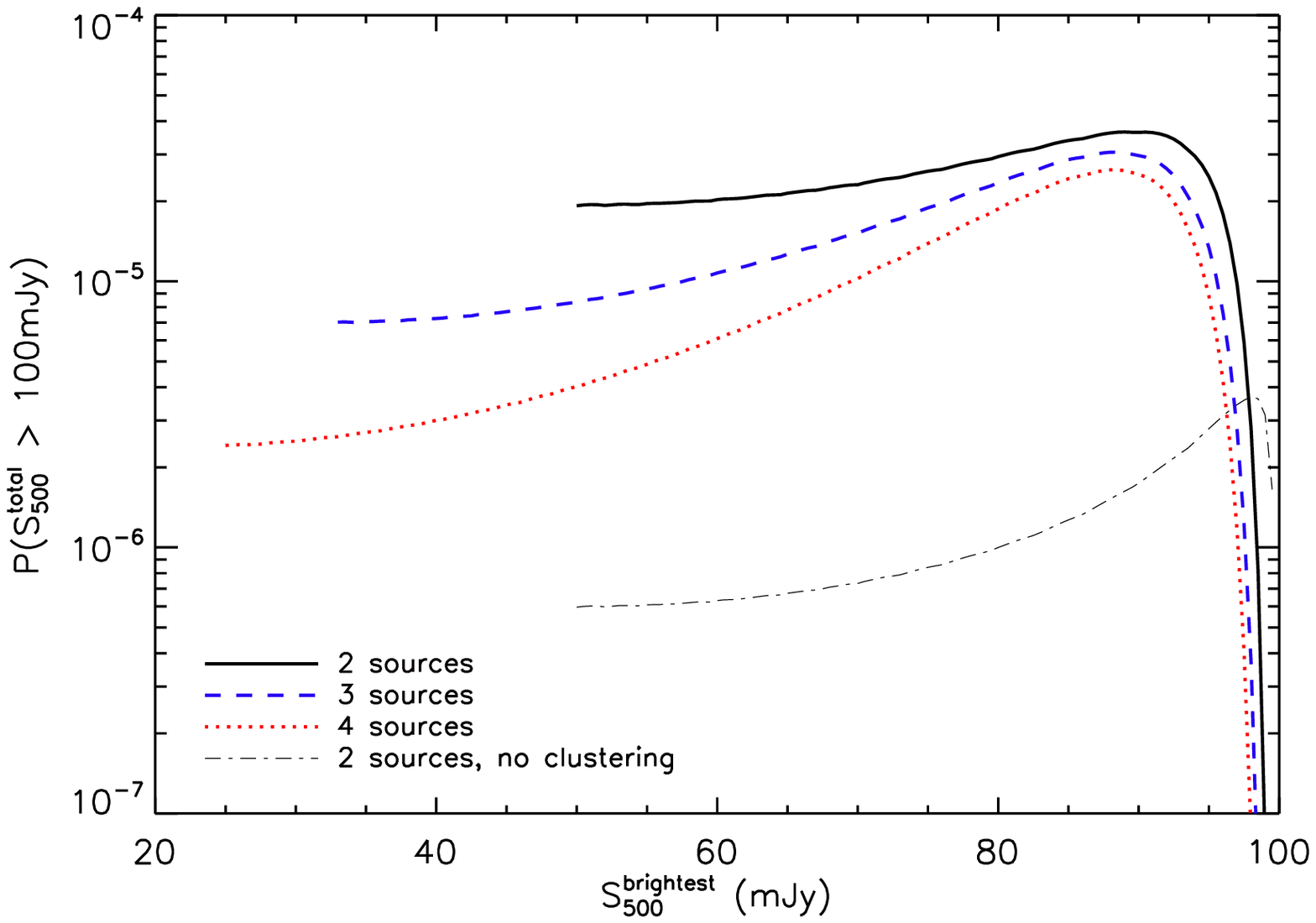}
  \caption{ Probability of multiple sources, with flux densities
    totaling more than 100\,mJy, blending in the \herschel\ beam, as
    a function of $S_{500}^{\rm brightest}$, the minimum 500\,\micron\
    flux density of the brightest component. Probabilities for two,
    three and four sources with clustering from \citet{Cooray10} are
    shown, in addition to the unclustered, two source case.  Blending
    in the \hermes\ catalogs is unlikely to produce sources that mimic
    gravitationally lensed SMGs.  }
  \label{fig:blend_comp}

\end{figure}

Thus far, the model results presented for bright \herschel\
500\,$\mu$m sources only include individual sources and strongly
lensed SMGs. It is possible that multiple faint sources could be
blended in the SPIRE beam and mimic bright, gravitationally lenses
SMGs. This is the same process that creates confusion noise in
submillimeter data. When dealing with catalogs it effectively shifts
sources from lower to higher flux bins such that the total number of
bright sources is higher than reality.

Sources that are separated by 18\arcsec\ or greater are reliably
deblended in the \hermes\ catalogs (Wang et al.\ in prep.) used here.
Therefore, we consider sources that are located within 18\arcsec\ of
each other as blended.  We use the intrinsic cumulative number counts
(Fig.~\ref{fig:numcount}) and begin by calculating the probability
that two, three, or four sources of equal flux density (or greater)
are located within 18\arcsec\ of each other
(Fig.~\ref{fig:blend_obs}); in this way we are able to consider the
blending rate as function of apparent (observed) flux density. We then
remove the requirement for an equal flux density cutoff in the
statistics and instead consider the probability that any two, three,
or four sources with total flux density greater than $100$\,mJy are
blended (Fig.~\ref{fig:blend_comp}).

The cumulative intrinsic 500\,\micron\ number counts from
Fig.~\ref{fig:numcount} are used to calculate the Poissonian
probability that randomly distributed sources of equal, or greater,
flux density are located within 18\arcsec\ of each other.  The effect
of the clustering of SMGs is included by using $w(\theta)$ as measured
for \hermes\ sources with $S_{500}>30$\,mJy; in this case
$w(18\arcsec)=5.7\pm2$ \citep{Cooray10}.  Clustering boosts the
probability of blending because the sources in a clustered
distribution are more likely to reside close to other sources. Indeed,
Fig.~\ref{fig:blend_obs} shows that, amongst sources with observed
$S_{500}>20$\,mJy, blending is $\sim30$ times more prevalent in the
clustered population.

Fig.~\ref{fig:blend_obs} shows that the probability of two 50\,mJy, or
brighter, sources being blended into a single 100\,mJy, or brighter,
source is $9.7\times10^{-6}$.  For three or four blended sources the
probabilities are $2\times10^{-6}$ and $6\times10^{-7}$, respectively.
At the $S_{500}\ge80$\,mJy corresponding to our supplementary sample of
lens candidates (see Appendix), the probability is an order of
magnitude higher at the level of $7\times10^{-5}$. However, even in
this case one in $\sim14,000$ of these candidates is expected to be
blended, while for the brighter flux density cut of our primary sample,
one in $\sim100,000$ is expected to be blended. 

We also consider the possibility that galaxies of unequal flux
densities are blended in the \hermes\ catalogs. Fig.~\ref{fig:blend_comp} shows the
probability that two, three or four sources with total
$S_{500}\ge100$\,mJy are blended, as a function of the flux density of
the brightest component.  It is apparent from
Fig.~\ref{fig:blend_comp} that the blending of intrinsically bright
sources ($S_{500}\ga95$\,mJy) with intrinsically faint sources is
unlikely; this is due to the intrinsic rarity of such
luminous sources.  The probability of a $>$ 100\,mJy source
being a blend of any two fainter galaxies is always
$<5\times10^{-5}$. 

These results are expected since our selection limit of
$S_{500}\ge100$\,mJy is $\sim15\times$ the \herschel\ point-source
confusion noise.  We note that blending contamination at a level
  of $\gtrsim1\%$ contamination requires clustering of
  $w(18\arcsec)\gtrsim180$ -- $\sim30\times$ higher than that
  extrapolated from larger scale observations. Therefore, our results
  are robust to the uncertainty in the SMG clustering. We conclude
that the blending of sources is rare amongst sources selected with
$S_{500}>100$\,mJy, and none of the lens candidates presented in this
paper are likely to be blends of unlesed sources.


\section{Follow-up data}
\label{sec:followup}

The confirmation and further study of the \hermes\ gravitational lens
candidates requires additional data. To that end, we have undertaken
extensive multi-wavelength follow-up programs.  Unfortunately, due to
scheduling constraints and source visibility, the ancillary data
coverage is non-uniform across the sample of 13 principal sources. We
summarize the follow-up programs here, and in Section~\ref{sec:indi}
we use the available ancillary data to investigate the individual
candidates and show that at least nine of them are strongly gravitationally
lensed.

\subsection{High-resolution imaging}
\label{sec:highres}

High-spatial resolution submillimeter interferometry,
primarily at 880\,\micron, is from the Submillimeter Array (SMA).  These
data include multiple array configurations, from sub-compact to very
extended and the synthesized beamsize range from
$\sim0.3-3\arcsec$ (FWHM).  The data shown here were initially taken as
part of a Director's Discretionary Time (DDT) follow-up program of HerMES
lensed sources (PI: A.\ Cooray) and later as part of a large multi-semester
program of lensed {\it Herschel} sources (PI: R.\ S.\ Bussmann).  

SMA observing conditions were typically very good, with low
atmospheric opacity ($\tau_{\rm 225~GHz} < 0.08$) and good phase
stability.  Targets were typically observed for $1-3$~hours
(on-source), depending on weather conditions and the brightness of the
target.  We used track-sharing to observe multiple targets in a given
night and maximize {\it uv} coverage.  All observations used an
intermediate frequency coverage of 4--8 GHz and provide a total of 8
GHz bandwidth (considering both sidebands).  For time-variable gain
(amplitude and phase) calibration, we used the nearest quasar with
$S_{880}>0.5$\,Jy (in some cases we used the average solution provided
by the two nearest quasars).  We typically used either the blazar
3C~279 or 3C~84 as the primary bandpass calibrator, while a planetary
moon (often Titan) was used as the absolute flux calibrator.  The
Multichannel Image Reconstruction, Image Analysis, and Display
(MIRIAD) software \citep{Sault95} {\sc invert} and {\sc clean} tasks
were used to invert the {\it uv} visibilities and deconvolve the dirty
map, respectively.  Natural weighting was typically chosen to obtain
maximum sensitivity.  Photometry is measured from SMA data using the
{\sc casa}\footnote{http://casa.nrao.edu} task, {\sc imfit}.

Two of the lensed SMG candidates were observed with the Jansky Very
Large Array (JVLA; \citealt{Perley11}) to obtain high-resolution radio
interferometry at 1.4 and 7\,GHz (Section~\ref{sec:indi}). The data
were taken as part of a larger follow-up program of lensed {\it
  Herschel} SMGs (program 11A-182; PI: R.\ Ivison) and were reduced
using {\sc aips} following the procedures described in
\citet{Ivison11}, to achieve a resolution of $<0.25\arcsec$ (FWHM) in
both cases.

We have also obtained high-resolution near-infrared imaging with the
Near Infrared Camera-2 (NIRC2) and laser guide star adaptive-optics
system (LGSAO) on Keck-II \citep{Wizinowich06}, or with the Wide Field
Camera 3 (WFC3) on the {\it Hubble Space Telescope} (\hst). The latter
observations are aimed at the subset of lensed candidates for which AO
observations are not feasible due to the lack of near-by bright stars
for tip-tilt corrections.

The Keck-II/NIRC2 LGSAO imaging (programs C213N2L; PI: H.\ Fu and
U034N2L; PI: A.\ Cooray) uses the $K_{\rm s}$-band filter
(2.2\,\micron) and integration times of 40 and 60\,minutes per
target. The resulting 5$\sigma$ point source detection limit is
$\sim25.6$\,mag for a 0.1\arcsec\ radius aperture. The estimated
Strehl ratios with the LGSAO system are 15 to 25\% at the target
positions. The images reach $\sim$0.1\arcsec\ spatial resolution in
the best cases (Figure~\ref{fig:postage}). These Keck-II/NIRC2 data
are reduced following the standard procedures using a customized
Interactive Data Language ({\sc idl}) pipeline (see \citealt{Fu12}).
The image astrometry is determined relative to SDSS and {\it Spitzer}
images and the flux scale is calibrated with bright stars and galaxies
detected in the 2-Micron All Sky Survey (2MASS).

The candidate lensed SMGs were observed as part of an \hst/WFC3 cycle
19 snapshot program (PI: M.\ Negrello). Data were taken with the F110W
filter (1.1\,\micron) and on-source integration times were at least
four minutes per target. Longer integrations of eight minutes were
used for sources with red SPIRE colors ($S_{500}>S_{350}$), which
indicate that they may be the highest redshift galaxies. The data were
reduced with {\sc
  MultiDrizzle}\footnote{http://stsdas.stsci.edu/multidrizzle}, to
produce images with 0.06\arcsec/pixel and spatial resolution of
$\sim0.14\arcsec$. The typical $5\sigma$ point-source detection limit
is $\sim23.5$\,mag.

When listing the coordinates of the foreground lensing
galaxy in Table~\ref{tab:detail} we make use of either \hst/WFC3 or
Keck-II/NIRC2 imaging. The locations of the background lensed SMG image
components are from the SMA or JVLA data.  The exact data
used for the study of each lensed SMG candidate is discussed
in Section~\ref{sec:indi}.

\subsection{Redshift measurements}
\label{sec:zdata}

Spectroscopic redshifts of the SMGs were measured using data from
multiple facilities, with the aim of observing several CO rotational
transitions \citep[e.g.][]{Weiss09b, Lupu10, Scott11, Riechers11} as
the detection of at least two emission lines is required for a secure
redshift determination. The data are primarily from a large program on
the Combined Array for Research in Millimeter-wave Astronomy (CARMA;
PI: D.\ Riechers), which is supplemented with observations with the
Institut de Radioastronomie Millimétrique (IRAM) Plateau de Bure
Interferometer (PdBI; PIs: A.\ Omont and D.\ Reichers) and the
Zpectrometer spectrometer \citep[][]{Harris07} on the 100-m diameter
GBT (PI: A.\ Harris).

CO redshifts and 3mm continuum photometry are adopted from Riechers et
al. (in preparation) and were obtained through ``blind'' detection of
CO emission lines observed in wide-band frequency scans of the 3mm
atmospheric window with CARMA.  Observations were predominantly taken
in the compact D-array configuration, where the CARMA beam is
5\arcsec\ FWHM. Therefore, the CARMA data are primarily used for
spectral line identification and continuum photometry.  Further
observations were obtained at 1cm and 2mm with the GBT/Zpectrometer
and the IRAM PdBI, which typically target additional CO transitions to
confirm the CARMA redshifts.

In Table~\ref{tab:cand} we provide the CO redshifts and highlight SMGs
with only a single millimeter emission line detection. This results in
a degeneracy between the identification of the line transition and the
redshift. In such cases we make use of photometric redshifts derived
from submillimeter photometry (as described in Section~\ref{sec:nz})
to determine the most likely line identification, assuming that it is
a CO transition. This is a reasonable assumption because CO lines are
the most luminous at the observed frequencies. Additional observations
are required to confirm the redshifts of these sources.

Redshifts of the foreground lensing galaxies (Table~\ref{tab:detail}
and Section~\ref{sec:indi}) are primarily from the public SDSS
database; spectroscopic measurements are used where available and
photometric redshifts otherwise. In cases where SDSS data are
unavailable, the foreground detector is undetected or the photometry
is blended, we have obtained redshifts from optical spectroscopy with
Low Resolution Imaging Spectrometer (LRIS) on Keck-I (PI: C.\ Bridge) and the Optical
System for Imaging and low-Intermediate-Resolution Integrated
Spectroscopy (OSIRIS) on the Gran Telescopio Canarias (GTC; PI: I.\ P\'erez-Fournon).

\subsection{Multi-band photometry}

\spitzer/MIPS 24, 70 and 160\,\micron\ flux densities of the candidate
lensed SMGs are retrieved from archival data, including \spitzer\
Wide-Area Infrared Extragalactic Survey (SWIRE) \citep{Lonsdale03} and
the \spitzer\ Deep, Wide-Field Survey (SDWFS) \citep{Ashby09}.  These
are listed in Table~\ref{tab:allwaves}, in addition to 870\,\micron\
photometry from SMA interferometry (described in
Section~\ref{sec:highres}) and millimeter continuum data from
spectroscopic surveys (Section~\ref{sec:zdata}). The 2 and 2.3\,mm
flux densities are extracted from our PdBI data, and CARMA provides
the 3\,mm data. Previous studies \citep{Conley11, Ikarashi11}
published Caltech Submillimeter Observatory (CSO)/Z-Spec 1\,mm
continuum flux densities for two sources (HXMM02 and HLock01) so these
are also included in Table~\ref{tab:allwaves}.

We also present 1.2\,mm flux densities from the Max-Planck Millimetre Bolometer
(MAMBO) instrument on the IRAM 30-m telescope (collaboration led by A.\
Omont and I.\ P{\'e}rez-Fournon). The observations are reduced and
analyzed using the method described in \citet{Omont03}. The MAMBO beam
is 10.6\arcsec\ (FWHM) at 1.2\,mm and therefore these data cannot be
used for resolved studies, since fully sampled maps were not obtained.


\section{Individual lens candidates}
\label{sec:indi}

\begin{figure*}
  \centering
  \includegraphics[width=16cm]{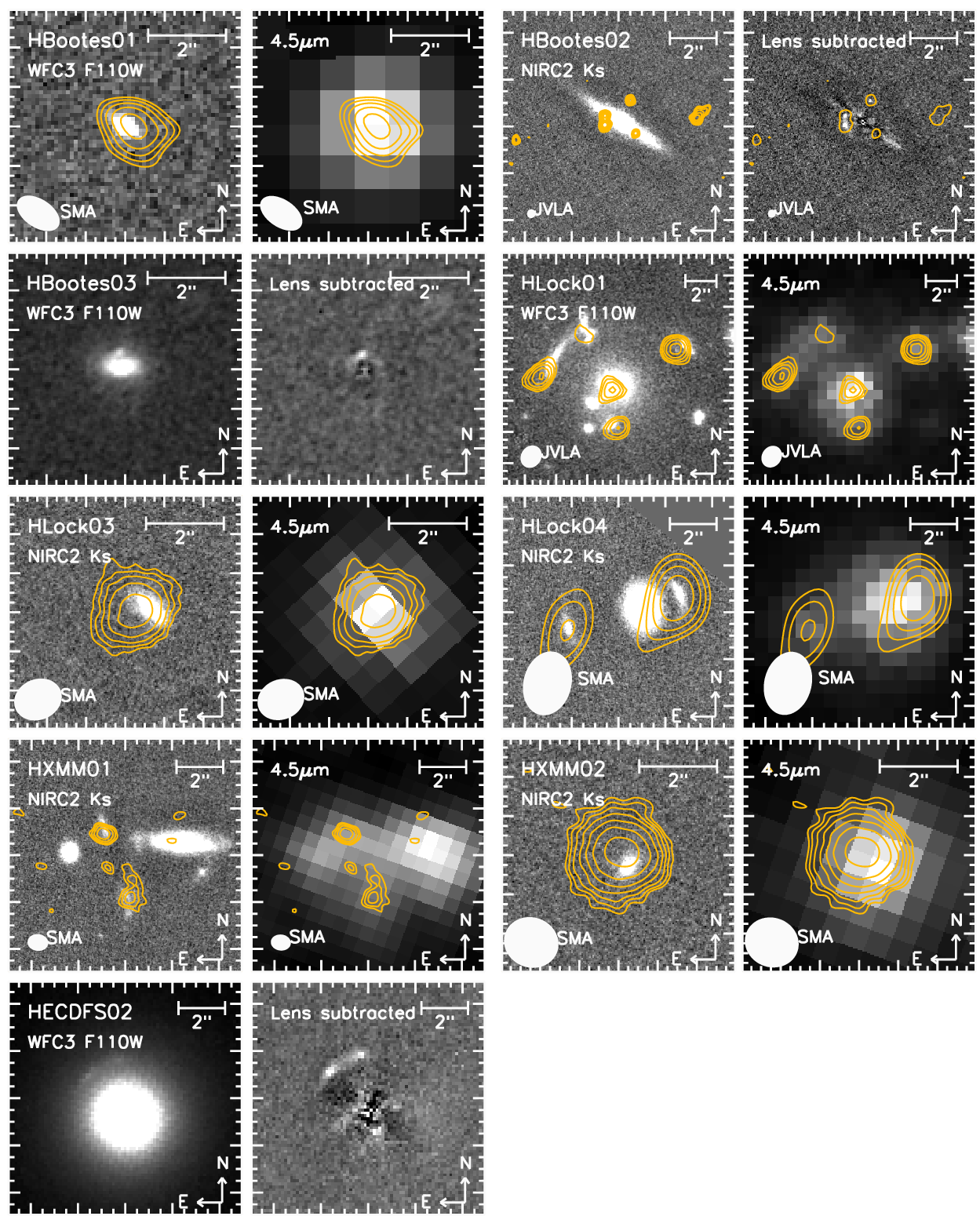}
  \caption{
Images of the nine confirmed gravitationally lensed SMGs from
our sample.
    The left-hand panels show near-infrared high-resolution \hst/WFC3 F110W
    or Keck-II/NIRC2 $K_S$-band data (as
    labeled).
    Archival \spitzer\ IRAC 4.5\,\micron\ imaging is presented in the right-hand
    panels. For the three cases (H\bootes02, H\bootes03 and HECDFS02) where \spitzer\
    IRAC imaging is unavailable we instead show the near-infrared data
    with the foreground lens emission subtracted in the right-hand panel.
The contours on each image are submillimeter (SMA) or  radio (JVLA)
 interferometry (as labeled). JVLA data are at 7\,GHz for H\bootes02 and
 1.4\,GHz for HLock01.  
Contour levels begin at $3\sigma$ and increase by a factor of
    $\sqrt 2$ at each step  (expect for the right-hand panel of
    H\bootes01, where only the $3\sigma$ contours are shown for
    clarity).
    The contours typically trace emission from the background
    submillimeter source, whereas the near-infrared images
    are either dominated by the foreground lens, or contain emission
    from both the foreground and background sources. The exception is
    HLock01, in which the JVLA data traces both the lensed background
    SMG and radio emission in the central lensing galaxy.  A
    $2\arcsec$ scale-bar is shown in the top left-hand corner of each
    image. }
  \label{fig:postage}
\end{figure*}

We next discuss the 13 candidate lensed SMGs identified in
Section~\ref{sec:sample}. Nine of these sources are shown to be
gravitationally lensed. Three of these sources, H\bootes03
\citep{Borys06}, HLock01 \citep{Conley11}, and HXMM02
\citep{Ikarashi11}, have been discussed previously. The four other
sources require additional observations before their lensing nature
can be confirmed.

\begin{figure*}
  \centering
  \includegraphics[width=14cm]{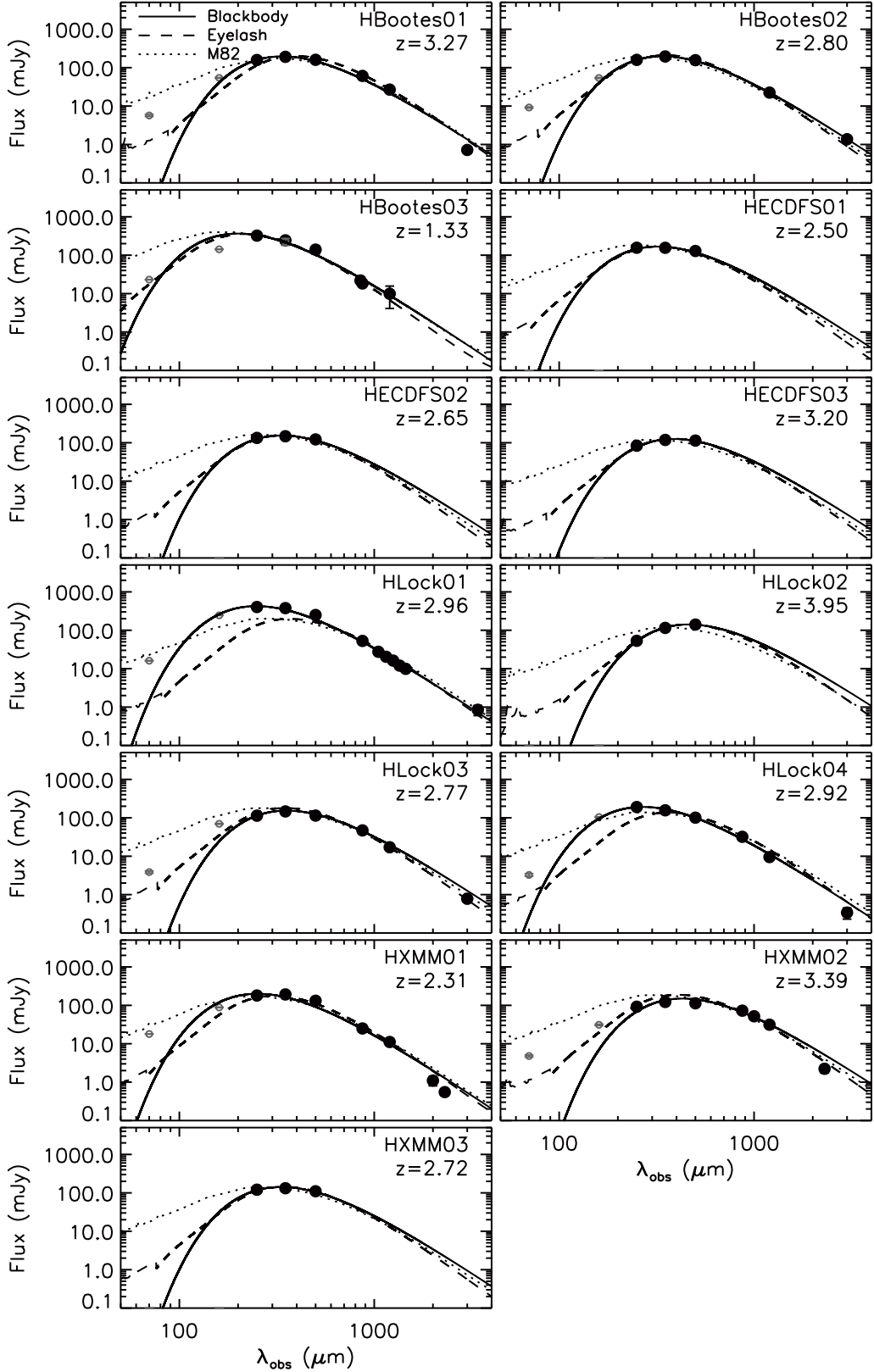}
  \caption{ 
Observed-frame far-infrared SEDs of the 13 \hermes\ candidate
gravitationally lensed SMGs. Photometry from
multiple components is integrated, such that the total flux densities
are presented for each source.
We show
the best-fit optically-thin modified blackbody SEDs, with fixed $\beta=1.5$, corresponding to the
far-infrared luminosities and temperatures presented in
Table~\ref{tab:detail}. The SEDs of M82 \citep{Silva98} and the ``Cosmic
Eyelash'' \citep[\eyelash;][]{Ivison10b, Swinbank10} are redshifted and rescaled to the observed
photometry, and shown for comparison.  
We also show 70 and 160\,\micron\ photometry from archival \spitzer\
MIPS surveys, and SHARC-II 350\,\micron\ observations of H\bootes03
\citep{Borys06}, although these data are not used in the fitting.  }
  \label{fig:seds}
\end{figure*}

Optical and infrared imaging of eight of the nine confirmed lenses are
presented in Fig.~\ref{fig:postage}; the final source not shown here
-- H\bootes03 -- has been discussed extensively in the literature
\citep{Borys06, Desai06, Iono06, Iono06b, Swinbank06, Sturm10,
  HaileyDunsheath10}, with multi-wavelength imaging available in
several of these papers.  The far-infrared SEDs of all 13 candidates
are displayed in Fig.~\ref{fig:seds}, with photometry from \hermes,
our follow-up programs, and archival studies.

The basic properties of the nine confirmed lens systems are summarized
in Table~\ref{tab:detail}. This include the lensing magnification
factor, $\mu$, determined from modeling the best available data. The
lens modeling is performed following the procedure outlined in
Section~2.2 of \citet{Gavazzi11}; we use the code {\sc sl\_fit}, which
have been previously used to study galaxy-scale strong lensing
\citep{Gavazzi07,Gavazzi08}. The foreground lensing potential is
modeled as a cored isothermal ellipsoid. For simplicity in the
modeling and to avoid computational costs associated with pixelized
inversion techniques \citep{WarrenDye03,Treu04,Suyu06}, we make use of
an analytical description of the source plane light distribution by
assuming a Gaussian radial profile with elliptical shape (e.g.,
\citealt{Marshall07,Bolton08b}). Fitting is performed by 
minimizing the difference between the model and the data; free
parameters include the shape of the background SMG and the foreground
mass profile. 

Due to the potential for differential magnification
\citep[e.g.][]{Gavazzi11, Fu12, Hezaveh12, Serjeant12}, we note that these
magnification estimates are only valid for the wavelength at which
they are derived. Systems in which differential magnification could be
important, and one case where modeling at both 2.2 and 880\,\micron\
shows that differential magnification is limited, are highlighted
below.

Two of the lens systems presented here, (HLock01,
Section~\ref{sec:l1}, \citealt{Gavazzi11} and HXMM01,
Section~\ref{sec:x1}, Fu et al.\ in prep.) have complex potentials
with multiple deflectors. The primary focus of this paper is broad
statistics and therefore we do not present individual lens models or
detailed properties of the foreground lensing galaxy or galaxies
here. Such details, including analysis of the foreground lensing
galaxies associated with {\it Herschel} SMGs, will be presented
elsewhere (Gavazzi et al.\ in preparation).  Here, we summarize the
basic properties of the nine confirmed lens systems, focusing
primarily on the background SMGs. Five of the candidates from the
supplementary sample are described in Appendix~\ref{app:sources}.


\subsection{H\bootes01}
\label{sec:b2}

H\bootes01 is the second brightest $500\,\micron$ source in our sample,
with $S_{500}=160\pm33$\,mJy. It is a confirmed gravitationally
lensed SMG, and has a far-infrared SED is similar to M82, with $T_{\rm D}\sim40$\,K
for fixed $\beta=1.5$ and $\mu L_{\rm IR}\sim6\times10^{13}{\rm L_{\sun}}$
(Fig.~\ref{fig:seds}; Table~\ref{tab:detail}).  SMA observations in the extended and compact array configurations detect
a single bright source, with $S_{870}=61\pm3$\,mJy. H\bootes01 has a
1.4\,GHz flux density of $0.26\pm0.04$\,mJy \citep{deVries02}, which is
consistent with that expected from star-formation, as calculated from the
far-infrared/radio correlation with the method described in
Section~\ref{sec:sample}.

Fig.~\ref{fig:postage} shows SMA contours on \hst/WFC3 F110W
  imaging and SDWFS 4.5\,\micron\ images.  The submillimeter emission
is slightly offset from a source in the \hst\ data and
  coincident with an IRAC source. The \hst\ source is also
  detected in the NOAO Deep Wide-Field Survey (NDWFS)
  \citep{Jannuzi99} and SDSS. It is red in the optical, and has
  $r=21.71$\,mag and $z_{\rm phot}=0.59\pm0.08$ in SDSS. \cco\ and
\dco\ are detected in our CARMA data, placing the submillimeter source
at $z=3.274$.  Therefore, it is clear that the optical and
  near-IR emission, and the submillimeter emission are dominated by
two distinct sources at different redshifts.  Despite the existence of
two sources at different redshifts and the offset between the
  near-IR and far-IR emission, there are no arcs or other
morphological indications of gravitational lensing in the SMA
data. Therefore the amplification from lensing is expected to be
  small: $\mu\lesssim5$.

The configuration of H\bootes01 is similar to HXMM02
(Section~\ref{sec:x6}; \citealt{Ikarashi11}) and HLSJ091828.6+514223
\citep{Combes12}. All three galaxies are gravitationally lensed, as
determined from redshift information, and have small ($\la1\arcsec$)
offsets between the submillimeter and optical sources.  However,
despite high-resolution submillimeter interferometry there are no
morphological indications of lensing and we are unable to determine
the precise lensing magnification.


\subsection{H\bootes02}
\label{sec:b1}

In the \hermes\ SPIRE data H\bootes02 is photometrically similar to
H\bootes01, with $S_{500}=157\pm33$\,mJy and an SED that peaks in the
350\,\micron\ band. A single emission line is detected in CARMA
observations, which is confirmed to be \cco\ at $z=2.804$ with
  the detection of \aco\ by the GBT. SED fitting of the far-IR data
  show that H\bootes02 has $T_{\rm D}=34\pm1$\,K and $\mu L_{\rm
    IR}=3.8^{+0.4}_{-0.3}\times10^{12}{\rm L_{\sun}}$, for fixed
  $\beta=1.5$ (Table~\ref{tab:detail} and Fig.~\ref{fig:seds}).

Optical data reveal that H\bootes02 is gravitationally lensed by the
central region of an edge-on disk galaxy at $z=0.414$
(Fig.~\ref{fig:postage}), which, due to the presence of dust-lane and its inclination angle, most likely has a higher mass density than would be estimated from its optical magnitude. The redshift of the deflector is determined
from multiple emission and absorption lines observed in long-slit
spectroscopy with OSIRIS on the GTC. We have also obtained a JVLA
7\,GHz map of H\bootes02, which reveals four images of the background
source, which are created due to gravitational lensing distortion. The
images are not resolved in the 0.25\arcsec\ JVLA beam, and the total
flux density is $2.3\pm0.2$\,mJy at 7\,GHz.

Lens modeling of the radio data indicate that the system has an
Einstein radius of $0.80\pm0.04\arcsec$ and is magnified by
$\mu_{\rm JVLA}\sim23$ at 7 GHz. Thus, H\bootes02 is the most highly magnified
source in our sample. The large magnification is due to the small size
of the radio emission region, which is $\le0.008\arcsec$ ($\le65$\,pc
at $z=2.8$) in the source plane. Since the brightness temperature is
much greater than $10^4$\,{\sc k} these data show that an AGN is
powering the radio emission.

H\bootes02 is also detected with a flux density of $12.36\pm0.50$\,mJy at
1.4\,GHz \citep{deVries02}, although its components are unresolved in those
data. The observed radio spectral index between 1.4 and 7~GHz is
$\alpha=-1.04$. For $\alpha=-1.04$ the expected flux density from
star-formation is only $\sim0.07$\,mJy and $\sim0.01$\,mJy at 1.4 and
7\,GHz, respectively (based on the far-infrared/radio correlation;
Section~\ref{sec:sample}).  Therefore, the bright radio flux densities in
H\bootes02 confirm that an AGN dominates the emission at these
wavelengths. 

H\bootes02 has also been observed with the SMA at 870\micron\ in
extended and compact array configurations, resulting in a map with a
beam of $\sim0.7\arcsec$, in which a partial ring structure is
resolved. Accounting for the larger SMA beam, the submillimeter
emission appears to be more extended than the radio emission, and the
flux ratios between the components differ between the two
datasets. This suggests that differential lensing is important in
H\bootes02, and that the submillimeter data trace the star-formation
and the radio data trace the AGN.  Indeed, lens modeling of the SMA
data determines that the submillimeter emission from H\bootes02 has
$\mu_{\rm SMA}=10.1\pm1.6$  and a half-light radius of
  $0.15\pm0.03\arcsec$ ($1.2\pm0.2$\,kpc at $z=2.8$).  Note that the
  submillimeter magnification (derived from the SMA data) is
  significantly lower than that at radio wavelengths (from the JVLA
  data), which is indicative of differential lensing resulting from
  two different emission regions.

H\bootes02 is also point-like in Keck $K$-band data, which indicates
that the AGN may also be dominating the near-infrared emission from
H\bootes02. However, lens modeling is not performed on those data
because the foreground deflector contains a complex dust lane. A more
reliable analysis is available from the JVLA and SMA data, which
are not complicated by the dust lane in the foreground galaxy.
We note that the best-fit lens models based on the radio and SMA
  data require deflectors with an SIE profile and high ellipticity, as
  indicated by the near-IR imaging.  A detailed analysis of the
  multi-wavelength observations of H\bootes02 will be presented in
  Wardlow et al. (in prep.).

We note that there are two additional sources detected in the JVLA data,
approximately 3\arcsec\ and 5\arcsec\ west and east of the center of
the lensing galaxy, respectively. Both of these sources are resolved,
although they are not detected in optical or near-infrared imaging, so
their nature is unclear. It is possible that they are relic jets from H\bootes02.


\subsection{H\bootes03}
\label{sec:b10}

H\bootes03, also known as MIPS~J142824.0+352619, is a gravitationally
lensed source that was discovered in \spitzer\ MIPS imaging of the
\bootes\ field \citep{Borys06}. Its has been subject to extensive
analysis and follow-up observations, including broad-band submillimeter
photometry from SHARC-II (350\,\micron), SCUBA (850\,\micron;
\citealt{Borys06}), and SMA (890\,\micron) with $2.5\arcsec$ resolution
\citep{Iono06}.  Optical, near- and mid-infrared spectroscopy revealed
the presence of a background infrared source at $z=1.325$ and a
foreground optical galaxy at $z=1.034$ \citep{Desai06, Borys06}.  These
data confirmed the nature of H\bootes03 as a gravitationally lensed
starburst galaxy, and the existing data are consistent with our new
\hermes\ SPIRE photometry (see Fig.~\ref{fig:seds}).
We have fit the new SPIRE and existing submillimeter photometry with
an optically-thin modified blackbody with fixed $\beta=1.5$, to yield
$T_D=41\pm1$\,K and $\mu L_{\rm
  IR}=(5.6\pm0.5)\times10^{13}L_{\sun}$. 

\spitzer\ IRS spectroscopy shows that the mid-infrared emission is
dominated by star-formation, with no evidence for the an AGN
\citep{Desai06}. There are no morphological indicators of lensing in
the submillimeter or radio data and \citet{Borys06} used size
arguments and positional offsets between the radio/submillimeter and
optical sources to estimate that the lensing magnification,
$\mu\la10$.  We have since obtained \hst/WFC3 F110W imaging of
  H\bootes03, in which a second, faint, near-IR source is present
  (Fig.~\ref{fig:postage}). This second source is $\sim0.4\arcsec$ to
  the north of the $z=1.325$ galaxy and its position is consistent
  with it being the background SMG. The small separation between these
  sources, the point-like nature of the second source, and the absence
  of counter-images further supports the hypothesis that H\bootes03 is
  only magnified by a small factor. Near-infrared integral field
spectroscopy of the H$\alpha$ emission, does not contain AGN
signatures and shows that the emission is spatially extended, although
similar effects are not observed in $H$ or $K$-band continuum data
\citep{Swinbank06}. On the basis of the IFU data, \citet{Swinbank06}
argue that it is possible that H\bootes03 has $\mu\gg10$ for
rest-frame optical emission, although the requisite precise alignment
of the system makes this unlikely.

Numerous far-infrared and submillimeter emission lines have been
detected in H\bootes03, including \bco\ and \cco\ \citep{Iono06b}.
The \bco\ emission is not spatially extended in the $1.3\arcsec$ beam,
which \citet{Iono06b} use to argue that $\mu\la8$ for gas emitting
region.  Recently, the \cii\ fine-structure transition was detected
\citep{HaileyDunsheath10, Stacey10}, and \oi\ and \oiii\ were observed
with \herschel\ PACS \citep{Sturm10}.  The emission line ratios in
H\bootes03 are similar to those observed in the nuclei of local
starburst galaxies, indicating that H\bootes03 may be a high-redshift,
high-luminosity analog of those galaxies \citep{HaileyDunsheath10,
  Sturm10}.


\subsection{HECDFS01}

HECDFS01 is near the edge of the \hermes\ area, which means that it is
located outside of the main ECDFS field in a region without deep
data. It has not yet been observed by any of our follow-up programs and
as such the only available data are from \hermes\ and shallow, all-sky
surveys. Therefore, we are unable to confirm whether HECDFS01 is indeed
a gravitationally lensed SMG. The SPIRE colors are flat, giving a
submillimeter photometric redshift of $z\sim2.50$, although this is not
well-constrained. We note that the galaxy 2dFGRS~S401Z151 at $z=0.222$
is only 32\arcsec\ from the SPIRE centroid, although we cannot confirm
whether it is related to the submillimeter emission.


\subsection{HECDFS02}
\label{sec:e105}

HECDFS02, like HECDFS01, is in a region without deep archival
data. However, follow-up imaging with \hst/WFC3 in the F110W filter
reveals gravitational arcs around a bright galaxy
(Fig.~\ref{fig:postage}). 
The SPIRE fluxes of HECDFS02 peak in the 350\,\micron\
band, which indicates that it is at $z\sim2.65$. Due to the
  position of HECDFS02 outside of the main ECDFS area there is
  inefficient data to determine a photometric redshift for the
foreground lens. However, we have used the \citet{Kormendy77} relation
for the $J$-band \citep{deVries00} to estimate that the lensing galaxy
is at $z\sim0.1$.
 The Einstein radius is $1.678\pm0.003\arcsec$
and modeling of the \hst\ data show two emitters in the source plane,
which may be two star-forming regions in a single SMG, or may be a 
  merger or interaction of two galaxies. The
amplification from lensing is $\mu_{\rm F110W}=3.79\pm0.02$ at
1.1\,\micron.  However, differential lensing may be important, so this
value may not apply to the submillimeter data.


\subsection{HECDFS03}

HECDFS03 is also outside the deep survey fields and only the
  follow-up data available are F110W \hst/WFC3 imaging from our
  snapshot program. There are no arcs or other lensing features in the
  \hst\ data, and although the stucture of the detected sources appear
  to resemble a group or small cluster, we cannot verify whether 
  HECDFS03 is gravitationally lensed. The SPIRE colors of HECDFS03
indicate that the submillimeter emission originates at $z\sim3.20$.


\begin{deluxetable*}{lccccccccc}
\tablewidth{0pt}
\tablecaption{Confirmed \hermes\ gravitationally
  lensed galaxies
 \label{tab:detail}}
\tablehead{\colhead{Source} & \colhead{${\rm RA}_{\rm SMG}^a$} & \colhead{${\rm Dec}_{\rm SMG}^a$}
  & \colhead{${\rm RA}_{\rm lens}^b$} & \colhead{${\rm Dec}_{\rm lens}^b$} &
  \colhead{$z_{\rm CO}$$^c$} & \colhead{$z_{\rm opt}$$^d$} & \colhead{$\mu^e$} &
  \colhead{$\mu L_{\rm IR}$$^f$} & \colhead{$T_{\rm D}$$^g$} \\
 & (J2000) & (J2000) & (J2000) & (J2000) & & & & \colhead{($10^{13}{\rm L_{\sun}}$)} & \colhead{(K)} }
\startdata
H\bootes01  & $14^{\rm h}33^{\rm m}30\fs83$ & $+34\degr54\arcmin39\farcs9$ & $14^{\rm h}33^{\rm m}30\fs84$ & $+34\degr54\arcmin40\farcs0$ & $3.274$  & $0.59\pm0.08$ & $\lesssim5$ (O) & $5.6\pm0.5$ & $41\pm1$  \\
H\bootes02  & $14^{\rm h}28^{\rm m}25\fs54$ & $+34\degr55\arcmin46\farcs9$ & $14^{\rm h}28^{\rm m}25\fs47$ & $+34\degr55\arcmin46\farcs8$ & $2.804$ & 0.414 & $\sim23$ (R) &$3.8_{-0.3}^{+0.4}$ &  $34\pm1$ \\ 
                     & $14^{\rm h}28^{\rm m}25\fs54$ & $+34\degr55\arcmin47\farcs2$ & \nodata & \nodata & \nodata & \nodata & \nodata & \nodata & \nodata  \\ 
                     & $14^{\rm h}28^{\rm m}25\fs45$ & $+34\degr55\arcmin47\farcs9$ & \nodata & \nodata & \nodata & \nodata & \nodata & \nodata & \nodata  \\
                     & $14^{\rm h}28^{\rm m}25\fs43$ & $+34\degr55\arcmin46\farcs5$ & \nodata & \nodata & \nodata & \nodata & \nodata & \nodata & \nodata  \\ 
H\bootes03$^h$  & $14^{\rm h}28^{\rm m}24\fs06$ & $+35\degr26\arcmin19\farcs8$ & $14^{\rm h}28^{\rm m}24\fs08$ & $+35\degr26\arcmin19\farcs5$ & $1.325$  & 1.034 & $\la10$ (O) & $1.9_{-0.2}^{+0.3}$ & $37\pm1$ \\
HECDFS02   &  $03^{\rm h}37^{\rm m}32\fs53$  & $-29\degr53\arcmin51\farcs8$   &  $03^{\rm h}37^{\rm m}32\fs40$  &  $-29\degr53\arcmin53\farcs6$  &  ?  & ?   &  $3.8\pm0.02$ (O)  & $2.6\pm0.6$  & $35\pm2$   \\
                     &  $03^{\rm h}37^{\rm m}32\fs42$ & $-29\degr53\arcmin51\farcs1$ & \nodata & \nodata & \nodata & \nodata & \nodata & \nodata & \nodata  \\ 
HLock01      & $10^{\rm h}57^{\rm m}50\fs46$ & $+57\degr30\arcmin28\farcs5$ & $10^{\rm h}57^{\rm m}50\fs96$ & $+57\degr30\arcmin25\farcs7$ & $2.958$  & $0.60\pm0.04$ & $10.9\pm0.7$ (O) & $12.6\pm0.5$ & $51\pm1$ \\ 
                     & $10^{\rm h}57^{\rm m}51\fs54$ & $+57\degr30\arcmin26\farcs8$ & \nodata & \nodata & \nodata & \nodata & \nodata & \nodata & \nodata  \\ 
                     & $10^{\rm h}57^{\rm m}51\fs19$ & $+57\degr30\arcmin29\farcs1$ & \nodata & \nodata & \nodata & \nodata & \nodata & \nodata & \nodata  \\ 
                     & $10^{\rm h}57^{\rm m}50\fs91$ & $+57\degr30\arcmin23\farcs8$ & \nodata & \nodata & \nodata & \nodata & \nodata & \nodata & \nodata  \\ 
HLock03      & $10^{\rm h}57^{\rm m}12\fs26$ & $+56\degr54\arcmin58\farcs7$ & $10^{\rm h}57^{\rm m}12\fs22$ & $+56\degr54\arcmin58\farcs6$ & $2.771^i$ & ? & $3.0^{+1.3}_{-1.4}$ (O) & $2.8\pm0.2$ & $34\pm1$ \\
HLock04      & $10^{\rm h}38^{\rm m}26\fs60$ & $+58\degr15\arcmin42\farcs6$ & $10^{\rm h}38^{\rm m}26\fs76$ & $+58\degr15\arcmin42\farcs3$ &    ?     & $0.58\pm0.04$& $6.2\pm0.1$ (O) & $5.2\pm0.6$ & $47\pm1$ \\ 
             &                               &                              &                             &                               &           &              & $5.3^{+1.3}_{-1.1}$ (D) &  & \\
                     & $10^{\rm h}38^{\rm m}27\fs19$ & $+58\degr15\arcmin41\farcs3$ & \nodata & \nodata & \nodata & \nodata & \nodata & \nodata & \nodata  \\
HXMM01       & $02^{\rm h}20^{\rm m}16\fs65$ & $-06\degr01\arcmin41\farcs9$ & $02^{\rm h}20^{\rm m}16\fs73$ & $-06\degr01\arcmin42\farcs7$ & $2.307$ & $0.654/0.502^j$ & $\sim 2.1^{j}$ (O) & $3.2\pm0.3$ & $43\pm1$  \\ 
                     & $02^{\rm h}20^{\rm m}16\fs58$ & $-06\degr01\arcmin44\farcs3$ & $02^{\rm h}20^{\rm m}16\fs40$ & $-06\degr01\arcmin42\farcs3$ & \nodata & ... & $\sim 1.6^{j}$ (O) & \nodata & \nodata \\ 
HXMM02       & $02^{\rm h}18^{\rm m}30\fs67$ & $-05\degr31\arcmin31\farcs5$ & $02^{\rm h}18^{\rm m}30\fs69$ & $-05\degr31\arcmin31\farcs9$ & $3.395$ & $1.35$ & $1.5^{+1.0}_{-0.4}$ (O) & $3.6_{-0.2}^{+0.3}$ & $33\pm1$ 
\enddata
\tablecomments{ 
$^a$ SMG coordinates are the centroids of high-resolution submillimeter
or radio interferometry, or optical arcs (as discussed in the text). Multiple coordinates
are listed where there are multiple images. 
$^b$ Coordinates of the foreground lenses are measured from optical
imaging. Multiple coordinates are listed for XMM01, where there are
two deflectors.
$^c$ $z_{\rm CO}$ is the redshift of the SMG, as traced by CO emission
lines (Riechers et al., in preparation). 
$^d$ $z_{\rm opt}$ is the redshift of the lens, as traced by optical
spectroscopic or photometric redshifts. 
$^e$ $\mu$ is the magnification factor from gravitational lensing. We denote the high-resolution image used to determine lensing magnification with R for radio (JVLA 7 GHz),
O for rest-frame optical (Keck-II or {\it HST} image), or D for dust emission with SMA 880 $\mu$m.
$^f$ $\mu L_{\rm IR}$ is the {\it apparent} rest-frame 8-1000\,\micron\ infrared
luminosity of the SMG. 
$^g$ $T_{\rm D}$ is the dust temperature of the SMG, obtained by fitting an
optically-thin modified
blackbody, with $\beta=1.5$ to all the available 250--3000\,\micron\
photometry (Table~\ref{tab:allwaves}). 
$^h$ H\bootes03 was identified as lensed by \citet{Borys06}, and has
been the subject of numerous additional studies (see Section~\ref{sec:b10})
$^i$ The CO redshifts of these sources are single line redshifts guided
by the submillimeter photometric redshift. 
$^j$ The two $z_{\rm opt}$ values are for the two foreground galaxies.
Furthermore magnification factors for HXMM01 are given separately for the two lensed components as the source plane is made up of two merging SMGs (see Section~6.11).
}
\end{deluxetable*}


\subsection{HLock01}
\label{sec:l1}

HLock01 is the brightest gravitationally lensed \hermes\ source, with
250, 350 and 500\,\micron\ flux densities of $403\pm7$, $377\pm10$ and
$249\pm7$\,mJy, respectively (Table~\ref{tab:cand}).  Due to its
extreme brightness, HLock01 was subject to a significant follow-up
effort and it was the first confirmed \hermes\ lensed SMG
\citep{Conley11}. 880\,\micron\ SMA interferometry resolved the SPIRE
source into four components \citep{Conley11} which were further
resolved in Keck-II/NIRC2 $K$-band imaging.  The redshift of HLock01
was established from CO emission lines as $z=2.958$ \citep{Scott11,
  Riechers11}, and \citet{Gavazzi11} used the NIRC2 $K$-band data to
show that the rest-frame optical emission is magnified by a factor of
$\mu=10.9\pm0.7$ by a small group at $z\sim0.6$. The
  the central group galaxy has a photometric redshift of $0.60\pm0.04$
  \citep{Oyaizu08}; photometric or spectroscopic redshifts are
  unavailable for the remaining group members due to their
  faintness. Instead, the group nature of the deflector is confirmed by the
  Einstein radius: $R_{\rm Ein}=4\farcs10\pm0\farcs02$, corresponding
  to a velocity dispersion, $\sigma_v=483\pm16$~km~s$^{-1}$ \citep{Gavazzi11}.

Thorough SED fitting of HLock01 was performed by \citet{Conley11}, who
find that it contains optically thick warm dust, with $T_{\rm
  D}=88\pm3$\,K for $\beta=1.95\pm0.14$ (note that for consistency with
the rest of our sample table~\ref{tab:detail} gives $T_{\rm D}$ for an
optically-thin modified blackbody with $\beta=1.5$). \citet{Scott11}
analyzed the CO spectral line energy distribution (SLED) of HLock01,
including observations of the $J$=1$\to$0, 3$\to$2, 5$\to$4, 7$\to$6,
8$\to$7, 9$\to$8 and 10$\to$9 transitions. They found that a single
warm, moderate density gas model adequately describes the data,
although a second denser component may also be present. Finally,
\citet{Riechers11} examined the dynamics of the gas in HLock01, by
studying the resolved CO lines. They showed that the \eco\ emission
exhibits resolved velocity structure, consistent with HLock01 being a
gas-rich merger at $z\sim3$.

Since the publications of \citet{Conley11}, \citet{Gavazzi11},
\citet{Scott11} and \citet{Riechers11} we have obtained \hst\ WFC3
F110W and JVLA 1.4\,GHz observations of HLock01, which are shown in
Fig.~\ref{fig:postage}. The radio data have $\sim1.1\arcsec$
resolution and isolate the four lensed images of HLock01 with higher
accuracy than the existing SMA data \citep{Conley11}, and show that
the central lensing galaxy is also radio-bright. The total flux
density in the
four radio images of HLock01 is $0.97\pm0.05$\,mJy, which is $\sim4$
times higher than expected from the far-infrared/radio correlation, if
$q_{\rm IR}=2.40$ (\citealt{Ivison10a}; see Section~\ref{sec:sample}
for details).  Indeed, HLock01 requires $q_{\rm IR}=1.8\pm0.4$ if
both the far-infrared and the radio emission are powered by
star formation. The luminous radio emission from HLock01, coupled with
its warm dust temperature ($T_{\rm D}=88.0\pm2.9$\,K; \citealt{Conley11}),
suggests that HLock01 harbors both AGN and star-formation activity.

The JVLA data also reveal a radio source $\sim75\arcsec$ north of
HLock01, which is composed of two bent lobes, indicative of an FR~II
source. Since bent FR~II sources are most likely to be located in
galaxy groups and clusters \citep[e.g.][]{Stocke79}, this observation adds weight to the
analysis of \citet{Gavazzi11} that the foreground lens is a small
group or cluster of galaxies.


\subsection{HLock02}

We have obtained \hst/WFC3 F110W imaging of HLock02. No lensing
features are apparent in the data and we do not have submillimeter or
radio interferometry to precisely pinpoint the source of the
submillimeter emission.  HLock02 has red SPIRE colors, indicative of a
source at $z\sim4$ (Fig.~\ref{fig:seds}), so the absence of lensing
features in the WFC3 image may be because the SMG is below the
detection threshold of these data.  Therefore, we cannot determine
whether or not HLock02 is gravitationally lensed.


\subsection{HLock03}

HLock03 has $S_{500}=114\pm8$\,mJy with submillimeter emission peaking
in the SPIRE 350\,\micron\ band. Continuum emission is detected with
the SMA at 890\,\micron, MAMBO at 1.2\,mm and CARMA at 3mm
(Fig.~\ref{fig:seds}). A single emission line is detected in CARMA
spectroscopy, which, on the basis of the submillimeter colors, we
attribute to \cco\ at $z=2.771$. If the emission is instead \bco\ then
HLock03 would be at $z=1.514$, although the continuum SED makes this
solution unlikely.

SMA interferometry from the compact and very extended array
configurations, has a combined beam of $1.18\times0.97\arcsec$ and
identifies HLock03 as a single submillimeter source
(Fig.~\ref{fig:postage}). The submillimeter emission is coincident
with an IRAC source, but offset by $\sim0.5\arcsec$ from the galaxy
that is detected in both the Keck-II/NIRC2 $K_{\rm S}$-band and
\hst/WFC3 F110W imaging (Table~\ref{tab:detail}). The $K_{\rm S}$-band
source is faint and therefore its redshift is unknown.

The most likely interpretation of these data is that the submillimeter
and IRAC data are tracing a background source at $z=2.771$ which is
gravitationally lensed by the $K_{\rm S}$-band galaxy. In this case
the absence of prominent detected lensing features means that the
magnification of the SMG is likely to be small, and indeed lens
modeling of the $K_S$-band image determines that $\mu=3.0^{+1.3}_{-1.4}$
for rest-frame optical emission.  Therefore, HLock03 is intrinsically
luminous, with $L_{\rm IR}\sim10^{13}\L_{\sun}$ and ${\rm
  SFR}\sim1700\,M_{\sun}{\rm yr}^{-1}$ \citep{Kennicutt98}.


\subsection{HLock04}

Gravitational lensing is confirmed in HLock04, and we have obtained
extensive follow-up observations, including Keck-II/NIRC2
$K_{\rm S}$-band and \hst/WFC3 F110W imaging, SMA interferometry and GBT and
CARMA spectroscopy.  Gravitationally lensed arcs are
evident in the Keck-II, \hst\ and SMA data with an Einstein radius of
$2.46\arcsec$ (Fig.~\ref{fig:postage}). The foreground lensing galaxy is
detected in SDSS, and has a photometric redshift of $z=0.60\pm0.02$
from that survey. Line emission from the lensed SMG  was not detected
in broad-band submillimeter spectroscopy, so the redshift of the SMG is
unknown, although the far-infrared photometry indicates $z\sim2$. 

HLock04 is unique in our sample, in that the lensing morphology is
detected with high significance in both the near- and far-infrared
wavelength regimes, enabling separate lens modeling of both the
stellar and dust emission. \citet{Serjeant12} showed that differential
lensing can be significant for lensed SMGs, due to
the irregular distribution of stars, AGN, gas and dust in source
plane. Indeed,  differential magnification has been identified in two  lensed
\herschel\ SMGs \citep{Gavazzi11, Fu12}, due to
source-plane offsets between the gas, dust, and stellar components. 

In the case of HLock04 the SMA data at 880 $\mu$m  trace the dust emission, which is
magnified by $\mu=5.32^{+1.28}_{-1.06}$. The stellar emission (traced
by the $K_{\rm S}$-band data) is magnified by $\mu=6.17\pm0.03$ and has
$R_{\rm eff}= 0.171\pm0.004\arcsec = 1.33\pm0.03$\,kpc in the source
plane.  The SMA data constrain the size of the emitting dust to
$R_{\rm eff}<0.5\arcsec$ ($<3.9$\,kpc). Therefore,
the stellar and dust models are consistent and we find no significant
differential magnification in HLock04, at least for dust and stellar emission.


\subsection{HXMM01}
\label{sec:x1}

SMA interferometry, coupled with optical imaging shows that HXMM01 is
gravitationally lensed by two galaxies to the east and west of the
submillimeter emission (Fig.~\ref{fig:postage}). Due to this unique
configuration we have obtained extensive follow-up observations of
HXMM01, which will be published in detail in Fu et al.\ (in
preparation); here we summarize those results.

Keck-I/LRIS spectroscopy reveals that the eastern and western lensing
galaxies are at $z=0.655$ and $z=0.502$, respectively, and are thus
not physically associated.  CO spectroscopy of the CO $J=$1$\to$0,
3$\to$2 and 4$\to$3 transitions shows that HXMM01 is at
$z=2.307$. We have also observed HXMM01 with the F110W filter on
  \hst/WFC3; these data are deeper, but lower resolution, than the
  Keck-I/LRIS imaging. The \hst\ imaging is consistent with the Keck
  data, but also reveal the presence of extended emission around the
  eastern lensing galaxy. This extended emission is inconsistent with
  being lensed emission from the either of the two submillimeter
  sources, although it may be lensed emission from a third
  submillimeter-faint galaxy. It is also possible that this extended
  emission originates from spiral structure in the foreground galaxy.

HXMM01 is a unique case in that the existing follow-up data clearly
show that it is made up of two individual, although interacting, SMGs
in the source plane. For example, H$\alpha$ is detected in
Keck-II/NIRSPEC long-slit spectroscopy with a $700\,\rm{km s^{-1}}$
velocity difference between the northern and southern components. The
\aco\ data also exhibit resolved velocity structure between the two
components of the same velocity difference, further supporting a
scenario in which the two peaks of SMA emission are two distinct
sources separated by $2.8\arcsec$ (23\,kpc at $z=2.3$).  Lens modeling
with two separate sources shows that the norther source is lensed by a
factor of $\sim$ 2.1 and the southern source by a factor of $\sim$
1.6, for a total of $\sim$ 3.7. No multiple images are predicted.  The
optical H$\alpha$ and [\ion{N}{2}] emission lines are broadened in
both components of HXMM01 and may trace AGN emission in both galaxies.


\subsection{HXMM02}
\label{sec:x6}

HXMM02 is the brightest source in 1.1\,mm AzTEC observations of the
Subaru/{\it XMM-Newton} Deep Field (SXDF), and as such is also known as
SXDF1100.001 or Orochi \citep{Ikarashi11}. The source is also bright at
250, 350 and 500\,\micron, with respective flux densities of $190\pm7$,
$192\pm8$ and $132\pm7$\,mJy in the three SPIRE bands.

\citet{Ikarashi11} derived a photometric redshift of $z_{\rm
  phot}=3.4^{+0.7}_{-0.5}$ for HXMM02, using continuum measurements
at 880--1500\,\micron\ and in the radio at 20 and 50\,cm.  We have
since obtained additional submillimeter spectroscopy and detected
\aco\ (GBT; Inoue et al., in prep.), \cco\ (CARMA; Riechers et al., in
prep.), \dco\ (PdBI) and \eco\ (CARMA; Riechers et al., in prep.),
which pinpoint HXMM02 at $z=3.395$. We note that there is an
additional \dco\ measurement obtained with the NRO 45-m telescope
\citep{Iono12}.

HXMM02 is in the SXDF region of the XMM-LSS SWIRE field and therefore
deep Subaru SuprimeCam and UKIRT WFCAM imaging are available
\citep{Furusawa08, Warren07}. \citet{Ikarashi11} showed that a faint
optical galaxy, with $z_{\rm phot}=1.39\pm0.01$, is located
$0.5\arcsec$ from the submillimeter centroid. We have since obtained
near-infrared spectroscopy with the Infrared Spectrometer And Array
Camera (ISAAC) on the Very Large Telescope (VLT) and detect H$\alpha$
at $z=1.33$.  The positional offset between the optical and the
submillimeter sources, coupled with the redshift difference, is
indicative of a SMG being gravitationally lensed by the optical
galaxy.

HXMM02 has also been imaged with NIRC2 and the Keck-II AO system
(Fig.~\ref{fig:postage}).  No arcs or other morphological indications
of gravitational lensing are observed, which is similar to
H\bootes01 and HLSJ091828.6+514223 \citep{Combes12}.  Modeling of the
SMA data derives a magnification of $\mu=1.5^{+1.0}_{-0.4}$ for
HXMM02. The apparent 8--1000\,\micron\ luminosity of HXMM02 is $\mu
L_{\rm IR}=(3.6^{+0.3}_{-0.2})\times10^{13}\,{\rm L_{\sun}}$
(Fig.~\ref{fig:seds}), thus the modest amplification from lensing
means that the intrinsic luminosity (magnification corrected) is
$L_{\rm IR}=(2.4^{+0.6}_{-1.5})\times10^{13}\,{\rm L_{\sun}}$. HXMM02
is an extreme source, even once the effect of gravitational lensing
has been accounted for. If the far-infrared emission is dominated by
star-formation it has intrinsic (lensing-corrected)
$\rm{SFR}\sim4000\,M_{\sun}{\rm yr}^{-1}$ \citep{Kennicutt98}.


\subsection{HXMM03}

A single line is tentatively detected in CARMA spectroscopy of HXMM03, which, due
to the submillimeter photometric redshift, we consider to be \cco\ at
$z=2.72$ (Riechers et al., in prep.). High-resolution submillimeter and radio interferometry is not
available and therefore the SMG cannot be pinpointed. No gravitational lensing features
are detected in the \hst/WFC3 F110W imaging, and without an
accurate location for the submillimeter emission we are unable to
determine whether or not HXMM03 is gravitationally lensed.


\section{Conclusions}

We used \herschel-SPIRE photometry from the \hermes\ survey to
investigate the nature and prevalence of SMGs that are magnified by
strong galaxy-galaxy gravitational lensing, where we consider
  strong lensing as that with magnification factor, $\mu\ge2$. The main results are:

\begin{enumerate}

\item We have identified 13 candidate strongly lensed SMGs
  with $S_{500}\ge100$\,mJy in 94.8\,deg$^2$ of blank-field
  \hermes\ data ($0.14\pm0.04\,{\rm deg}^{-2}$). We also identified a
  supplementary sample of 29 candidates ($0.31\pm0.06\,{\rm deg}^{-2}$)
  with $S_{500}=80$--100\,mJy, which have a higher rate of contamination
  from intrinsically luminous galaxies.

\item Extensive follow-up data for nine (70\%) of the sources showed
  that all nine are strongly gravitationally lensed, with magnification
  factors of $\mu\sim2$--23. Our sample, in combination with data from
  \atlas\ \citep{Negrello10}, demonstrates that $\ga80\%$ of candidates
  with $S_{500}\ge100\,$mJy are strongly lensed.

\item We showed that the candidate gravitationally lensed sources have
  red submillimeter colors, indicative of a high-redshift
  population. Indeed, follow-up data confirm that most of the lensed
  galaxies are at $z=2$--4, with apparent 8--1000\,\micron\
  luminosities of $\mu L_{\rm IR}>10^{13}\,{\rm L_{\sun}}$. Thus, if any of
  these sources are not magnified by gravitational lensing then they
  trace some of the most extreme episodes of star formation in the
  Universe.

\item We created a simple statistical model of the gravitational
  lensing of SMGs by a distribution of foreground masses. The model
  reproduces the observed 500-\micron\ number counts in \hermes\ and,
  independent of the analysis of follow-up data, predicts that
  32--74\% of our candidates are strongly gravitationally lensed.

\item The model predicts that the mean magnification of
  strongly-lensed \herschel-selected galaxies is a factor of $\sim7$
  and $\sim15$ for galaxies with $S_{500}=100$ and 200\,mJy,
  respectively.  Gravitationally lensed SMGs are predicted to have
  broad distribution of intrinsic (unlensed) 500-\micron\ flux
  densities with a peak at $\sim5$\,mJy, and 65\% of sources being
  intrinsically fainter than 30\,mJy at 500\,\micron. Thus, samples of
  gravitationally lensed SMGs enables the detailed study of
  high-redshift, star-forming galaxies, $\sim65\%$ of which would
  otherwise be to faint to detect.
 
\end{enumerate}

While this paper is based on existing HerMES data, planned
extragalactic imaging programs with \herschel-SPIRE will cover
$\sim1000\,{\rm deg}^2$ prior to the end of the mission. Thus, with the
method described here, \herschel\ will identify
$\sim150$ lensed galaxy candidates at $z\sim2$--4. With a sample of that
size and a selection function that is easily described, it may be
possible to perform new fundamental cosmological tests
\citep[e.g.][]{Cooray10b}.  
Furthermore, these sources are ideal candidates for high-resolution
follow-up studies with ALMA and other facilities, enabling us to study
the detailed physical conditions in intrinsically faint, star-forming galaxies at
high-redshift. 


\acknowledgments

We thank Elisabetta Valiante for providing the redshift
  distributions from her model and an anonymous referee for thoughtful
  feedback.

SPIRE has been developed by a consortium of institutes led by Cardiff
Univ. (UK) and including: Univ. Lethbridge (Canada); NAOC (China); CEA,
LAM (France); IFSI, Univ. Padua (Italy); IAC (Spain); Stockholm
Observatory (Sweden); Imperial College London, RAL, UCL-MSSL, UKATC,
Univ. Sussex (UK); and Caltech, JPL, NHSC, Univ. Colorado (USA). This
development has been supported by national funding agencies: CSA
(Canada); NAOC (China); CEA, CNES, CNRS (France); ASI (Italy); MCINN
(Spain); SNSB (Sweden); STFC, UKSA (UK); and NASA (USA).

This research has made use of data from the HerMES project
(http://hermes.sussex.ac.uk/). HerMES is a Herschel Key Programme
utilizing Guaranteed Time from the SPIRE instrument team, ESAC
scientists and a mission scientist. HerMES is described in
\citet{Oliver12}.  The data presented in this paper will be released
through the \hermes\ Database in Marseille, HeDaM
(http://hedam.oamp.fr/HerMES/)

JLW, AC, FdB, CF, SK, HF, and JC acknowledge partial support from NSF
CAREER AST-0645427 (to AC at UCI).  SJO, LW, and AS acknowledge
support from the Science and Technology Facilities Council [grant
number ST/F002858/1] and [grant number ST/I000976/1] at U. of Sussex.
AF, GM, LM, and MV were supported by the Italian Space Agency (ASI
``Herschel Science'' Contract I/005/07/0). DR acknowledges support
from NASA through a Spitzer Space Telescope grant.

Support for Program numbers GO-12194 and GO-12488 were provided by
NASA through a grant from the Space Telescope Science Institute, which
is operated by the Association of Universities for Research in
Astronomy, Incorporated, under NASA contract NAS5-26555.

Some of the data presented herein were obtained at the W.M. Keck
Observatory, which is operated as a scientific partnership among the
California Institute of Technology, the University of California and
the National Aeronautics and Space Administration. The Observatory was
made possible by the generous financial support of the W.M. Keck
Foundation. The authors wish to recognize and acknowledge the very
significant cultural role and reverence that the summit of Mauna Kea
has always had within the indigenous Hawaiian community. We are most
fortunate to have the opportunity to conduct observations from this
mountain.

The Submillimeter Array is a joint project between the Smithsonian
Astrophysical Observatory and the Academia Sinica Institute of
Astronomy and Astrophysics and is funded by the Smithsonian Institution
and the Academia Sinica.

Support for CARMA construction was derived from the Gordon and Betty
Moore Foundation, the Kenneth T. and Eileen L. Norris Foundation, the
James S. McDonnell Foundation, the Associates of the California
Institute of Technology, the University of Chicago, the states of
California, Illinois, and Maryland, and the National Science
Foundation. Ongoing CARMA development and operations are supported by
the National Science Foundation under a cooperative agreement, and by
the CARMA partner universities.

The National Radio Astronomy Observatory is a facility of the National
Science Foundation operated under cooperative agreement by Associated
Universities, Inc.

Based on observations carried out with the IRAM Plateau de Bure
Interferometer. IRAM is supported by INSU/CNRS (France), MPG (Germany)
and IGN (Spain).

This research has made use of the NASA/IPAC Extragalactic Database
(NED) which is operated by the Jet Propulsion Laboratory, California
Institute of Technology, under contract with the National Aeronautics
and Space Administration.

Based on observations made with the Gran Telescopio Canarias (GTC), installed in the Spanish Observatorio
del Roque de los Muchachos of the Instituto de Astrofísica de Canarias, in the island of La Palma.
The GTC observations are part of the International Time Programme 2010--2011
(PI: Perez­‐Fournon).

{\it Facilities:} \facility{Herschel (SPIRE)}, \facility{CARMA},
\facility{GBT (Zpectrometer)}, \facility{VLA}, \facility{Keck:II
  (NIRC2)}, \facility{Keck:I (LRIS)}, \facility{SMA},
\facility{IRAM:Interferometer}, \facility{IRAM:30m}, \facility{HST
  (WFC3)}, \facility{Spitzer (IRAC, MIPS)}, \facility{GTC (OSIRIS)}




\appendix

\section{Supplementary sources}
\label{app:sources}

\begin{figure*}
  \centering
  \includegraphics[width=16cm]{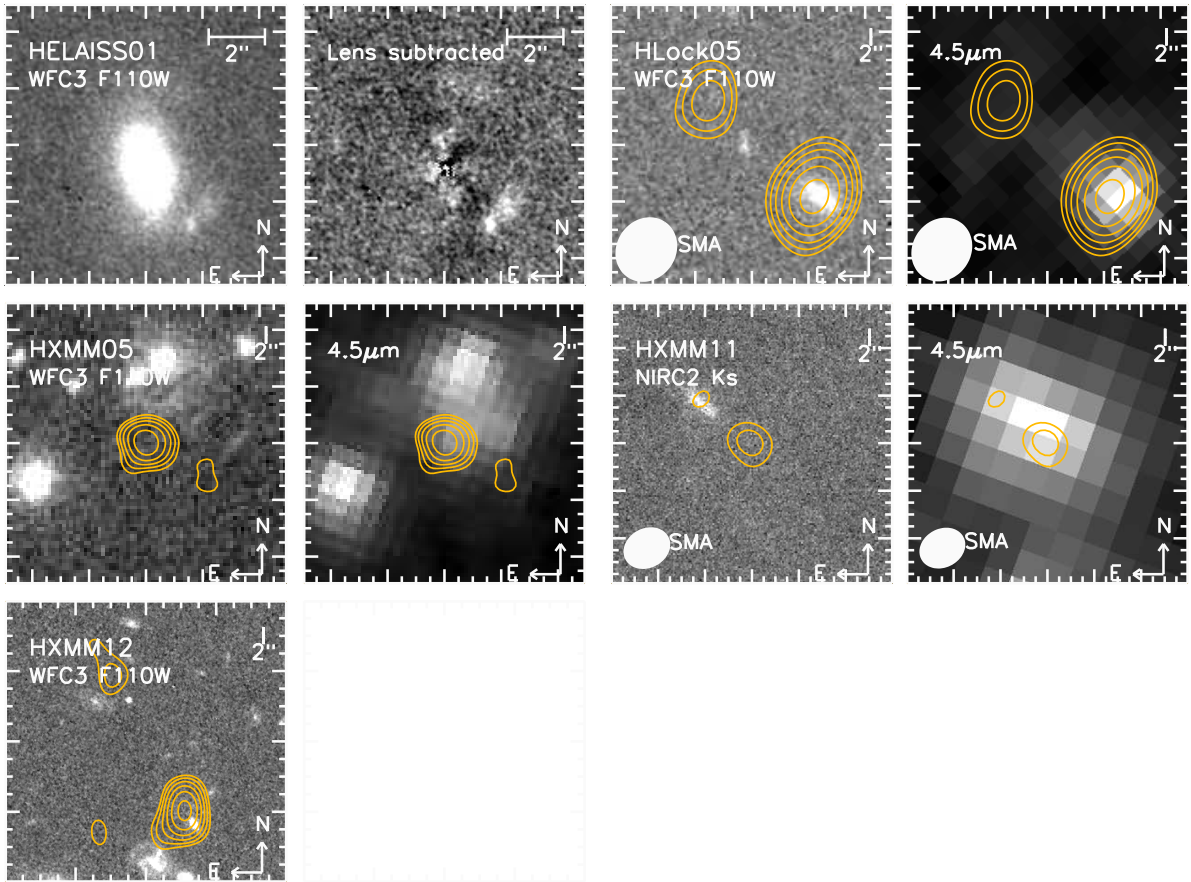}
  \caption{
Images of the five supplementary sources with significant follow-up data (appendix~\ref{app:sources}).
Data and contours are as in Fig.~\ref{fig:postage}.  }
  \label{fig:post_supple}
\end{figure*}

\begin{figure*}
  \centering
  \includegraphics[width=14cm]{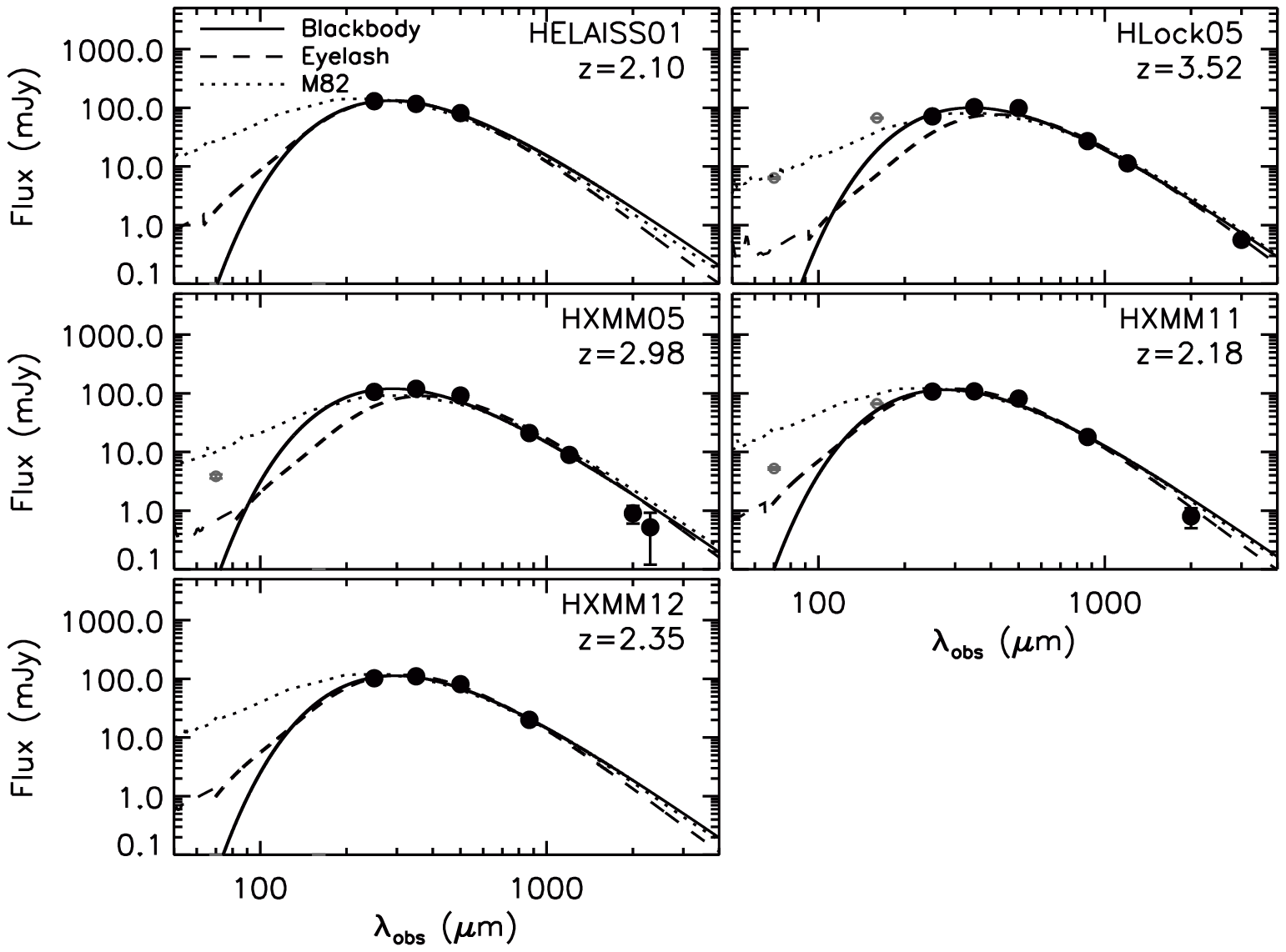}
  \caption{ Observed-frame far-infrared SEDs of the five supplementary sources with significant
    follow-up data (appendix~\ref{app:sources}).  As in
    Fig.~\ref{fig:seds} the far-infrared photometry is from
    \hermes\ SPIRE data and our extensive follow-up programs
    (Section~\ref{sec:followup}) and we show the best-fit
    optically-thin modified blackbody SEDs ($\beta=1.5$), together
    with M82 \citep{Silva98} and the ``Cosmic Eyelash''
    \citep[\eyelash][]{Ivison10b, Swinbank10}.
The 70 and 160\,\micron\ photometry are not used in the fitting.  }
  \label{fig:seds_supple}
\end{figure*}

In Table~\ref{tab:cand} we present 29 supplementary candidate
gravitationally lensed SMGs with $S_{500}=80$--100\,mJy. These 29
sources have lower fidelity than the principal sample of 13 candidates
with $S_{500}\ge100$\,mJy.  Five of the supplementary lensed
candidates have sufficient follow-up data for detailed study. These
five sources are discussed here and images and SEDs are presented in
Figs~\ref{fig:post_supple} and \ref{fig:seds_supple}, respectively.
Of the five supplementary candidates, two are confirmed to be
gravitationally lensed (HELAISS01 and HXMM11), one is intrinsically luminous
(HXMM05), one is likely to be composed of multiple sources blended in
the \herschel\ beam (HXMM12). The final source (HLock05) is likely to
be gravitationally lensed, but we cannot confirm this, or exclude
blending as its origin.

\subsection{HELAISS01}

We have obtained \hst/WFC3 F110W imaging of HELAISS01 (Fig.~\ref{fig:post_supple}), which reveals
the presence of arc-like structures around a foreground
source. No other follow-up data is available and unfortunately,
HELAISS01 is located near the edge of the field and the only archival
data available is SWIRE 70\micron\ imaging in which nothing is
detected. The SPIRE emission from HELAISS01 has a submillimeter
photometric redshift of $z\sim2.10$ (Fig.~\ref{fig:seds_supple}).


\subsection{HLock05}

Contours from SMA compact configuration observations of HLock05 are
shown overlaid on \hst/WFC3 F110W imaging in
Fig.~\ref{fig:post_supple}. Two submillimeter sources are detected at
870\,\micron; one of these sources is coincident with an IRAC and
F110W galaxy; the other is below the detection threshold of both
datasets. There is an optical source between the two SMA peaks, which
may be a foreground lens, although it is faint.  We have obtained
CARMA spectroscopy of HLock05, in which one emission line is clearly
detected, which is most likely \dco\ at $z=3.52$.

The two submillimeter sources identified with the SMA may have
different colors -- the southernmost source is optically bright and
detected by IRAC, but the northernmost sources is undetected in the
F110W data and marginally detected at 4.5\,\micron. If these were two
images of the same lensed source the flux ratios in the SMA data are
such that a detection of the northernmost source is expected in both
the \hst\ and the IRAC data. Due to the depths of the data the
apparent color difference is marginal and insufficient to excluding a
lensing origin for this sources, particularly because differential
magnification \citep[e.g.][]{Fu12}, or a SMG merger may be involved.
We are therefore unable to conclusively determine the nature of
HLock05. It is most likely to be a gravitationally lensed complex
source at $z=3.52$, although it may also be a blend of two
unassociated SMGs.


\subsection{HXMM05}

We have obtained submillimeter interferometry of HXMM05 with the SMA
and high-resolution F110W imaging with \hst/WFC3. As shown
in Fig.~\ref{fig:post_supple}, a single unresolved source is detected
in the SMA data. The submillimeter source is detected in the
IRAC data, and it is offset by $\sim2\arcsec$ from a resolved optical
and near-IR galaxy. The SMG redshift is determined to be
$z=2.985$ with the detection of CO emission lines by CARMA and
PdBI. The redshift of the optical galaxy is unknown, but the
separations between the two sources, and the absence of a lensed SMG
counter-image, is sufficient to rule out significant
magnification. Therefore, we conclude that HXMM05 is a single,
unlensed but intrinsically luminous SMG. It has $L_{\rm
  IR}=(3.2\pm0.4)\times10^{13}\,{\rm L_{\sun}}$ (${\rm
  SFR}\sim5500\,{\rm M_{\sun}yr}^{-1}$; \citealt{Kennicutt98}) and
$T_{\rm D}=45\pm1$\,K, for $\beta=1.5$.


\subsection{HXMM11}

HXMM11 is the only source in the supplementary candidate list for which
we currently have evidence indicating that it is strongly
gravitationally lensed. The configuration is shown in
Fig.~\ref{fig:post_supple} and the SED is in
Fig.~\ref{fig:seds_supple}.
Two faint sources are resolved in SMA interferometric observations, the
fainter of which is coincident with a faint $K_{\rm S}$-detected source. The
combined centroid of the two SMA galaxies is coincident with an IRAC
source, which is most likely the foreground lens. However, the
resolution of IRAC is insufficient for this to interpretation to be
robust. 
\cco\ at $z\sim2.18$ is observed by CARMA  and
corresponding \aco\ is detected by the GBT. 


\subsection{HXMM12}

Two SMGs, separated by $\sim11\arcsec$, are detected in SMA
interferometry of HXMM12 and no central lensing galaxy is visible in
the \hst/WFC3 F110W imaging. On the basis of this spatial configuration
we conclude the HXMM12 is most likely to be a blend of the two SMGs in
the \herschel-SPIRE beam. No spectroscopy is available for this source,
and if it is a blend of multiple SMGs then the submillimeter
photometric redshift is unreliable.


\section{Data tables}
\label{app:tables}

\clearpage
\LongTables
\begin{deluxetable*}{llccccc}
\tabletypesize{\footnotesize}
\setlength\tabcolsep{0.2mm}
\tablecolumns{11}
\tablecaption{Candidate strongly gravitationally lensed SMGs in
  \hermes\ blank-fields
\label{tab:cand}}
\tablehead{\colhead{ } & \colhead{Source} & \colhead{$S_{250}$}
  & \colhead{$S_{350}$} & \colhead{$S_{500}$}
  & \colhead{$z_{\rm CO}$$^a$}  & \colhead{$z_{\rm opt}$$^b$} \\
\colhead{ } &\colhead{ }  & \colhead{(mJy)}       &
\colhead{(mJy)}     &  \colhead{(mJy)}  & 
\colhead{} & \colhead{} }
\startdata
H\bootes01         & 1HERMES S250 J143330.8+345439 & $  158 \pm  6 $ & $  191 \pm  7 $ & $  160 \pm 33 $ & 3.274    & $0.59\pm0.08$  \\
H\bootes02         & 1HERMES S250 J142825.5+345547 & $  159 \pm  6 $ & $  196 \pm  7 $ & $  157 \pm 33 $ & 2.804& 0.414  \\
H\bootes03         & 1HERMES S250 J142823.9+352619 & $  323 \pm  6 $ & $  244 \pm  7 $ & $  140 \pm 33 $ & 1.325& 1.034   \\
HECDFS01           & 1HERMES S250 J033141.6-295317 & $  156 \pm  6 $ & $  156 \pm  4 $ & $  129 \pm  5 $ & \nodata  & 0.222 \\
HECDFS02           & 1HERMES S250 J033732.4-295352 & $  133 \pm  6 $ & $  148 \pm  4 $ & $  122 \pm  5 $ & \nodata  & \nodata \\ 
HECDFS03           & 1HERMES S250 J032443.0-282133 & $   84 \pm  6 $ & $  118 \pm  4 $ & $  114 \pm  5 $ & \nodata  & \nodata \\
HLock01              & 1HERMES S250 J105750.9+573026 & $  403 \pm  7 $ & $  377 \pm 10 $ & $  249 \pm  7 $ & 2.957    & $0.60\pm0.04$ \\ 
HLock02             & 1HERMES S250 J104050.5+560652 & $   53 \pm  7 $ & $  115 \pm 10 $ & $  141 \pm  7 $ & \nodata  & \nodata\\ 
HLock03              & 1HERMES S250 J105712.2+565457 & $  114 \pm  7 $ & $  147 \pm 10 $ & $  114 \pm  8 $ & 2.771$^c$& \nodata \\
HLock04              & 1HERMES S250 J103826.6+581542 & $  191 \pm  7 $ & $  157 \pm 10 $ & $  101 \pm  7 $ & \nodata  & $0.61 \pm 0.02 $    \\
HXMM01                & 1HERMES S250 J022016.5-060143 & $  180 \pm  7 $ & $  192 \pm  8 $ & $  132 \pm  7 $ & 2.307    & $0.6545$    \\ 
                             &                               &                     &                     &                     &          & $0.5018$    \\ 
HXMM02               & 1HERMES S250 J021830.5-053124 & $   92 \pm  7 $ & $  122 \pm  8 $ & $  113 \pm  7 $ & 3.390    &  $1.35$   \\
HXMM03               & 1HERMES S250 J022135.1-062617 & $  121 \pm  7 $ & $  132 \pm  8 $ & $  110 \pm  7 $ & 2.72$^c$ & \nodata \\
\cutinhead{Supplementary sample}     
HADFS01             & 1HERMES S250 J044153.9-540350 & $   80 \pm  6 $ & $  103 \pm  6 $ & $   93 \pm  6 $ & \nodata  & \nodata  \\
HADFS02             & 1HERMES S250 J043619.3-552425 & $  110 \pm  6 $ & $  102 \pm  6 $ & $   87 \pm  5 $ & \nodata  & \nodata  \\
H\bootes04        & 1HERMES S250 J142650.6+332942 & $  142 \pm  6 $ & $  134 \pm  6 $ & $   95 \pm 33 $ & \nodata  & \nodata  \\
H\bootes05       & 1HERMES S250 J144013.0+350825 & $  162 \pm  6 $ & $  128 \pm  6 $ & $   81 \pm 33 $ & \nodata  & $0.24 \pm 0.05 $   \\
HECDFS04          & 1HERMES S250 J033210.8-270535 & $   73 \pm  6 $ & $   86 \pm  4 $ & $   85 \pm  5 $ & \nodata  & \nodata \\
HECDFS05          & 1HERMES S250 J032636.3-270044 & $  155 \pm  6 $ & $  132 \pm  4 $ & $   85 \pm  7 $ & \nodata  & \nodata \\
HECDFS06          & 1HERMES S250 J032603.3-291803 & $   90 \pm  6 $ & $   56 \pm  4 $ & $   82 \pm  5 $ & \nodata  & \nodata \\
HEGS01               & 1HERMES S250 J142201.4+533213 & $   74 \pm  6 $ & $   99 \pm  5 $ & $   90 \pm  6 $ & \nodata  & $0.53 \pm 0.24 $   \\ 
HELAISS01        & 1HERMES S250 J002906.2-421419 & $  129 \pm  6 $ & $  116 \pm  5 $ & $   81 \pm  6 $ & \nodata  & \nodata \\
HFLS01              & 1HERMES S250 J172612.0+583742 & $  108 \pm  7 $ & $  124 \pm  6 $ & $   99 \pm  7 $ & \nodata  & \nodata \\
HFLS02              & 1HERMES S250 J171450.8+592633 & $  164 \pm  7 $ & $  1484 \pm  6 $ & $   87 \pm  6 $ & \nodata  & $0.56 \pm 0.02 $   \\
HFLS03              & 1HERMES S250 J170607.7+590921 & $   98 \pm  7 $ & $  106 \pm  6 $ & $   81 \pm  6 $ & \nodata  & $0.16 \pm 0.17 $   \\
HLock05$^d$     & 1HERMES S250 J103618.3+585456 & $   71 \pm  7 $ & $  102 \pm 10 $ & $   99 \pm  8 $ & 3.520$^c$& $0.49 \pm 0.12 $  \\
HLock06         & 1HERMES S250 J104549.2+574511 & $  136 \pm  7 $ & $  128 \pm 10 $ & $   97 \pm  9 $ & 2.991$^c$& $0.20 \pm 0.02 $    \\
HLock07         & 1HERMES S250 J105007.4+571653 & $   96 \pm  7 $ & $  104 \pm 10 $ & $   87 \pm  8 $ & \nodata  & \nodata \\
HLock08         & 1HERMES S250 J105551.2+592845 & $  142 \pm  7 $ & $  119 \pm 10 $ & $   85 \pm  8 $ & 1.699$^c$& $0.38 \pm 0.11 $    \\ 
HLock09        & 1HERMES S250 J105311.0+564205 & $   50 \pm  7 $ & $   76 \pm 10 $ & $   82 \pm  8 $ & \nodata  & $0.14\pm0.02$ \\
HLock10        & 1HERMES S250 J103217.6+583113 & $   90 \pm  7 $ & $   98 \pm 10 $ & $   81 \pm  8 $ & \nodata  & \nodata \\
HLock11         & 1HERMES S250 J104140.3+570857 & $   98 \pm  7 $ & $  113 \pm 10 $ & $   81 \pm  8 $ & \nodata  & $0.49 \pm 0.12 $    \\
HXMM04          & 1HERMES S250 J022021.7-015328 & $  144 \pm  7 $ & $  137 \pm  8 $ & $   94 \pm 11 $ & \nodata  & $0.21 \pm 0.14 $    \\
HXMM05$^d$       & 1HERMES S250 J022547.8-041750 & $  106 \pm  7 $ & $  119 \pm  8 $ & $   92 \pm  7 $ & 2.985    & \nodata   \\
HXMM06          & 1HERMES S250 J021433.0-041823 & $   93 \pm  7 $ & $  108 \pm  8 $ & $   87 \pm  7 $ & \nodata  & \nodata \\
HXMM07          & 1HERMES S250 J021918.4-031051 & $   91 \pm  7 $ & $  104 \pm  8 $ & $   86 \pm  7 $ & \nodata  & $0.42 \pm 0.07 $    \\ 
HXMM08          & 1HERMES S250 J022626.1-061722 & $   68 \pm  7 $ & $   77 \pm  8 $ & $   85 \pm  7 $ & \nodata  & \nodata \\
HXMM09          & 1HERMES S250 J022029.2-064845 & $  127 \pm  7 $ & $  115 \pm  8 $ & $   84 \pm  7 $ & \nodata  & $0.21 \pm 0.09 $    \\
HXMM10          & 1HERMES S250 J023146.5-035132 & $  142 \pm  7 $ & $  122 \pm  8 $ & $   83 \pm  7 $ & \nodata  & $0.57\pm0.07$  \\
HXMM11$^d$            & 1HERMES S250 J022201.6-033340 & $  107 \pm  7 $ & $  108 \pm  8 $ & $   81 \pm  7 $ & 2.179& \nodata   \\
HXMM12$^d$            & 1HERMES S250 J023006.0-034152 & $  102 \pm  7 $ & $  110 \pm  8 $ & $   81 \pm  7 $ & \nodata  &  \nodata   \\
HXMM13               & 1HERMES S250 J022141.4-070321 & $   76 \pm  7 $ & $   91 \pm  8 $ & $   81 \pm  7 $ & \nodata  & $0.38 \pm 0.08 $  
\enddata                                                                             
\tablecomments{
The primary source sample has $S_{500}>100$\,mJy and is expected to
contain 0--5 interlopers (section~\ref{sec:modpredict}); these sources are discussed in detail in section~\ref{sec:indi}. The supplementary sample consists of fainter
sources ($S_{500}=80$--100\,mJy) and is therefore less reliable. 
$^a$ Redshift of the submillimeter source, measured from the rotational
transitions of CO (Riechers et al., in prep.).
$^b$ Optical redshift, which typically traces the foreground lens.
$^c$ These source redshifts are measured from a single CO emission line
and are guided by the submillimeter photometric redshift.
$^d$ The galaxies are discussed in further detail in appendix~\ref{app:sources}.
}
\end{deluxetable*}

\clearpage
\LongTables
\begin{landscape}
\begin{deluxetable*}{lcccccccccc}
\tablewidth{0pt}
\tabletypesize{\footnotesize}
\setlength\tabcolsep{0.2mm}
\tablecolumns{11}
\tablecaption{Multiwavelength far-infrared and radio photometry of candidate strongly gravitationally lensed SMGs in \hermes 
\label{tab:allwaves}}
\tablehead{\colhead{ } & \colhead{$S_{24\micron}$} &
  \colhead{$S_{70\micron}$} & \colhead{$S_{160\micron}$} & 
  \colhead{$S_{870\micron}$} & \colhead{$S_{1\rm{mm}}$} &
  \colhead{$S_{1.2\rm{mm}}$} & \colhead{$S_{2\rm{mm}}$} & \colhead{$S_{2.3\rm{mm}}$}
  & \colhead{$S_{3\rm{mm}}$} &\colhead {$S_{1.4\rm{GHz}}$}  \\
\colhead{ } & \colhead{(mJy)}       &
\colhead{(mJy)}     &  \colhead{(mJy)}  &  
\colhead{(mJy)}     &  \colhead{(mJy)}  &  
\colhead{(mJy)}     &  \colhead{(mJy)}  & 
\colhead{(mJy)}     &  \colhead{(mJy)}  &
\colhead{(mJy)}}
\startdata  
H\bootes01     & $0.20 \pm 0.01 $ & $ 5.69 \pm 0.31 $  & $ 54.2 \pm 1.2 $ & $61.0\pm3.0$ & \nodata & $26.8\pm1.5$& \nodata & \nodata & $0.72\pm0.14$  & $0.26\pm0.04$ \\ 
H\bootes02     & $0.60 \pm 0.01 $  & $9.16 \pm 0.29 $   & $ 53.9 \pm 1.2 $ & \nodata & \nodata & $22.4\pm1.0$ & \nodata & \nodata & $1.37\pm0.22$  &  $11.7 \pm 0.5 $ \\
H\bootes03     & $0.28 \pm 0.01 $  & $23.1 \pm 0.30 $  & $ 142.3 \pm 1.31 $ & $18.4\pm2.5$ & \nodata & $9.9\pm5.8$ & \nodata & \nodata & \nodata & $0.73\pm0.05$ \\ 
HECDFS01       & $0.58 \pm 0.02$ & \nodata & \nodata  & \nodata & \nodata & \nodata &\nodata  & \nodata &\nodata  & \nodata \\
HLock01$^{a}$    & $ 1.24 \pm 0.02 $  & $ 16.1 \pm 0.24 $ & $ 244.4 \pm 1.39 $ & $52.8\pm0.6$ & $27.5\pm0.6$ & \nodata           & \nodata & \nodata & $0.85\pm0.25$$^a$ &\nodata    \\
HLock03           & $ 0.25 \pm 0.01 $ & $3.86 \pm 0.24 $ & $ 69.6 \pm 1.22 $ & $47.0\pm1.3$       & \nodata           & $17.1\pm1.6$ & \nodata & \nodata & $0.78\pm0.11$        &\nodata     \\
HLock04           & $0.44 \pm 0.01$   & $ 3.25 \pm 0.23 $  & $ 104.6 \pm 1.27$ & $32.1\pm1.5$     & \nodata           & $9.5\pm0.9$   & \nodata & \nodata & $0.34\pm0.11$ &\nodata    \\ 
HXMM01             & $0.57 \pm 0.02 $ & $17.9 \pm 0.03 $ & $88.3 \pm 1.04$  & $25.1\pm1.1$  & \nodata & $11.1\pm0.9$ & $1.1\pm0.3$ & $0.55\pm0.1$ & \nodata     &\nodata    \\ 
HXMM02             &  $0.12 \pm 0.01 $ & $4.79 \pm 0.26 $ & $31.1 \pm 1.14 $ & $72.6\pm2.2$ & $51.9\pm1.2$& $31.3\pm0.8$ & \nodata       & $2.23\pm0.5$& \nodata   & $ 1.2 \pm 0.5$ \\ 
\cutinhead{Supplementary sample}
HADFS01        & $0.81 \pm 0.06$     & $2.97 \pm 0.64 $  & \nodata & \nodata & \nodata & \nodata & \nodata & \nodata & \nodata &\nodata   \\
H\bootes04     & $0.37 \pm 0.01 $   & $6.84 \pm 0.22 $  & $ 57.9 \pm  1.2 $ & \nodata & \nodata & \nodata & \nodata & \nodata & \nodata  & $11.4 \pm 0.5$ \\
HECDFS04      & $ 0.15 \pm 0.01 $ & $3.49  \pm 0.23 $ & $ 23.2  \pm 1.3 $ & \nodata & \nodata & \nodata & \nodata & \nodata  & \nodata &\nodata \\
HECDFS05      & \nodata &\nodata&\nodata&\nodata&\nodata&\nodata&\nodata&\nodata&\nodata & $ 1.2 \pm 0.3 $ \\
HEGS01           & $0.40 \pm 0.01 $   & \nodata & \nodata &  \nodata & \nodata & \nodata & \nodata & \nodata & \nodata &\nodata \\
HFLS01            & $0.88 \pm 0.06 $  & $ 5.38 \pm 0.4 $ & \nodata & \nodata & \nodata & \nodata & \nodata & \nodata & \nodata   & $1.8 \pm 0.5$ \\ 
HFLS02            & $1.05 \pm 0.06 $  & $ 7.13 \pm 0.46 $ & \nodata & \nodata & \nodata & \nodata & \nodata & \nodata & \nodata  &\nodata \\ 
HFLS03            & \nodata &\nodata&\nodata&\nodata&\nodata&\nodata&\nodata&\nodata&\nodata & $ 0.8 \pm 0.5 $  \\
HLock05$^b$      & $ 0.31 \pm 0.01 $ & $6.36 \pm 0.23 $ & $ 67.0 \pm 1.23 $ & $26.9\pm0.93$        & \nodata           & $11.3\pm0.9$ & \nodata & \nodata & $0.56\pm0.11$      &\nodata      \\
HLock06           & $ 0.66 \pm 0.02 $ & $22.6 \pm 0.25 $ & $130.0 \pm 1.30 $ & \nodata            & \nodata           & \nodata           & \nodata & \nodata & $0.57\pm0.15$ &\nodata    \\
HLock07           & $0.25 \pm 0.01 $ & \nodata & $54.8 \pm 1.23  $  & \nodata       & \nodata           & $5.4\pm0.8$   & \nodata & \nodata & \nodata      &\nodata    \\   
HLock08           & \nodata           & $35.6 \pm 0.32 $ & \nodata  & \nodata            & \nodata           & \nodata           & \nodata & \nodata & \nodata         &\nodata  \\
HLock11           & $ 0.26 \pm 0.01 $ & $6.02 \pm 0.24 $ & $ 41.3 \pm 1.25 $ & \nodata            & \nodata           & \nodata           & \nodata & \nodata & \nodata    &\nodata     \\
HXMM05$^b$      & $0.37 \pm 0.01 $ & $3.85 \pm 0.29 $ & \nodata  & $21\pm1.0$  & \nodata  & $8.9\pm0.9$   & $0.9\pm0.3$ & $0.52\pm0.4$ & \nodata   &\nodata    \\ 
HXMM07         & $0.46 \pm 0.01 $ & $4.60 \pm 0.23 $ & $28.85 \pm 0.99 $  & \nodata & \nodata & \nodata & \nodata & \nodata & \nodata    &\nodata    \\ 
HXMM09         & \nodata &\nodata&\nodata&\nodata&\nodata&\nodata&\nodata&\nodata&\nodata & $ 1.0 \pm 0.4 $  \\
HXMM11$^b$      & $0.39 \pm 0.01 $ & $5.22 \pm 0.23 $ & $66.2 \pm 1.03$ &  $18\pm1.0$  & \nodata & \nodata           & $0.8\pm0.3$ & \nodata & \nodata    &\nodata    \\ 
HXMM12$^b$      & \nodata & \nodata & \nodata  &  $20\pm1.0$  & \nodata & \nodata           & \nodata         & \nodata & \nodata  &  $7.4 \pm 0.5$ 
\enddata                                                                             
\tablecomments{
The primary source sample has $S_{500}>100$\,mJy and is expected to
contain 0--5 interlopers; these sources are discussed in detail in section~\ref{sec:indi}. The supplementary sample consists of fainter
sources ($S_{500}=80$--100\,mJy) and is therefore less reliable. Sources that are unobserved or undetected at all of these
wavelengths are not included in this table. 
$^a$ The flux for HLock01 is at 3.4mm, not 3mm as for the other sources. 
$^b$ Theses sources are discussed in further detail in appendix~\ref{app:sources}.
}
\end{deluxetable*}


\clearpage
\end{landscape}


\end{document}